\documentclass[apj,twocolumn,twocolappendix,numberedappendix]{openjournal}
\usepackage{newtxtext,newtxmath,enumitem,multirow,ctable,verbatim}
\usepackage[T1]{fontenc}
\usepackage[dvipsnames]{xcolor}
\usepackage[breaklinks,colorlinks,citecolor=blue,urlcolor=blue]{hyperref}

\newcommand{\Alf}{{Alfv\'en}}
\newcommand{\bhat}{\hat{\bf b}}

\newcommand{\paperone}{Paper {\small I}}
\newcommand{\papertwo}{Paper {\small II}}
\newcommand{\paperthree}{Paper {\small III}}

\newcommand{\orcidauthor}[3]{\author{\href{http://orcid.org/#1}{#2$^{#3}$}}}

\shorttitle{CR X-rays and Radio in Perseus}
\shortauthors{Hopkins et al.}

\begin{document}

\title{\vspace{-0.8cm}An Inverse-Compton-Boosted Cool Core Unifies Perseus's Radio and X-ray Halos\vspace{-1.5cm}}

\orcidauthor{0000-0003-3729-1684}{Philip F. Hopkins}{1*}
\orcidauthor{0000-0002-1616-5649}{Emily M. Silich}{1}
\orcidauthor{0000-0002-8213-3784}{Jack Sayers}{1}
\orcidauthor{0000-0002-7484-2695}{Sam B. Ponnada}{2}
\orcidauthor{0000-0002-1159-4882}{Isabel S. Sands}{1}
\affiliation{$^{1}$Division of Physics, Mathematics, and Astronomy, California Institute of Technology, Pasadena, CA 91125, USA}
\affiliation{$^{2}$Department of Physics and Astronomy,
Chalmers University of Technology,
SE-412 96 Gothenburg, Sweden}

\thanks{$^*$E-mail: \href{mailto:phopkins@caltech.edu}{phopkins@caltech.edu}},

\begin{abstract}
Perseus is the brightest X-ray strong cool-core (SCC) cluster, with a bright central radio and $\gamma$-ray source as well as extended low-frequency radio mini and giant halos. It is also the archetype of the cooling flow (CF) problem with an X-ray luminosity implying naive mass cooling rates orders-of-magnitude larger than observed cold gas masses or star formation rates. 
In \citet{hopkins:2025.cr.ic.clusters.ideapaper}, we suggested that ancient ($\gtrsim$\,Gyr-old) cosmic ray (CR) halos (ACRHs), injected by the central source, would produce thermal-like soft X-ray inverse-Compton (CR-IC) emission ``boosting'' the CC and alleviating the CF problem. 
We apply the same analytic models to Perseus, and show that a simple model of CRs injected by NGC 1275 (+satellites) simultaneously accounts for the ``excess'' CF luminosity and radio minihalo (+giant halo). The models are consistent with Perseus's soft X-ray surface brightness and X-ray inferred density, temperature, pressure, entropy, metallicity, cooling time, and mass deposition rates; $\gamma$-ray spectra; spatially-resolved hard X-ray spectra; and radio surface brightness and spectral index data, from kpc-Mpc radii. These also reproduce independent constraints on the magnetic field strengths and mass/potential models of the cluster.
The evolution of the minihalo spectral index and surface brightness are predicted by an aging population of CRs which dominates the apparent SCC luminosity via CR-IC, and match well the observed hard X-ray slopes. Furthermore, the (more extended) ``giant'' low-frequency halo properties can be predicted by the sum of ACRHs around the lower-luminosity radio galaxies distributed throughout the cluster, which dominate the diffuse synchrotron at $\gtrsim 100\,$kpc.
Re-acceleration is neither needed nor particularly important in these models, and the implied streaming/advection/diffusion speeds are consistent with buoyant advection.
Previous claims of upper limits to non-thermal X-rays and CR pressure relied on strong assumptions which are not valid at the CR energies of interest, e.g.\ a  power-law spectrum of CRs. 
This could resolve many historical puzzles about Perseus, and makes new predictions for future observations.
\end{abstract}


\maketitle

\section{Introduction}
\label{sec:intro}

``Strong cool core'' (SCC) galaxy clusters are an important laboratory and testbed for our understanding of ICM/CGM plasma physics, galaxy formation, and AGN or supermassive black hole (SMBH) ``feedback'' potentially regulating galaxy quenching \citep[e.g.][]{mcnamara:2007.agn.cooling.flow.review.cavity.jet.power.vs.xray.luminosity.scalings.emph.compilation}. Perseus is the brightest and nearest X-ray SCC, both the archetype or prototypical example of the class, and in many ways the best-studied and constrained example of a SCC. As such, it presents many theoretical and observational puzzles which have still defied consensus. 

Like almost all SCCs, it embodies the ``cooling flow'' (CF) problem: the apparent X-ray cooling luminosity, if all the X-ray emission is interpreted as thermal emission, is in excess of $10^{45}\,{\rm erg\,s^{-1}}$, and this implies a mass deposition rate of $\sim 1000\,{\rm M_{\odot}\,yr^{-1}}$ rapidly cooling to much lower temperatures and losing pressure support within the SCC radius of $\lesssim 100\,$kpc. But, the orders-of-magnitude smaller observed mass of cooler gas (as well as star formation and other constraints) imply physical cooling/mass deposition rates two orders of magnitude smaller \citep[e.g.][]{fabian:1994.cluster.cooling.flows.review,fabian:2002.classical.cooling.flow.problem.obs.definition.missing.lum.profile.of.mdot.vs.r.wrong}. 

This discrepancy is generally attributed to AGN feedback: the BCG in Perseus, NGC 1275 (3C 84), is also an extremely luminous radio and $\gamma$-ray source \citep[see e.g.\ the recent literature compilation in][]{sinitsyna:2025.ngc.1275.radio.through.gamma.ray.compilation}, with inferred CR lepton production/acceleration rate $\gtrsim 10^{45}\,{\rm erg\,s^{-1}}$ \citep{bottcher:2013.blazar.modeling.almost.all.blazars.better.fit.by.leptonic.cr.models.not.hadronic,tavecchio:2014.jet.leptonic.luminosity.ngc.1275.2e45,keenan:2021.jet.leptonic.power.1e41to1e45.easily.produced.from.modest.agn.bursts.or.steady.jets,hodgson:2021.perseus.jet.minimum.kinetic.luminosity.gamma.rays,foschini:2024.blazar.agn.jet.power.favor.leptonic.large.power.energy.much.more.than.kinetic.lobe.cavity.power} within the compact radio core ($\ll$\,kpc). It has long been noted that energetically, the AGN appears suspiciously similar to the X-ray cooling luminosity, and the most popular interpretation is that the system is self-regulating with $\sim 99\%$ of the cooling being offset by the AGN in quasi-steady-state. But it has proven challenging to understand how the energy from the AGN is coupled and how it could so precisely offset the cooling (with so little fractional imbalance), especially given the enormous mismatch in scales of SMBH accretion versus SCC, the fact that cooling appears to be highly clumpy and inhomogeneous and locally unstable, and the fact that the jets and cavities around NGC 1275 are anisotropic and non-volume filling \citep[e.g.][and references therein]{beckmann:2019.cluster.molecular.gas.sims.produce.too.clumpy.if.all.mhd.wrong.detailed.properties}. 

Meanwhile, it is also known that Perseus exhibits a radio ``mini-halo,'' like many other SCCs \citep{gitti:2002.perseus.minihalo.xray.inverse.compton}. Recent observations with LOFAR have shown that this minihalo is fundamentally a low-frequency system, which transitions from a flat-spectrum to ultra-steep source with increasing distance from NGC 1275 out to the SCC radius, but also that it appears to be embedded in a larger, much lower surface-brightness low-frequency giant radio halo extending to $\sim$\,Mpc \citep{vanweeren:2024.perseus.giant.radio.halo.filled.electrons.just.tiny.fraction.high.energy,groeneveld:2026.investigatingradioemissionperseus}. Models for these minihalos have struggled to explain their origins and power sources, as well as to reconcile them with apparent X-ray properties that often seem at odds \citep[e.g.][]{zuhone:2013.turbulent.reaccel.secondary.electrons.as.potential.minihalo.explanation.radio.emission,brunetti.jones:2014.cr.diffusion.clusters.minihalo.giant.halo.review.many.solutions.including.faster.transport.and.local.sources,gitti:2016.radio.minihalos.coolcore.clusters.candidates.review,bravi:2016.minihalo.luminosity.strong.corr.xray.luminosity.clusters,richard.laferriere:2020.minihalos.agn.closely.correlated.agn.energize.minihalos.produce.them.but.shortlived,giancintucci:2024.example.large.radio.minihalo.related.to.icm.sloshing.potential.disturbance.reenergizing,keshet:2024.radio.minihalos.relics.unified.with.steepening.spectra.models.for.vstream.300.to.3000.kms.model}. 

Recently, \citet{hopkins:2025.cr.ic.clusters.ideapaper} suggested that many of these apparent puzzles in SCCs could be explained if ``ancient'' cosmic ray halos (ACRHs) accelerated by the central radio source and streaming outwards contribute non-negligibly to the X-ray emission from the SCC via inverse-Compton (IC) scattering cosmic microwave background (CMB) photons. \citet[][henceforth \paperone]{hopkins:2025.crs.inverse.compton.cgm.explain.erosita.soft.xray.halos} emphasized that as CRs stream from a central source, losses to IC, synchrotron, bremsstrahlung and Coulomb+ionization processes produce a CR spectrum with substantial curvature (not a power-law) regardless of the injection spectrum, and this in turn produces soft X-ray IC scattering from the leptons which contain most of the CR energy ($\sim 0.1-1\,$GeV) to soft X-ray energies $\sim$\,keV with a remarkably thermal-like spectrum and slowly-falling surface brightness (SB) profile out to $\sim 100\,$kpc radii (where losses would truncate the emission). While \paperone\ focused on this CR-IC around Milky Way and M31-mass galaxies (where it would resemble relatively ``hot'' gas at relatively large radii, compared to the low-mass-halo virial temperature or radius), \citet[][\papertwo]{hopkins:2025.cr.ic.clusters.ideapaper} noted that in clusters, this emission would resemble relatively ``cool'' gas in the cluster centers, i.e.\ CCs. There and in more detailed modeling in \citet[][\paperthree]{hopkins:2025.cr.ic.clusters.detailedobs}, it was noted that the SB profile and effective CR-IC X-ray temperatures strongly resemble the profiles of SCCs, that this makes unique predictions in the Sunyaev-Zeldovich effect in SCC centers for which there appears to be tentative evidence already \citep{romero:2017.cluster.pressure.profiles.highres.sz.xray.cool.cores.show.central.pressure.deficit}, and that this provides an immediate and almost-trivial explanation for the close correlation between apparent AGN/radio/jet/cavity power and X-ray-inferred ``cooling luminosity'' or mass deposition rate, as the two simply trace the same injected lepton population. In \paperthree, a broader range of properties and correlations of the SCC population as a whole were explored, and it was shown that these can naturally be reproduced by these ACRHs, while corresponding predictions were made for radio and $\gamma$-ray observations. Because of the same loss effects, the radio halos corresponding to the same ACRHs tend to be low-frequency, with their compact cores corresponding to recently-observed low-frequency LOFAR sources, but often (for more distant clusters) these are restricted to only the brightest emission regions. In Perseus, the combination of its extreme luminosity and close proximity means that these models can be put to the test far more rigorously, as observations have detected diffuse, low-frequency radio emission out to $\sim$\,Mpc scales, while strong constraints on diffuse $\gamma$-ray and very hard X-ray emission exist across a range of energies. 

In this paper, we therefore consider Perseus as a case study of the models in \paperone-\paperthree. Specifically we apply these models to the Perseus cluster, attempting to vary the more uncertain input assumptions, to explore whether the predictions above for the SCC population in ACRH models can be applied to a case with much more complete and deep multi-wavelength observations. We consider predictions for the spectra from $10^{6}-10^{26}$\,Hz, at cluster-centric distances from kpc through Mpc, to ask whether a simple ACRH model can indeed explain this wealth of constraints and observed phenomenology. We find that not only can these simple ACRH models explain the X-ray apparent SCC and cooling flow, but simultaneously that they provide a new interpretation of radio minihalos and giant halos in relaxed SCC systems.

In \S~\ref{sec:model} we describe the model and equations integrated (\S~\ref{sec:equations}) and assumptions and variations considered (\S~\ref{sec:variations}). \S~\ref{sec:results} shows the resulting predictions, for CR and pan-wavelength emission (\S~\ref{sec:basics});  soft X-rays (\S~\ref{sec:obs.xr}) including radial profiles of X-ray inferred properties (\S~\ref{sec:obs.xr:soft}), dependence on model assumptions (\S~\ref{sec:obs.xr:var}) and central metallicity discrepancies (\S~\ref{sec:obs.xr:z}); harder emission (\S~\ref{sec:obs.hard}) in X-rays (\S~\ref{sec:obs.hard:xr}) and $\gamma$-rays (\S~\ref{sec:obs.hard:gamma}); infrared/optical/UV emission (\S~\ref{sec:obs.uvoir}); and radio (\S~\ref{sec:obs.radio}) including the mini-halo (\S~\ref{sec:obs.radio:minihalo}) and giant halo (\S~\ref{sec:obs.radio:gianthalo}) and independent magnetic field constraints (\S~\ref{sec:B}); implications for ``non-thermal pressure'' constraints and mass models (\S~\ref{sec:mass}) including constraints that do not apply to the models here (\S~\ref{sec:mass:unconstrained}) and those that do including mass/potential reconstruction (\S~\ref{sec:mass:mass}) and SZ (\S~\ref{sec:sz}). In \S~\ref{sec:differences} we discuss key differences from previous models and assumptions in improved data (\S~\ref{sec:differences:data}), why reacceleration plays a minimal role in these models (\S~\ref{sec:differences:reaccel}) and how this compares to previous claimed constraints on CR-IC from hard X-rays (\S~\ref{sec:differences:cric}) or CR pressure constraints from $\gamma$-rays (\S~\ref{sec:differences:gamma}). We summarize and conclude in \S~\ref{sec:conclusions}.

\section{Model and Assumptions}
\label{sec:model}

We consider a simple model as defined in \papertwo, for the emission from Perseus. Specifically, we assume a spherically symmetric cluster with a specified gas density, temperature, and metallicity profile, with a central source (NGC 1275) injecting CRs, which diffuse and stream outwards suffering various losses. We then calculate the total emission including thermal+CR processes. 

\subsection{Equations Integrated}
\label{sec:equations}

For the CRs, we begin from the usual diffusion-advection (single-moment or Fokker-Planck-like) equation for the CR distribution function $f = dN/dp_{\rm cr}$ in CR momentum $p_{\rm cr}$ (for ultra-relativistic CRs, kinetic energy $E_{\rm cr}=p_{\rm cr}\,c$), assuming local flux equilibrium and a close-to-isotropic pitch angle distribution \citep{hopkins:m1.cr.closure,thomas:2021.compare.cr.closures.from.prev.papers}: 
\begin{align}
\label{eqn:onemoment.full} \frac{\partial f}{\partial t} &=  j_{0} + \nabla \cdot  \left[ \kappa_{\|} \bhat \bhat \cdot \nabla f - {\bf v}_{e} f \right]
 + 
\\
\nonumber &  
\frac{\partial }{\partial p_{\rm cr}}\left[ p_{\rm cr} \left( \mathcal{R} 
+ \frac{\nabla \cdot {\bf v}_{e}}{3} \right) f 
+ \frac{(v_{A}^{2} - \bar{v}_{A}^{2})}{9 \kappa_{\|}}   \frac{p_{\rm cr}^{4}\, \partial (f/p_{\rm cr}^{2})}{\partial {p_{\rm cr}}}
\right] \ , 
\end{align}
where $j_{0}$ represents sources/sinks, $\kappa_{\|}\equiv v_{\rm cr}^{2}/3\nu_{\rm cr}$ diffusion (in terms of CR velocity $v_{\rm cr} \sim c$ and mean scattering rate $\nu_{\rm cr}$), ${\bf v}_{e}\equiv \bar{v}_{A}\,\bhat + {\bf u}$ effective streaming (with \Alf{ic} streaming $\bar{v}_{A} \equiv (\nu_{+}-\nu_{-})/(\nu_{+}+\nu_{-})\,v_{A}$ defined in terms of the \Alf\ speed and scattering rate from forward(+) or backward(-) propagating modes, plus gas velocity ${\bf u}$), and $\mathcal{R}$ continuous/radiative losses.
This gives the classical ``convective'' (${\bf u}$), ``streaming'' ($\bar{v}_{A} \bhat$), and ``diffusive'' ($\kappa_{\|}$)-like flux terms , as well as the  ``adiabatic'' ($\nabla \cdot {\bf u}$), ``streaming loss'' ($\nabla \cdot \bar{v}_{A}\bhat$), and ``turbulent reacceleration'' ($(v_{A}^{2}-\bar{v}_{A}^{2})/9\kappa_{\|}$) loss/gain terms. We will simplify by assuming steady-state, spherical symmetry, $\bar{v}_{A} \approx v_{A}$, neglecting large compressive motions ($\nabla \cdot {\bf u}_{\rm gas}$ small), giving: 
\begin{align}
\label{eqn:onemoment} 0 &=  \frac{1}{r^{2}} \frac{\partial }{\partial r}  \left[ r^{2} \left( \kappa \,\frac{ \partial f}{\partial r} - v_{\rm st} f \right)\right]
 + \frac{\partial (p \mathcal{R}^{\prime} f ) }{\partial p} + S(r,p) \ .
\end{align}
Here $S(r,\,p)$ represents injection from the central source (NGC 1275); $\kappa$ is the effective diffusivity and $v_{\rm st}$ the effective ``streaming speed'' (which can represent \Alf{ic} or diffusive-like ``super-\Alf{ic}'' streaming, or super-diffusion, or advection/convective transport; \citealt{Zwei13,zweibel:cr.feedback.review,wiener:cr.supersonic.streaming.deriv,thomas.pfrommer.18:alfven.reg.cr.transport,ruszkowski.pfrommer:cr.review.broad.cr.physics,liang:2025.leaky.boxes.levy.flights.modeling.crs.transport}); and $\mathcal{R}^{\prime}$ represents radiative losses (combined with the streaming/turbulent reacceleration terms), for which we include all the processes tabulated in \citet{hopkins:cr.multibin.mw.comparison} (bremstrahhlung, ionization, Coulomb, inverse Compton, synchrotron, and annihilation). Most important are bremstrahhlung, Coulomb, inverse Compton, and synchrotron, with loss rates and spectra from each following \citet{1965AnAp...28..171G,blumenthal:1970.cr.loss.processes.leptons.dilute.gases,1972Phy....60..145G,ginzburg:1979.book,rybicki.lightman:1979.book}. Note the expressions we adopt include the full semi-relativistic and ultra-relativistic corrections, as well as Klein-Nishina terms, and self-absorption corrections for the emergent spectra. Once we specify $\kappa$, $v_{\rm st}$, $S(r,\,p)$, and $\mathcal{R}(p,\,r)$, Eq.~\ref{eqn:onemoment} is straightforward to numerically integrate.

\subsection{Default Assumptions \&\ Variations}
\label{sec:variations}

We assume the cluster has gas whose volume-weighted mean density $n(r)$, metallicity $Z(r)$, and volume-filling phase temperature $T(r)$ (in the absence of CRs), and magnetic field strength $B(r)$, are given according to the empirical functions below. We also assume cluster mass (virial mass $1.16 \times 10^{15}\,{\rm M_{\odot}}$ using the \citealt{bryan.norman:1998.mvir.definition} definition) and mass profile motivated by observations (\S~\ref{sec:mass:mass} below). We assume Solar abundance ratios and that the gas is fully-ionized throughout. The only other quantity needed is the radiation spectrum $ I_{\nu}(r,\,\nu)$, which we assume is given by the sum of the $z=0$ meta-galactic background compilation in \citet{cooray:2016.extragalactic.background.light.compilations.review,khaire:2019.extragalactic.background.light.spectra}, plus all the radiation field from the cluster itself. We add the cluster optical+UV (starlight+AGN) background following the light profile observed and compiled in \S~\ref{sec:obs.uvoir}, and compute the cluster gas thermal emission using APEC \citep{smith:2001.apec.methods,foster:2012.apec}, at each radius $r$, and include the CR emission itself from wavelengths $\sim 10^{3}-10^{28}$\,Hz. Because the CR losses and emergent spectra depend not just on this background, but on the CR emission itself (i.e.\ CRs can inverse-Compton scatter their own emitted radiation), we solve Eq.~\ref{eqn:onemoment} and the emergent spectra iteratively: first taking the emergent $I_{\nu}$ ignoring CR emission, then updating $I_{\nu}$ with the CR emission, plugging that back in to re-solve Eq.~\ref{eqn:onemoment} and update $I_{\nu}$, etc. In practice the effects on the CR spectra from this are very weak since the IC losses are dominated by the CMB at the radii of interest. But this does have some non-negligible effects on the predicted emergent spectra, from IC-scattering of higher-energy photons (e.g.\ up-scattering some CR-IC X-rays to $\gamma$-rays). 

We systematically vary a number of uncertain assumptions within the model, in order to explore which observations are more or less sensitive to those variations. This includes:

\begin{enumerate}

\item CR transport parameters (normalization). Specifically we consider models for $\kappa$ and $v_{\rm st}$ in Eq.~\ref{eqn:onemoment} normalized to 
(a) $(\kappa/{\rm kpc^{2}\,Myr^{-1}},\,v_{\rm st}/{\rm km\,s^{-1}}) = (2,\,40)$; (b) $=(0.3,\,100)$; (c) $=(3,\,20)$, 
at ridigity $R_{\rm cr} = 1$\,GV. This allows for a range relative diffusion-vs.-streaming and values motivated by CGM constraints around lower-mass galaxies \citep{karwin:2019.fermi.m31.outer.halo.detection,recchia:2021.gamma.ray.fermi.halos.around.m31.modeling,butsky:2022.cr.kappa.lower.limits.cgm,hopkins:2025.crs.inverse.compton.cgm.explain.erosita.soft.xray.halos}. 

\item CR transport energy/rigidity dependence. We assume $\kappa$ and $v_{\rm st}$ scale relative to their $1\,$GV value with: 
(a) $\propto 1 + (E_{\rm cr}/{\rm GeV})^{0.5}$, motivated by LISM CR transport models \citep{korsmeier:2022.cr.fitting.update.ams02,hopkins:cr.multibin.mw.comparison,dimauro:2023.cr.diff.constraints.updated.galprop.very.similar.our.models.but.lots.of.interp.re.selfconfinement.that.doesnt.mathematically.work,tovar:2024.inhomogeneous.diffusion.cr.spectra,silver:2024.cr.propagation.low.energies.new.data,recchia:2024.cr.spectral.modeling.features.bumps.wiggles,delatorre.luque:2024.gas.models.of.galaxy.key.for.scale.height.but.need.halo.for.crs,ramirez:2024.3d.struct.galaxy.gas.influences.cr.fitting.params.vs.2d}, 
(b) $\propto E_{\rm cr}^{0}$ (energy-independent), as in many streaming models \citep{wiener:cr.supersonic.streaming.deriv,thomas.pfrommer.18:alfven.reg.cr.transport,hopkins:cr.transport.constraints.from.galaxies,hopkins:2021.sc.et.models.incompatible.obs,ponnada:2023.synch.signatures.of.cr.transport.models.fire,fitzaxen:2024.cr.transport.into.gmcs.suppressed.starforge,sike:2024.cr.winds.pfrommer.model.launch.warm.gas,weber:2025.cr.thermal.instab.cgm.fx.dept.transport.like.butsky.study}, and 
(c) $\propto E_{\rm cr}^{1/3}$, in-between.

\item Injection spectrum/hardness. We define $S(r,\,p) \equiv \dot{E}_{\rm cr,\,\ell} \,S^{\prime}_{r}(r)\,S^{\prime}_{p}(p)$, where $\int_{0}^{\infty} 4\pi r^{2} S^{\prime}_{r} {\rm d} r = 1$ and $\int_{0}^{\infty} S^{\prime}_{p}\,{\rm d}p = 1$ so that $\dot{E}_{\rm cr,\,\ell}$ is the CR lepton injection rate from the source (acceleration/injection region, i.e.\ anything not diffuse gas), $S^{\prime}_{r}$ gives the spatial distribution and $S^{\prime}_{p}$ the momentum distribution (injection spectrum). We consider: 
(a) a power-law like $S^{\prime}_{p} \propto R_{\rm cr}^{a_{1}} / (1 + (R_{\rm cr}/R_{\rm cr,\,0})^{a_{2}})$ with $(a_{1},\,a_{2},\,R_{\rm cr,\,0}/{\rm GV}) = (1/2,\,2,\,3)$ (equivalent to $d \dot{N}_{\rm cr}/d E_{\rm cr} \propto E_{\rm cr}^{-\delta}$ with $\delta=1+a_{2}-a_{1} = 5/2$ at high energies, or a shallow synchrotron index $\alpha \sim 0.7$ at injection, with a weak cutoff below $R_{\rm cr,\,0}$ to prevent a low-energy divergence); 
(b) a fitted $S^{\prime}_{p}$ to observed LISM CR spectra (the result of the injection region itself having injection and losses in some quasi-steady-state), using the $e^{-}$ LISM (Solar-modulation-corrected) fit from \citet{bisschoff:2019.lism.cr.spectra}, roughly $(a_{1},\,a_{2},\,R_{\rm cr,\,0,\,GV})\sim(0,\,2.43,\,0.6)$; and
(c) an even harder ``pure acceleration'' (DSA-like) power-law $(a_{1},\,a_{2},\,R_{\rm cr,\,0,\,GV})\sim(0,\,2.2,\,10)$ \citep[e.g.][]{malkov:2001.shock.cr.dsa}. 
We have also considered a number of other spectra ``in between'' these choices but they make very little difference, as we show below.

\item Injection spatial distribution. We consider: 
(a) a point-like injection $S^{\prime}_{r} \propto \delta(r)$ (appropriate for any injection region smaller than $\ll1\,$kpc (given the scales we model here); as in e.g.\ \citealt{tavecchio:2014.jet.leptonic.luminosity.ngc.1275.2e45}); or 
(b) an extended injection region with $S^{\prime}_{r} \propto 1 - (R/R_{0})^{3/2}$ for $R \le R_{0}$, with $R_{0} \sim 10\,$kpc, such that most of the injection ($\int 4\pi r^{2}\,S(r)$) occurs around $R_{0}$, appropriate for e.g.\ injection in the extended but active 3C 84 jet termination shocks \citep{su:2025.crs.at.shock.fronts.from.jets.injection}. We have also varied the form of the ``cutoff'' or tapering function, to little effect.

\item Injection luminosity. We consider $\int S(p,\,r)\,dp\,4\pi r^{2}\,dr = \dot{E}_{\rm cr,\,\ell} = \dot{E}_{45}\,10^{45}\,{\rm erg\,s^{-1}}$ with 
(a) $\dot{E}_{45}=2$, 
(b) $\dot{E}_{45}=1$, 
and 
(c) $\dot{E}_{45}=3$, motivated by detailed models for the NGC 1275 AGN and compact radio source ($\ll$\,kpc; \citealt{bottcher:2013.blazar.modeling.almost.all.blazars.better.fit.by.leptonic.cr.models.not.hadronic,tavecchio:2014.jet.leptonic.luminosity.ngc.1275.2e45,keenan:2021.jet.leptonic.power.1e41to1e45.easily.produced.from.modest.agn.bursts.or.steady.jets,hodgson:2021.perseus.jet.minimum.kinetic.luminosity.gamma.rays,foschini:2024.blazar.agn.jet.power.favor.leptonic.large.power.energy.much.more.than.kinetic.lobe.cavity.power}). 

\item Background magnetic field strengths. For synchrotron, we need a model for the average field strength $B = \langle |{\bf B}| \rangle_{\rm Vol}(r)$ in the diffuse, volume-filling gas (not dense regions, satellite galaxies, the jet or shocks) at a given radius. Motivated by the observational constraints compiled in \S~\ref{sec:B}, we consider: 
(a) a power-law $B \propto r^{-1/2}$, with a floor, so $B^{2} = B_{1}^{2}\,(R/{\rm kpc})^{-1/2} + B_{0}^{2}$, with fiducial $(B_{1},\,B_{0})=(5,\,1)\,{\rm \mu G}$, (but both varied by a factor of $\sim 3$ in either direction); 
(b) a constant plasma $\beta \equiv c_{s}^{2}/v_{A}^{2} = 100$ (also varied from $=30-200$, the approximate range of our fiducial model), or $\beta=100$ added in quadrature with a ``floor'' at $B_{0} =1\,{\rm \mu G}$; or 
(c) constant $B$ varied from $0.7-2\,{\rm \mu G}$.

\item True gas density. CR-IC causes the gas density in the CC center to be over-estimated, so we must assume a true volume-filling hot/diffuse gas $n(r)$, for which we consider: 
 (a) a typical non-CC (NCC) profile as fit to X-ray observations, scaled to this halo mass, $n_{\rm true,\,NCC} \sim n_{0}\,[ (r/r_{c})^{-\alpha}\,(1+(r/r_{c})^{2})^{\alpha/2-3\beta}\,(1+(r/r_{s})^{3})^{-\epsilon/3}]^{1/2}$ (with [$n_{0}/{\rm cm^{-3}},\,\alpha,\,\beta,\,\epsilon,\,r_{c}/r_{500},\,r_{s}/r_{500}] \approx [0.005,\,0.08,\,0.35,\,2.6,\,0.07,\,0.6]$) as fit in \citet{vikhlinin:temperature.metallicity.profiles,ghirardini:2019.cluster.profiles.compilation.universal.fits} and adopted in \papertwo; 
 (b) a typical weak CC profile from the same references above, with parameters $\approx [0.0075,\,0.7,\,0.39,\,2.6,\,0.07,\,0.6]$, or 
 (c) the geometric mean of model (a) and (b) at each $R$. 

\item True gas temperature profile. Again we compare
(a) a typical NCC fit scaled to the Perseus virial temperature $T_{\rm true,\,NCC} \sim T_{\rm vir}^{0}\,(1+(r/0.45\,r_{500})^{2} + (r/r_{500})^{10})^{-0.3}$ with $T_{\rm vir}^{0} = 7.6\,$keV \citep[following][]{vikhlinin:temperature.metallicity.profiles,ghirardini:2019.cluster.profiles.compilation.universal.fits};\footnote{The extra truncation scaling as 
$[...+(r/r_{500})^{10}]^{-0.3}$, i.e.\ $r^{-3}$ at $\gg r_{500}$, is added here to represent the steep drop outside the virial shock as measured in \citet{urban:2014.azimuthally.resolved.perseus.profiles,zhu:2021.perseus.profiles.outskirts.virial.shock}.}
(b) a CC-like fit to Perseus, with the function from \citet{vikhlinin:temperature.metallicity.profiles}, $T_{\rm true} = T_{\rm true,\,NCC}\,(0.44+x)/(1+x)$ with $x\equiv(r/r_{\rm CR})^{3}$ ($r_{\rm CR}\approx 80\,$kpc): i.e.\ assuming there is some genuine cooler gas as predicted where the CR pressure is large \citep{ji:fire.cr.cgm,hopkins:2020.cr.outflows.to.mpc.scales,ji:20.virial.shocks.suppressed.cr.dominated.halos,hopkins:2020.cr.transport.model.fx.galform,butsky:2020.cr.fx.thermal.instab.cgm,weber:2025.cr.thermal.instab.cgm.fx.dept.transport.like.butsky.study}; 
or 
(c) the same as (a), but assuming that $1/2$ of the thermal cooling luminosity $L_{X}$ interior to the radius where the true physical cooling time $t_{\rm cool} < 10\,$Gyr in the cluster center is distributed according to a standard cooling-flow model ($dL_{X}/dT\sim$\,constant) with maximum temperature $T_{\rm max}(r) = T(r)$ (the temperature given by (a)) and minimum temperature $T_{\rm min}(r) = 0.1\,$keV.

\end{enumerate}

For each of the above, choice (a) represents the (somewhat arbitrary) ``fiducial'' choice about which we vary other parameters, but we have experimented with a number of other possible combinations that do not change our conclusions (noted throughout).

\section{Results}
\label{sec:results}

\begin{figure}
	\centering
	\includegraphics[width=0.98\columnwidth]{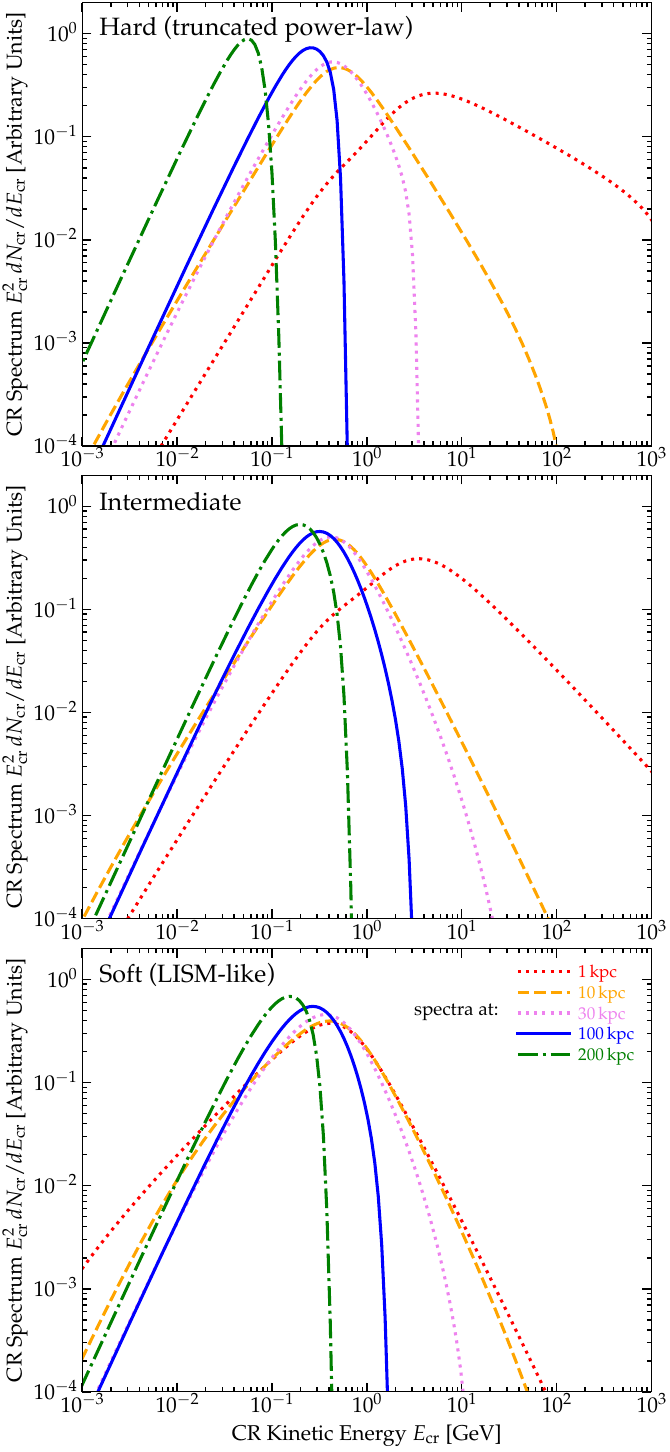}
	\caption{Model (\S~\ref{sec:model}) CR (leptonic) spectra, for CRs injected by NGC 1275 at the center of Perseus (accounting for streaming/advection+diffusion+losses), as a function of distance from the center $R$ (labeled). We renormalize each by total energy to compare spectral shapes, and show three representative models from \S~\ref{sec:variations} that span the range of assumed CR ``injection zone'' spectra emerging from the the central $1-10\,$kpc, from the hardest (a power-law $dN_{\rm cr}/dp_{\rm cr} dt \propto p_{\rm cr}^{-2.2}$, truncated below $\sim 10\,$GeV), to the softest (an LISM-like spectrum). Despite very different assumed injection spectra, all the models converge at $\gtrsim 10\,$kpc to similar shapes, owing to synchrotron+IC depleting high-energy CRs while Coulomb losses deplete very low-energy CRs. Outside the injection region (interior to which the spectrum is set ``by hand''), none resemble the pure power-law spectra often modeled in the past (\S~\ref{sec:basics}).
	\label{fig:crspec}}
\end{figure}

\subsection{CR and Basic Emission Properties}
\label{sec:basics}

Fig.~\ref{fig:crspec} shows CR spectra at different spherical radii $r$ in mock clusters, evolved as above. Rather than show every model variant from \S~\ref{sec:variations}, we focus on three extremes of the assumed injection spectra which bracket the qualitative range explored: (1) LISM-like, point-like injection; (2) harder power-law, point-like injection; (3) harder power-law, extended ($\sim10\,$kpc) injection. 
By construction, this produces significant differences in the CR spectrum at $\lesssim 1-10\,$kpc (the injection region). However, Fig.~\ref{fig:crspec} shows that losses very quickly drive these towards the same spectra as soon as they propagate outside the injection region. High-energy CRs lose their energy most rapidly, truncating the spectra, such that it is sharply peaked at $\sim 0.1-1\,$GeV at all radii $\gtrsim10\,$kpc. Coulomb losses also narrow the spectrum at low energies, but this effect is smaller. Curvature in the spectra clearly cannot be neglected at any energies outside the injection zone. 

Fig.~\ref{fig:crspec.reacc} shows that for these models, streaming losses and/or turbulent reacceleration have much smaller effects on the spectra compared to the dominant losses. 

\subsubsection{Weak Effects of Streaming Losses \&\ Turbulent Reacceleration} 
\label{sec:stream.reacc}

\begin{figure}
	\centering
	\includegraphics[width=0.98\columnwidth]{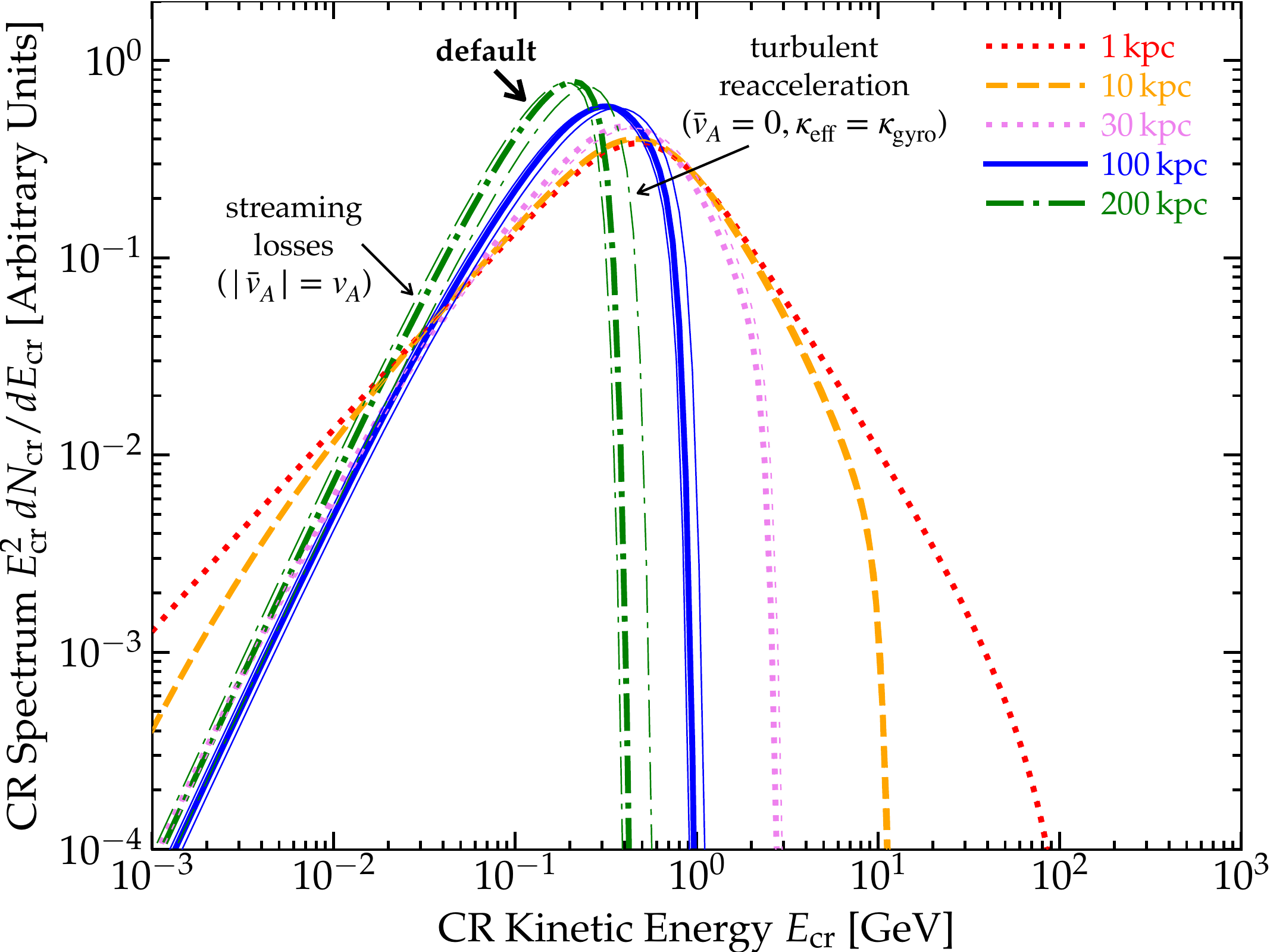}
	\caption{Example of the (very small) effects of including CR ``streaming losses'' (from asymmetric scattering in the comoving \Alf\ frame; \textit{lower-energy thin lines}) and ``turbulent reacceleration'' (from symmetric extrinsic scattering, if assumed; ; \textit{higher-energy thin lines}), on one model of CR spectra (as Fig.~\ref{fig:crspec}). The ``default'' assumption (\textit{thick lines}) ignores these effects owing to uncertainties in scattering microphysics. The results are similar for all models we consider here. At small radii near injection these have no effect, and even at the largest radii CRs have streamed, $\sim 200\,$kpc, their effect is very weak for the default parameters we assume here (\S~\ref{sec:stream.reacc}). 
	\label{fig:crspec.reacc}}
\end{figure} 

In Eq.~\ref{eqn:onemoment}, we neglect the ``streaming loss'' ($\nabla \cdot (\bar{v}_{A} \bhat)$) and ``turbulent reacceleration'' ($(v_{A}^{2}-\bar{v}_{A}^{2})/9\,\kappa_{\|,\,{\rm gyro-resonant}}$ terms), as these depend on quantities like $|\bar{v}_{A}|/v_{A}$ and the gyro-resonant symmetric vs.\ anti-symmetric scattering rates which remain uncertain (see \S~\ref{sec:differences:reaccel}). However they are also neglected because even if we take them to have their maximal theoretical values, they have very little effect on our predictions. 

Fig.~\ref{fig:crspec.reacc} illustrates this. First consider streaming losses. In spherical symmetry, keep all assumptions from Eq.~\ref{eqn:onemoment.full} but do not drop the streaming term, and retain the $\bar{v}_{A}\bhat$ part of ${\bf v}_{e}$ in $\nabla \cdot {\bf v}_{e}$ with $\bar{v}_{A} = {\bf v}_{A}$. Using $\nabla \cdot {\bf B} = 0$ and the definition of ${\bf v}_{A}$, we obtain the spherical streaming loss term, which is identical to an additional term in $\mathcal{R}^{\prime} \equiv \mathcal{R} + \mathcal{R}_{\rm st} + \mathcal{R}_{\rm reacc}$ with the form $\mathcal{R}_{\rm st} = f_{\rm st} (\langle |{\bf v}_{A} \cdot \hat{\bf r} |\rangle / 6\,r)\,{\rm d}\ln{\rho}/{\rm d}\ln{r}$, with $f_{\rm st} \equiv {\rm MIN}[1,\,|\bar{v}_{A}|/v_{A}]$. Taking this to have its maximal value ($f_{\rm st} \rightarrow 1$), which means $\mathcal{R}_{\rm reacc} \rightarrow 0$, we can then re-calculate the model with this effective loss term added. Note that this gives a streaming loss timescale of many \Alf\ crossing times (especially since the $\rho$ gradient is weak in the halo center), or $\gtrsim 10\,$Gyr at $\sim 100\,$kpc. 

Alternatively, we can take the maximum turbulent reacceleration term, which comes from assuming $\bar{v}_{A} \rightarrow 0$ (so our ``$v_{\rm st}$'' term must reflect advection/convection, not true streaming) and strictly gyro-resonant, extrinsic-turbulent scattering with $\kappa_{\|}=3\,\kappa$. Then in terms of the one-dimensional distribution function $f$, this term in Eq.~\ref{eqn:onemoment.full} becomes $\rightarrow v_{A}^{2} p_{\rm cr}^{3} /(9\,\kappa_{\|}) \partial (f/p_{\rm cr}^{2})/\partial \ln p_{\rm cr} \rightarrow v_{A}^{2} p_{\rm cr}^{3} /(27\,\kappa) \partial (f/p_{\rm cr}^{2})/\partial \ln p_{\rm cr}$, which is subsumed into $\mathcal{R}^{\prime}$ as $\mathcal{R}_{\rm reacc} = v_{A}^{2}/(27\,\kappa)\,(\partial \ln f / \partial \ln p_{\rm cr} - 2)$. Note that for the typical CR transport parameters here, this gives quite long reacceleration times $\sim 27\,\kappa/v_{A}^{2} >10\,$Gyr.

There is also an ``advective'' or ``adiabatic'' loss/gain term $\nabla \cdot {\bf u}_{\rm gas}$ which appears, which we can see immediately must have the same qualitative effects as the streaming loss or turbulent reacceleration terms on the spectra, depending on whether the gas velocity field is locally diverging (outflow) or converging (inflow) or turbulent. Here we have even less of a prior on the sign of this term, since it depends on the gas motions and their sign, not just on the direction of the CR pressure gradient (which is directed radially outwards). However, plausible effects are weak ($\sim 10\%$ on the location of the CR spectral peak, comparable to the effects in Fig.~\ref{fig:crspec.reacc}), given the observed velocities in Perseus, and we discuss this further in \S~\ref{sec:obs.xr:turb} as it relates to diagnostics of turbulence in the Perseus core.

As expected from the basic scalings above, these have little qualitative effect on the predicted CR spectra to leading order -- much smaller than the variations in transport parameters or $B$ or $n$ (through their effect on loss rates) in the different models we assume. Therefore, at least for these parameters, we are justified in neglecting them further. But we discuss in \S~\ref{sec:differences:reaccel} how this differs from historical minihalo models in which reacceleration plays an important role.

\begin{figure}
	\centering
	\includegraphics[width=0.99\columnwidth]{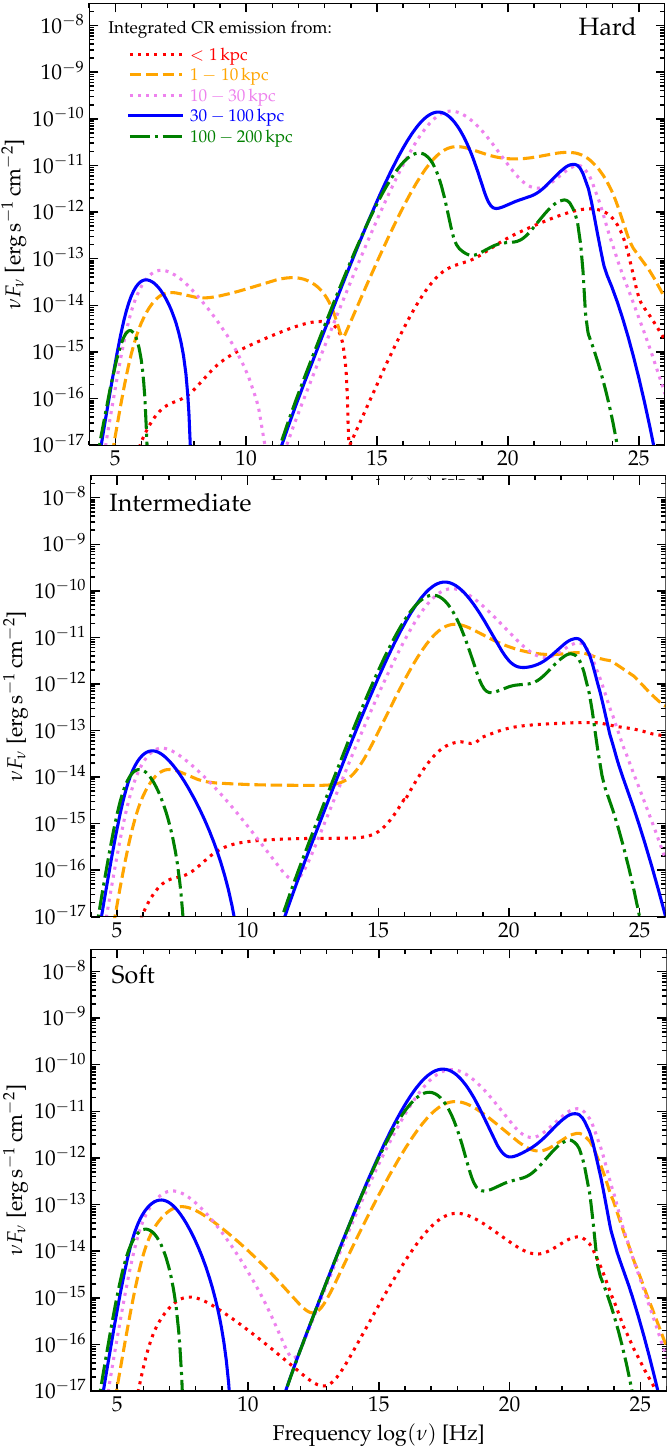}
	\caption{All-wavelength CR-emission-only spectra (\S~\ref{sec:basics}), integrating in projected radial annuli (labeled) in our model of Perseus, from the CR spectra in Fig.~\ref{fig:crspec}. 
	The three peaks in $1-100$\,MHz radio, $1-10\,$keV soft X-rays, and $\sim 10-100\,$MeV $\gamma$ rays come from synchrotron, IC (of CMB photons), and bremsstrahlung, respectively. 
	The strong curvature of CR spectra, with peaks at $\sim 0.1-1\,$GeV where losses are minimized, leads to soft, thermal-like X-ray spectra and extremely soft $\gamma$-ray and low-frequency radio emission. 
	\label{fig:allspec}}
\end{figure}

\subsubsection{CR Emission Spectra} 
\label{sec:basics:spectra}

Given the CR spectra calculated by the models (shown in Fig.~\ref{fig:crspec}), Fig.~\ref{fig:allspec} shows the corresponding all-wavelength spectra of the CR emission only (neglecting direct and stimulated thermal emission), calculated as \S~\ref{sec:variations}. We specifically show the total flux in different annular regions, which makes it easier to compare different spectra. As in \papertwo-\paperthree, we see three main peaks: soft X-ray emission with a thermal black-body-like spectrum driven by IC, synchrotron peaked at $\sim$\,MHz frequencies, and $\sim 10-100\,$MeV $\gamma$-rays primarily from relativistic bremstrahhlung. Note that the continuum emission in far-IR through UV wavelengths will not be detectable in general, as it is orders-of-magnitude less luminous than either the spectrum of NGC 1275 itself at these wavelengths \citep{sinitsyna:2025.ngc.1275.radio.through.gamma.ray.compilation}, or thermal free-free, or the diffuse ICL stellar emission \citep{kluge:2025.euclid.icl.properties.of.gc.and.icl.in.perseus}. The dominant signatures in these wavelengths will be indirect, via excitation of emission lines from CR ionization in multi-phase gas, and supporting more gas in different phases, which will be studied in detail in future work.

Consistent with the CR spectra in Fig.~\ref{fig:crspec}, outside the acceleration/injection region, the spectra are quite similar, driven to that similarity by losses. Within the injection zone, there can be larger differences, though the same qualitative arguments apply. As expected, making the CR injection spectra much harder correspondingly shifts the synchrotron, X-ray, and $\gamma$-ray spectra to higher energies and gives them a shallower power-law tail to high energies. Each of these will be examined and compared to observations in more detail below.

\begin{figure}
	\centering
	\includegraphics[width=0.99\columnwidth]{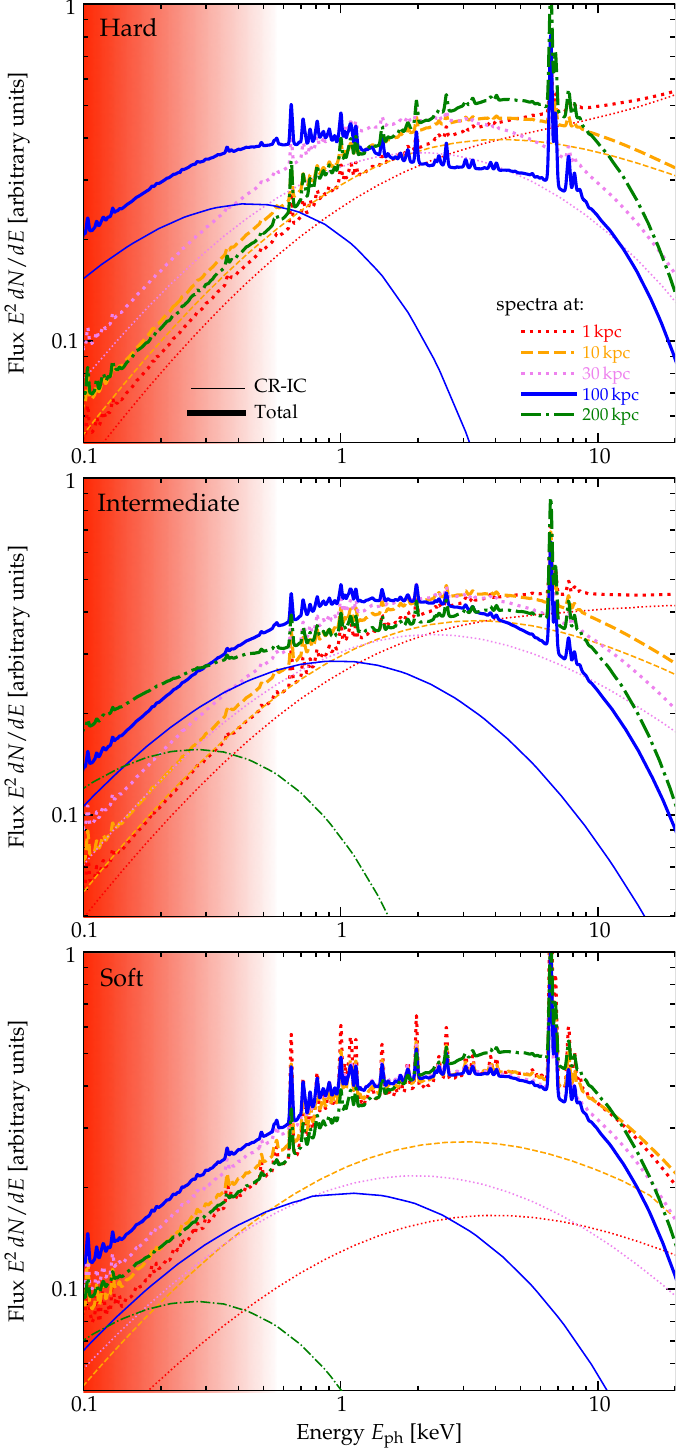}
	\caption{Examples of (un-absorbed) soft X-ray spectra (projected surface brightness or flux in a narrow radial annulus at $0.1-20\,$keV) from the models in Fig.~\ref{fig:allspec} (\S~\ref{sec:obs.xr}), including both CR emission (primarily CR-IC) and thermal (free-free+line) emission from a weak ``true'' CC (\S~\ref{sec:variations}). The CR spectral shape (Fig.~\ref{fig:crspec}) produces an IC spectrum which shows strong curvature and a hard cutoff, completely different from the often-assumed $E^{2}\,dN/dE \propto E^{+0.5}$ to $\propto E^{+1.5}$ (photon index $\Gamma_{X} =1.5-2.5$) power-law often assumed, much more similar to another thermal component with $T \lesssim 10\,$keV. The spectra are effectively indistinguishable from multi-component thermal spectra. 
	Red shading indicates where the spectrum becomes degenerate with small variations in the foreground absorbing column (given the observed $N_{H} \sim 10^{21}\,{\rm cm^{-2}}$ towards Perseus).
	\label{fig:xrspec}}
\end{figure}

Fig.~\ref{fig:xrspec} shows the X-ray spectral emissivity for the same model set and same radii, including both CR processes as in Fig.~\ref{fig:allspec}, plus the full thermal emission (including re-radiation of CR heating/excitation). 
We see that CRs contribute significantly to the total X-ray emission. But (1) the CR-IC X-ray spectra show strong curvature: only in the cases with the hardest injection spectra, at radii well within the injection zone ($r \lesssim 1\,$kpc) do we see any power-law-like CR-IC component (and even then the spectra show significant curvature below $\sim 2\,$keV). (2) While CR-IC contributes continuum, there is still significant line emission. As a result (3) these spectra are essentially identical to single or multi-temperature thermal spectra, for appropriate choice of temperature/density/metallicity. Indeed, in \paperthree, we show these can be essentially perfectly-fit (from $0.1-10$\,keV) with standard two-temperature cluster models.

\begin{figure}
	\centering
	\includegraphics[width=0.98\columnwidth]{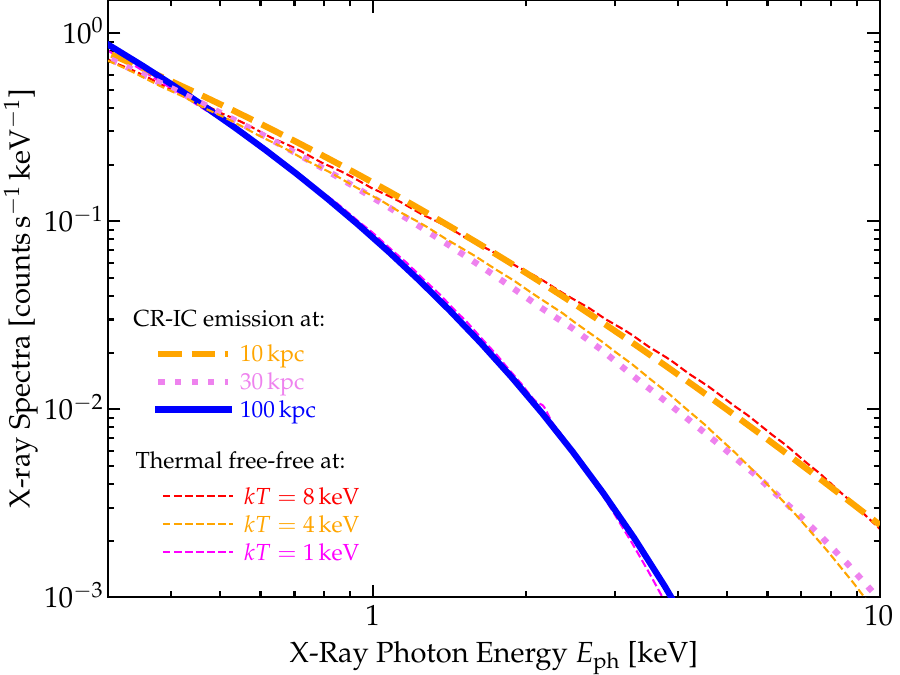}
	\caption{Further demonstration of the thermal-continuum-like character of the X-ray CR-IC emission of the spectra in Fig.~\ref{fig:xrspec}. Here we plot just the CR-IC emission at different radii outside the acceleration zone for the first model in Fig.~\ref{fig:xrspec}, over a typical observed range in observed units (arbitrarily normalized for comparison), compared to metal-free thermal emission spectra at the representative temperatures plotted. Note the latter are not fits, simply representative temperatures. 
	\label{fig:vsthermal}}
\end{figure} 

That connection is made more explicit in Fig.~\ref{fig:vsthermal}, which shows just the CR-IC emission contribution for one example outside of the injection zone (per the above, at these radii the various models are all similar qualitatively). We compare thermal spectra of metal-free gas at various temperatures shown. The X-ray CR-IC spectra are indeed effectively identical to thermal spectra for all practical purposes.

\begin{figure*}
	\centering
	\includegraphics[width=0.331\textwidth]{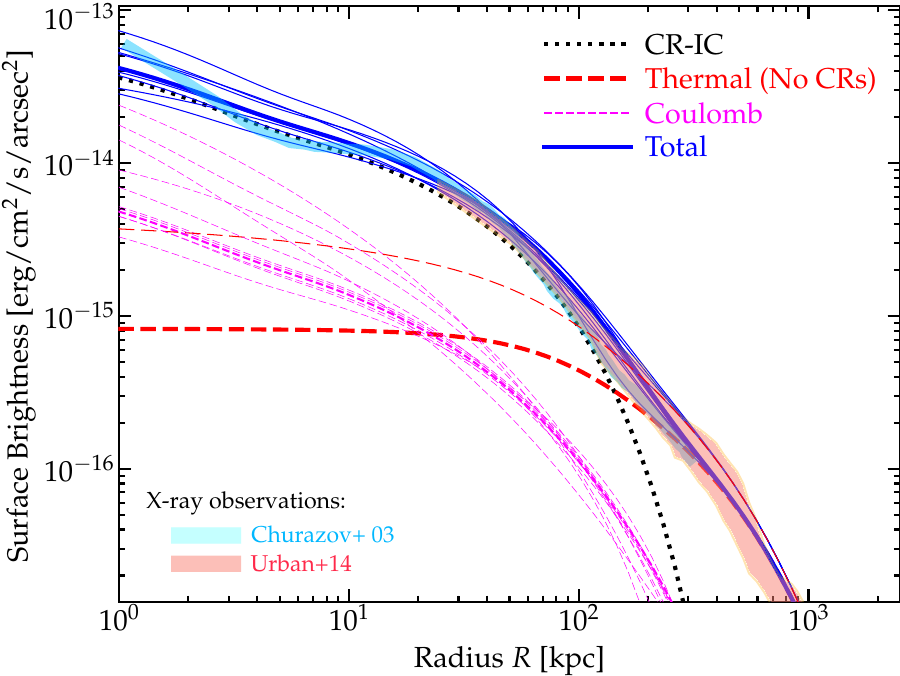}
	\includegraphics[width=0.331\textwidth]{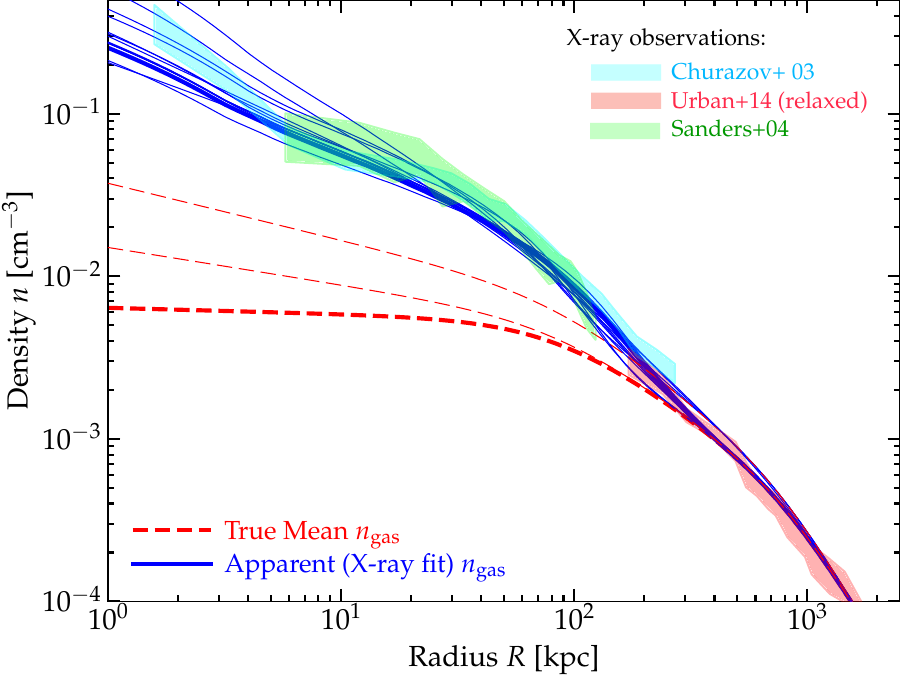}
	\includegraphics[width=0.331\textwidth]{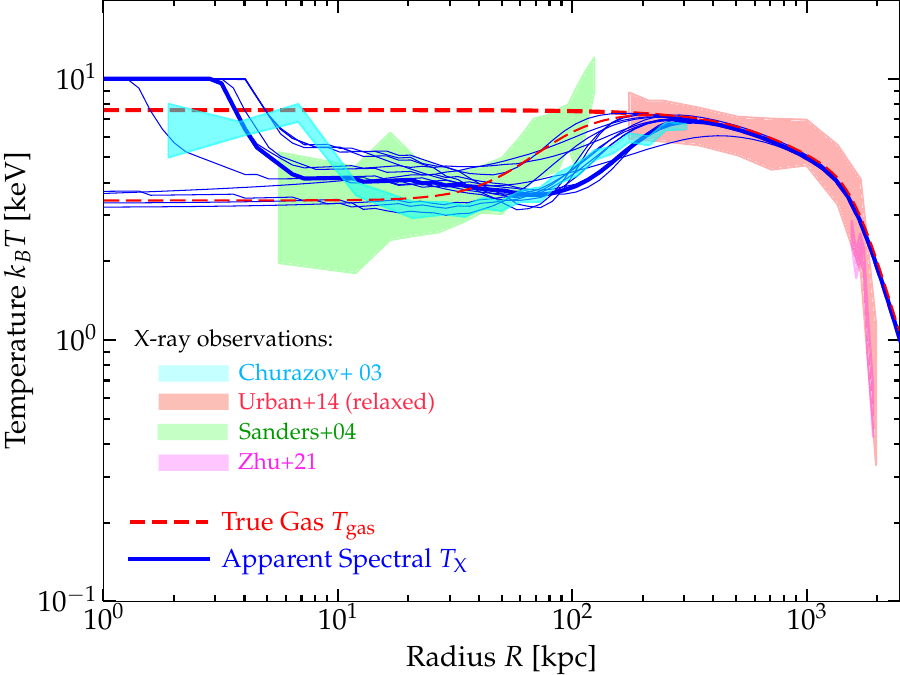}
	\caption{Radial profiles of soft X-ray properties in the models here  (\S~\ref{sec:obs.xr}), versus cluster-centric distance $R$ from NGC 1275. Predictions (\textit{blue solid lines}) are shown for all model variants (some lines overlap), with the nominal ``fiducial'' case (\S~\ref{sec:variations}) thicker, estimated as in \S~\ref{sec:obs.xr} given the total predicted X-ray spectra (Fig.~\ref{fig:xrspec}) that include both CR-IC and thermal emission. 
	We compare: 
	\textbf{(1)} soft X-ray surface brightness $I_{X}$; 
	\textbf{(2)} inferred gas density $n_{X}$ (assuming the emission is all thermal, and the gas volume-filling) from the standard emission measure; 
	\textbf{(3)} inferred gas temperature $T_{X}$ (same assumptions), from the spectral shape; 
	We also show ``true'' (thermal-only spectra, with no CR emission) values (\textit{red dashed}) assumed. 
	Wherever possible we compare Perseus X-ray observations showing the inter-quartile range of values observationally inferred at each $R$ (\textit{shaded}; \S~\ref{sec:obs.xr:soft}), though for some (e.g.\ \citealt{churazov:2003.perseus.profiles}) only the mean is available. 
	In these models, CR-IC significantly boosts the continuum interior to $\sim 100\,$kpc, explaining the apparent SCC (strong rise of $S_{X}$, $n_{X}$ towards the center). 
	\label{fig:xr.profiles}}
\end{figure*} 

\begin{figure*}
	\centering
	\includegraphics[width=0.331\textwidth]{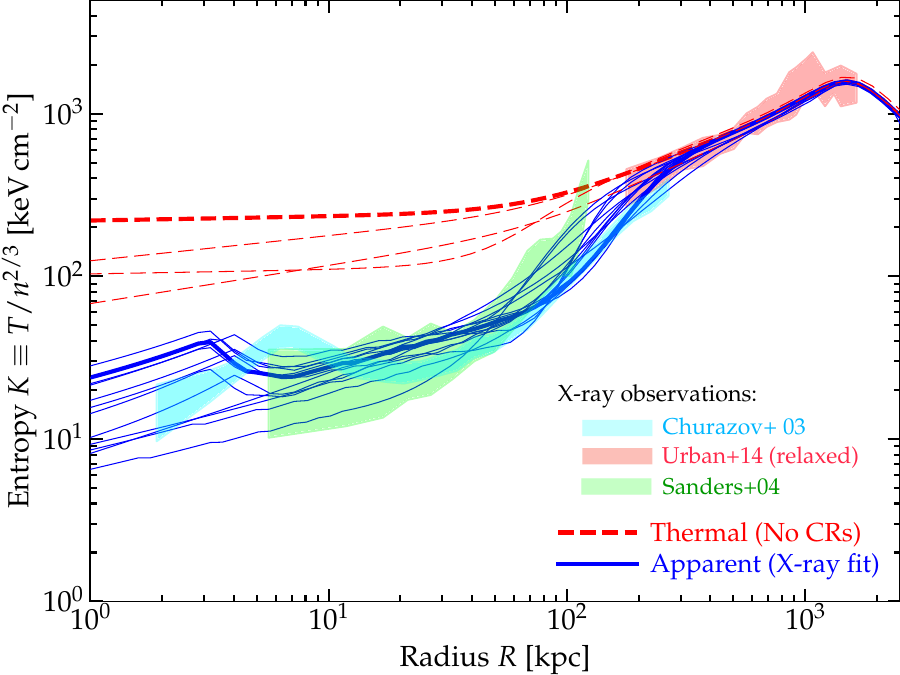}
	\includegraphics[width=0.331\textwidth]{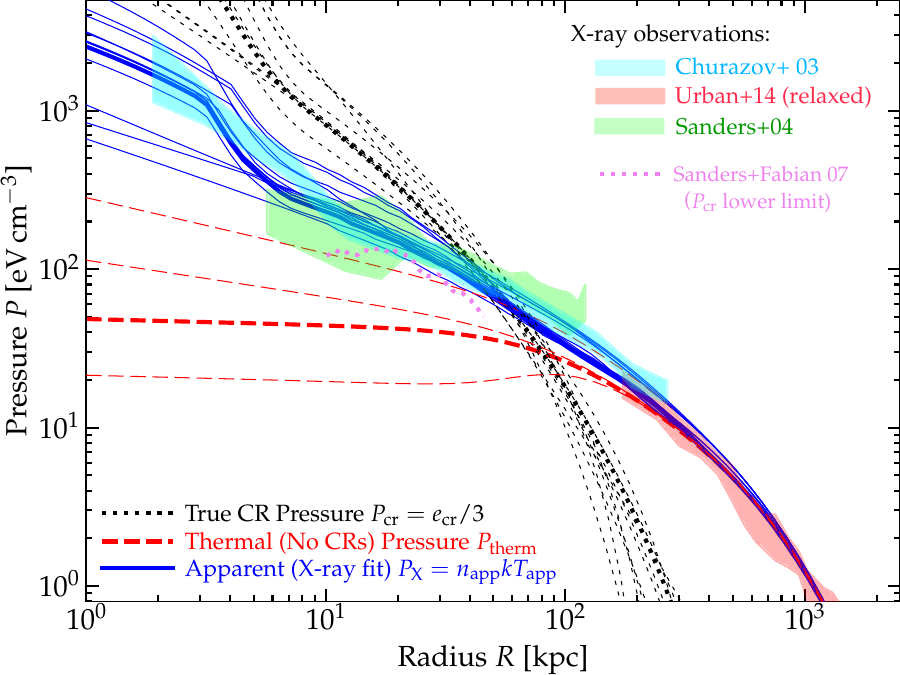}
	\includegraphics[width=0.331\textwidth]{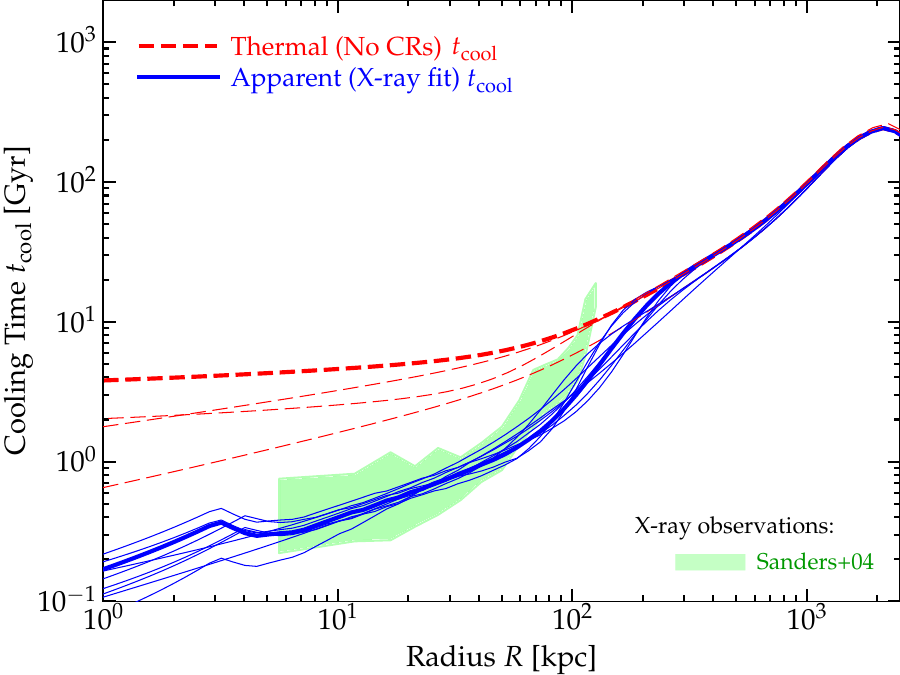}
	\includegraphics[width=0.331\textwidth]{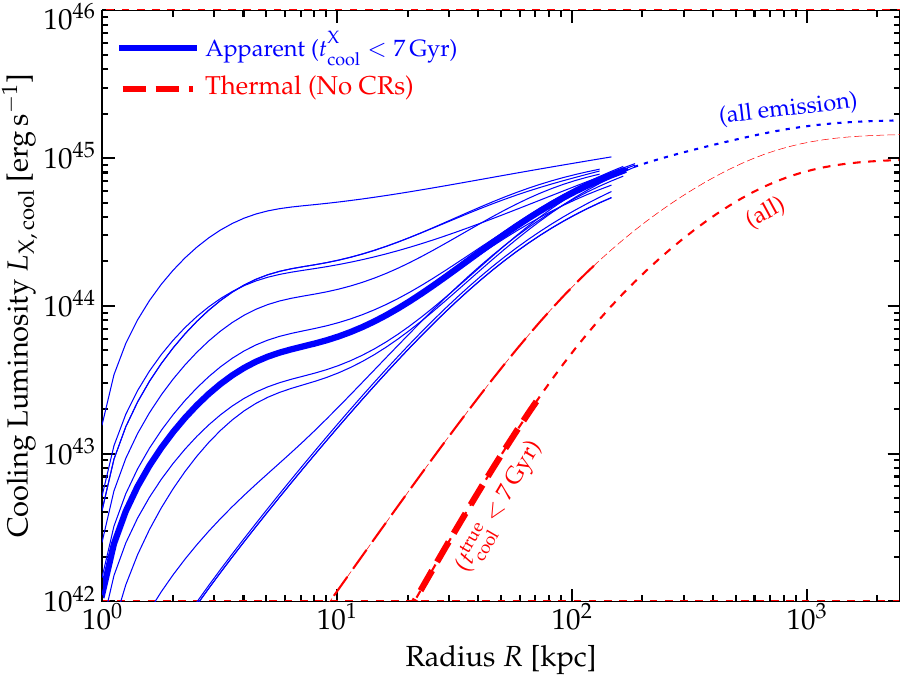}
	\includegraphics[width=0.331\textwidth]{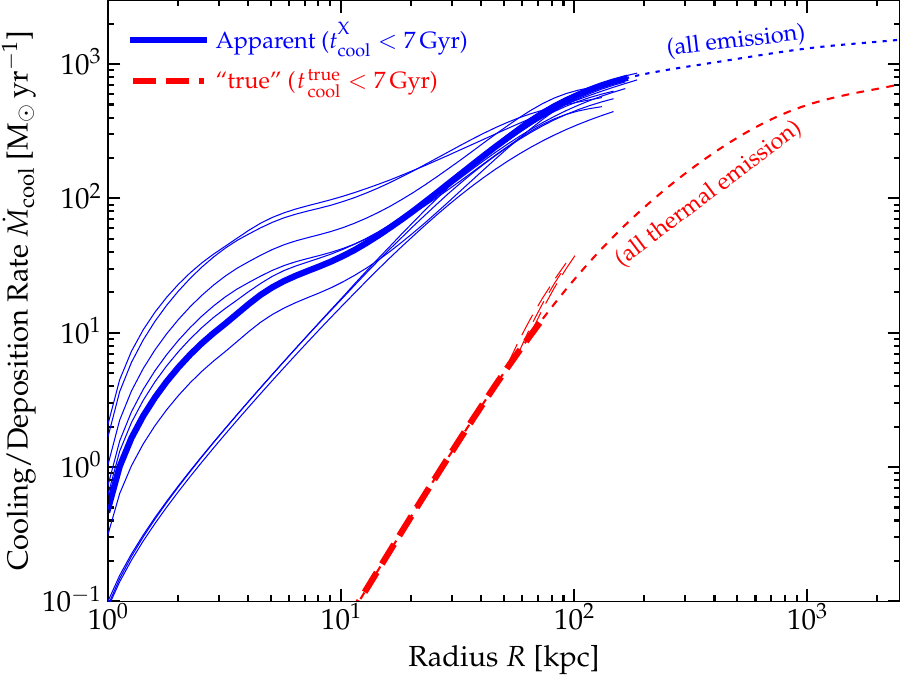}
	\caption{Radial profiles of soft X-ray properties (\S~\ref{sec:obs.xr}; as Fig.~\ref{fig:xr.profiles}). We show derived properties from those in Fig.~\ref{fig:xr.profiles}: 
	\textbf{(1)} entropy $K_{X} \equiv T_{X}/n_{X}^{2/3}$; 
	\textbf{(2)} pressure $P_{X} \equiv n_{X}\,k_{B} T_{X}$; 
	\textbf{(3)} cooling time $t_{{\rm cool},\,X}$ (from $T_{X}$, $n_{X}$, $Z_{X}$); 
	\textbf{(4)} cooling luminosity $L_{X,\,{\rm cool}}$ ($L_{X}(<R)$ interior to radii where $t_{{\rm cool},\,X} < 7\,$Gyr; total $L_{X}(<R)$ shown in \textit{thin dotted});
	\textbf{(5)} implied classical CF ``mass deposition rate'' $\dot{M}_{{\rm cool},\,X} \equiv \int_{0}^{r} (2\mu m_{p}/5 k_{B} T_{X})\,{\rm d} L_{{\rm cool},\,X}(r)$. 
The apparent SCC $L_{X,\,{\rm cool}}$ and $\dot{M}_{{\rm cool},\,X}$ are boosted by a factor of $\sim 100$ owing to CR-IC, which decreases the central $K_{X}$ and $t_{{\rm cool},\,X}$, while boosting the apparent $P_{X}$ (primarily through boosting the central surface brightness and therefore $n_{X}$).
	\label{fig:xr.profiles.deriv}}
\end{figure*}

\begin{figure}
	\centering
	\includegraphics[width=0.98\columnwidth]{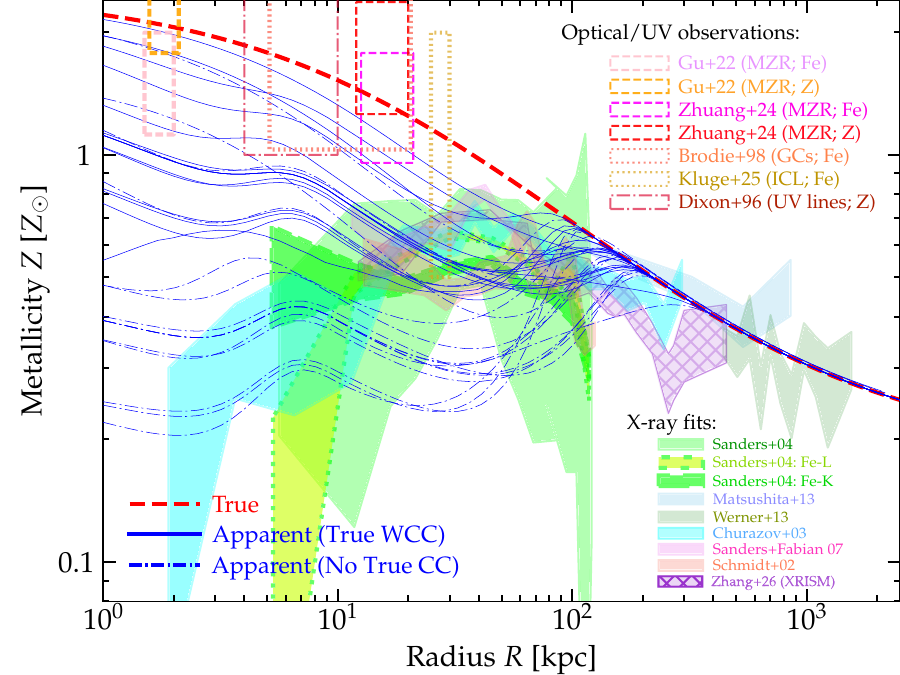}
	\caption{Radial profiles of soft X-ray properties (\S~\ref{sec:obs.xr}; as Fig.~\ref{fig:xr.profiles}). 
	We plot metallicity $Z_{X}$ (from the equivalent width of the $6-7$\,keV line emission), and compare independent (non-X-ray) observations of $Z$, which follow the assumed true $Z_{\rm true}$.  In these models, CR-IC explains why the central $Z_{X}$ is suppressed relative to independent (non X-ray) observations of $Z$ at the same radii ($\lesssim 10$\,kpc), because continuum is filled in by CR-IC, though the strength of this effect is most sensitive to the model details (owing to the sensitivity of the lines to the true $T$, $n$, CR heating rate, etc.).
	\label{fig:xr.profiles.Z}}
\end{figure}

\subsection{Comparison with Soft X-ray Observations: Cluster Profiles and Apparent ``Cooling Flow'' Properties}
\label{sec:obs.xr}

\subsubsection{Profiles of Soft X-ray Properties} 
\label{sec:obs.xr:soft}

In Figs.~\ref{fig:xr.profiles}, \ref{fig:xr.profiles.deriv}, \&\ \ref{fig:xr.profiles.Z}, we quantitatively compare to bulk X-ray-inferred properties in Perseus. 
Specifically, we compare the projected X-ray surface brightness (SB) $I_{X}$ (models and observations corrected to the same $0.1-10\,$keV band, using the best-fit spectral templates to remove instrumental sensitivity effects), and apparent X-ray inferred de-projected gas density $n_{X}$, temperature $T_{X}$, metallicity $Z_{X}$, and subsequent derived X-ray properties like entropy $K_{X} \equiv T_{X}/n_{X}^{2/3}$, pressure $P_{X} \equiv n_{X} k_{B} T_{X}$, cooling time $t_{{\rm cool},\,X} \equiv (3/2)\,n_{X} k_{B} T_{X} / n_{X}^{2} \Lambda(T_{X},\,Z_{X})$ (cooling tables from \citealt{wiersma:2009.coolingtables} as updated in \citealt{hopkins:fire2.methods,hopkins:fire3.methods}), cooling luminosity $L_{{\rm cool},\,X}(<r)$, and implied ``mass deposition rate'' $\dot{M}_{{\rm cool},\,X} \equiv \int_{0}^{r} (2\mu m_{p}/5 k_{B} T_{X})\,{\rm d} L_{{\rm cool},\,X}(r)$ (including all emission or just gas with $t_{{\rm cool},\,X} < 7$\,Gyr, a common observational definition). We define $n_{X}$, $T_{X}$, and $Z_{X}$ as in \papertwo-\paperthree, as the density, temperature, and metallicity which would produce the same soft X-ray surface brightness, mean hardness ratio, and Fe-K equivalent width in a single-phase, single-temperature APEC \citep{smith:2001.apec.methods} fit to the observed spectra. For all but $I_{X}$ we plot the 3D profiles and compare to de-projected observations (assuming perfect deprojection), compiled from \citet{schmidt:2002.perseus.core.metallicity.temperature.maps,churazov:2003.perseus.profiles,sanders:2004.perseus.profiles,sanders:2007.perseus.profiles.claimed.hardxr.cr.vs.thermal,werner:2013.perseus.metal.profile,matsushita:2013.perseus.metal.profile.to.rvir,urban:2014.azimuthally.resolved.perseus.profiles,zhu:2021.perseus.profiles.outskirts.virial.shock,zhang:2026.xrism.perseus.turb.zmetallicity.profile}. Wherever possible -- specifically from the measurements in \citet{sanders:2004.perseus.profiles,urban:2014.azimuthally.resolved.perseus.profiles,zhu:2021.perseus.profiles.outskirts.virial.shock} -- we show the variation in $n_{X}$, $T_{X}$, etc., \textit{at a given radius}, specifically the inter-quartile range of the mean fitted X-ray values across different azimuthal angles/wedges from the cluster center. For the others (e.g.\ \citealt{churazov:2003.perseus.profiles}), the smaller shaded range plotted simply reflects that the authors only quote the azimuthally-averaged mean values and uncertainties in the mean, not variance or range at a given $r$. 

Note that we are modeling a simple, spherically-symmetric, steady-state cluster with single-phase, single-temperature gas, so we do not expect a strict $\chi^{2}$ ``agreement'' with the observed X-ray spectra (indeed the observations are not even azimuthally symmetric in detail, and even our ``pure thermal'' models are far too simple to reproduce the full complexity of observed Perseus core X-ray spectra, including nonaxisymmetric features like the inner cavities which we discuss briefly in \S~\ref{sec:obs.radio:minihalo} as they are clearly distinct in the radio from the diffuse gas which we focus on here). Moreover the observations shown represent many independent observations of X-ray spectra of Perseus, at different depth/sensitivity, spatial resolution, wavelength range, projected radii, etc., using different instruments like Chandra, XMM, Suzaku, etc., each of which would have to be modeled completely independently (and each of which would require different forward modeling and de-projection). What we want to understand are the bulk properties of the X-rays, so therefore it is much more helpful, and tractable, to compare the radial profiles shown, which are equivalent to reproducing qualitatively similar spectra to those observed (see Fig.~\ref{fig:xrspec}). As discussed in \paperthree\ and \S~\ref{sec:obs.xr:temps}, the combination of spatially-resolved (XMM+Chandra) CCD spectroscopy of the Perseus core \citep{churazov:2003.perseus.profiles,sanders:2004.perseus.profiles}, plus the Hitomi spectrally-resolved (but spatially-integrated out to $\sim 100\,$kpc) microcalorimeter spectroscopy of the core region \citep{hitomi:2018.perseus.temperature.structure.hard.emission,hitomi:2018.perseus.core.turbulence.temperatures}, can be well-fit (to at least ``as good'' a formal fit quality as pure-thermal models) with some CR-IC contributions with effective soft X-ray continuum IC ``temperatures'' $T_{\rm IC}$ at $<100\,$kpc in the range $\sim 0.5-20\,$keV, equivalent to CR spectra peaked between $E_{\rm cr} \sim 0.3-3\,$GeV (including all examples in Fig.~\ref{fig:crspec}, at $\lesssim 100\,$kpc). 
The strongest constraints on the shape of the CR spectrum allowed by soft X-ray spectra come from radii where CR-IC and thermal emission contribute comparably ($\sim R_{\rm cool}$, in the models here), where the Hitomi spectra and the agreement between different line-ratio and continuum spectral temperature measurements require (if CR-IC is important) a CR spectrum whose spatially emission-weighted soft X-ray luminosity is dominated by a peak in $E_{\rm cr}^{3}\,dN_{\rm cr}/dE_{\rm cr}$ around $E_{\rm cr} \sim 0.7-1.5$\,GeV, again very similar (by construction) to all the models considered here. Of course, obtaining a formally good-fit to any observed spectrum would require (in thermal or mixed CR-IC+thermal) models not only forward-modeling the detailed instrument and measurement, but also allowing for the full physical degrees of freedom of the problem (i.e.\ fitting to a free and non-parametric distribution function of CR electrons for CR-IC, rather than assuming some power-law, plus a multi-temperature gas distribution), and assessing ``goodness of fit'' in a formal sense is challenging as even standard thermal-only multi-phase best-fit models fit to CCD or microcalorimeter spectra are known to change by much larger than the nominal statistical errorbars when fit to different spectral subregions or instruments (see \citealt{hitomi:2018.perseus.temperature.structure.hard.emission}). But provided these constraints are met, \paperthree\ (and e.g.\ Fig.~18 therein) shows that no X-ray spectral measurement alone (even with infinite signal-to-noise and sub-eV spectral resolution) can distinguish between a pure-thermal and mixed thermal+CR-IC origin of the soft X-rays. Hence our focus below on complimentary multiwavelength constraints that can break these degeneracies.

\subsubsection{Model Variations and Sensitivity}
\label{sec:obs.xr:var}

In Figs.~\ref{fig:xr.profiles}-\ref{fig:xr.profiles.Z}, we plot all of the model variants from \S~\ref{sec:variations} (where the lines are distinguishable; some lie almost exactly on top of each other in some panels); the ``fiducial'' case is the bold line. In addition to the apparent X-ray inferred values, we also show the ``true'' thermal value which would be given by our background model (\S~\ref{sec:model}; e.g.\ assumed initial ``true'' $n_{\rm gas}$, $T_{\rm gas}$, $Z_{\rm gas}$ of the gas) absent CRs. In some panels, we show additional information: for $I_{X}$, we show the SB from thermal absent CRs, plus CR-IC, plus re-radiated Coulomb-heating from the gas (which boosts the thermal emission); for $P_{X}$ we also show the true CR pressure $P_{\rm cr} = e_{\rm cr}/3$.

Given the simplicity of the models, it is remarkable that they appear to fit these profiles over the entire kpc-Mpc range. To be clear, we choose fiducial parameters to be reasonable, but nothing we present is fitted to the data in any way, and the parameters of the fiducial model are all order-of-magnitude plausible numbers used for general SCC clusters in \papertwo\ and \paperthree. But we discuss their values further below. Importantly, many of the predicted properties are surprisingly robust to the details of the model choices, and where there are sensitivities, these have an obvious explanation. For example, the sharp uptick in $T_{X}$ at $R < 10\,$kpc in some models (which then manifests in other parameters that depend on $T_{X}$) comes from the models which assume a hard CR injection spectrum spread over a large $10\,$kpc injection region, which (per \S~\ref{sec:basics} and Figs.~\ref{fig:allspec}-\ref{fig:xrspec}) produce, necessarily, harder X-ray spectra within their injection region. This appears to match the temperature behavior seen by \citet{churazov:2003.perseus.profiles}, but those authors argued the central region ($\lesssim 10-30\,$kpc) could be contaminated still by AGN emission (they saw hard X-rays in their spectra) -- of course we want to compare to fits that do include this. But even if future observations find that this was due to scattered light (which would be surprising given the Chandra PSF should correspond to $\sim 0.2$\,kpc spatial resolution here) or some other effect, this would argue instead for the models which lie at the opposite extreme (point-like injection models). The spread in models is also usually very comparable to the spread along different azimuthal angles observed. And of course in reality we do not expect spherical symmetry in detail, as the losses, injection (via jets), and transport physics will not be exactly symmetric. 

Like argued in \papertwo-\paperthree, in these models, the apparent SCC in Perseus is dominated by CR-IC. The ``true'' CC is very weak, with a true cooling luminosity between $\sim 10^{43}-10^{44}\,{\rm erg\,s^{-1}}$ in the models considered (as compared to $\sim 10^{45}\,{\rm erg\,s^{-1}}$ predicted), which implies a mass deposition rate more like $\sim 10\,{\rm M_{\odot}\,yr^{-1}}$ instead of $\sim 1000\,{\rm M_{\odot}\,yr^{-1}}$. If anything our models slightly over-predict the cooling luminosity and surface brightness at $\sim 100-200\,$kpc, but this can easily be adjusted if we were actually ``fitting'' by reducing $\dot{E}_{\rm cr,\,\ell}$ and $v_{\rm st,\,eff}$. 

\subsubsection{Weak Sensitivity to (and Constraining Power of) Actual Gas Temperatures}
\label{sec:obs.xr:temps}

Note the primary effect of CR-IC on the ``thermodynamic'' properties is through boosting the central SB, which in turn primarily boosts $n_{X}$ (which then increases $P_{X}$, $L_{X,\,{\rm cool}}$, $\dot{M}_{{\rm cool},\,X}$ and decreases $K_{X}$ and $t_{{\rm cool},\,X}$). The effects on the X-ray inferred temperature are relatively weak, and our predictions are largely agnostic to whether the true ``underlying'' gas temperature profile of Perseus is constant (as in a NCC) or decreasing (as in a CC). Even if the true CC is much weaker than would be naively inferred without CR-IC, the latter scenario (lower central $T_{\rm gas}$ interior to $\sim 80-100\,$kpc) is completely plausible in the CR-IC models, as (1) this is precisely the radius where Fig.~\ref{fig:xr.profiles.deriv} shows there should still be \textit{some} (albeit much weaker than inferred in thermal-only models) cooling and CC; and (2) this is also (not accidentally in the CR-IC models) roughly the radius where CR pressure becomes dominant, which will support more cool gas against sinking into the galaxy and runaway cooling (see references to simulation studies in \S~\ref{sec:variations}). 

In \paperthree, we show that CR-IC+thermal emission model spectra (a) can be extremely well-fit from $0.1-10$\,keV (including line ratios, thermal broadening, and continuum shape aspects) even at sub-eV spectral resolution by standard thermal-only models, with some metallicity bias (below) and small $\sim 10\%$ temperature bias (e.g.\ $4.1$ vs $4.5$\,keV, well within the range allowed by Hitomi+XRISM data, let alone CCD observations); (b) the sense of residuals even from single-temperature fits is very small and nearly-identical to that seen in microcalorimeter observations from Hitomi, with the best-fit line-ratio temperature increasing by $\sim 0.2\,$keV fitting lines from $\sim 2 \rightarrow 8$\,keV \citep{hitomi:2018.perseus.core.turbulence.temperatures,hitomi:2018.perseus.temperature.structure.hard.emission}; and (c) all of these effects are much smaller than the systematic uncertainties from using different atomic line databases, or different instrumental background models, or different temperature-distribution or deprojection priors as quoted in \citet{hitomi:2018.perseus.temperature.structure.hard.emission}. 
This lack of sensitivity is useful here, as it removes a potential degree of freedom from the models, but it also means that more detailed constraints on the temperature structure of Perseus or comparisons of e.g.\ different line and continuum-based kinetic temperature estimates as in \citet{hitomi:2018.perseus.core.turbulence.temperatures,hitomi:2018.perseus.temperature.structure.hard.emission} do not constrain the ratio of CR-IC to thermal emission in the CC (even if they spatially resolved the smaller scales of interest here, which they do not at present).

\subsubsection{Metallicity as an Indicator of CR-IC}
\label{sec:obs.xr:z}

Of all the profiles in Figs.~\ref{fig:xr.profiles}-\ref{fig:xr.profiles.Z}, the inferred metallicity $Z$ is clearly the most sensitive to exact model assumptions. There we compare models with single-temperature cooling, but also models with a true but very weak CF. Specifically per \S~\ref{sec:variations} following \paperthree, we take the thermal cooling emission from the model plus thermalized Coulomb heating that must be re-radiated, and where the cooling time is $<7$\,Gyr, we assume that half this emission is distributed in a CF model with maximum temperature $=T_{\rm true}$ of the single-temperature model, and minimum temperature $=0.1\,$keV. But recall these true CC luminosities are all well below the observed SCC luminosity. 

Because of the strong temperature sensitivity of the relevant lines, $Z_{X}$ is very sensitive to the true underlying multi-phase gas structure in the CC. But we also see there are large systematic uncertainties in $Z_{X}$ measured, differing by factors of $\sim 3-10$ at radii interior to the CC ($\lesssim 150\,$kpc). Nonetheless, the observations and models all show some suppression (by CR-IC filling in continuum) of $Z_{X}$ relative to the metallicities measured or inferred for Perseus from optical/UV diagnostics (to which our ``true'' $Z_{\rm true}$ is fit). Specifically we plot measurements of stellar metallicities in young globular clusters \citep{brodie:1998.perseus.gc.stellar.models,kluge:2025.euclid.icl.properties.of.gc.and.icl.in.perseus}, in the ICL \citep{kluge:2025.euclid.icl.properties.of.gc.and.icl.in.perseus}, from stars in the inner galaxy using either the relation between stellar mass $M_{\ast}$ or central stellar velocity dispersion $\sigma$ and metallicity in elliptical galaxy centers (interior to $R_{e}/8$, with range shown including the $99\%$ scatter in large samples of ellipticals with NGC 1275-like $M_{\ast}$ or $\sigma$, per \citealt{gu:2022.massive.galaxy.central.metallicity.mzr.stellar}; note either $M_{\ast}-Z$ or $\sigma-Z$ correlations give results differing by $<10\%$ for NGC 1275); similarly using the $M_{\ast}-Z$ relation for \textit{gas-phase} metallicities measured at closer to the effective radius in (\citealt{zhuang:2024.mzr.stellar.gas.different.species.starforming.and.non.very.extensive.compilation}; see also \citealt{andrews.martini:2013.gas.mzr.sdss.galaxies.direct.estimation}); or estimating gas-phase $Z$ from UV emission lines from the cooler gas within NGC 1275 \citep{dixon:1996.perseus.uv.obs.metallicity.in.cooling.gas}. 

We also identify whether the metallicity is specifically for iron (Fe), or for the sum of all species ($Z$). 
Notably, while these have non-negligible uncertainties individually, these non X-ray diagnostics (1) all agree with one another, and (2) are all super-Solar at $\ll 20\,$kpc, significantly above the X-ray inferred metallicities at the same radii (which are all sub-Solar), implying some continuum dilution by CR-IC. Interestingly, our ``best fit'' models are those with little ``true'' CC. 

\subsubsection{Surface Brightness, Velocity, and ``Density'' Fluctuations}
\label{sec:obs.xr:turb}

In subsonic, but super-\Alf{ic} (weakly-magnetized, plasma $\beta \gg 1$) turbulence as expected and observed in Perseus \citep{hitomi:2016.perseus.weak.turb}, one expects the generic relationship between density ($\rho$) and gas velocity (${\bf u}$) fluctuations $\delta \rho/\rho \sim \delta \ln{\rho} \sim \langle |\nabla \cdot {\bf u}\,\Delta t_{c} |^{2} \rangle \sim \xi_{t} \sim \eta_{t} \,\mathcal{M}_{c}^{m}$ where $\Delta t_{c}$ is some coherence/response time (e.g.\ the sound crossing time $\sim \lambda/c_{s}$), $\eta_{t}$ some $\mathcal{O}(1)$ constant, $\mathcal{M}_{c}$ the compressible sonic Mach number ($\delta u_{\rm rms} / c_{s}$) and $m\approx 1$ for modestly compressible turbulence (steepening to $m=2$ in the weakly-compressible, highly subsonic regime). Since the compressions are quasi-adiabatic, this implies associated temperature fluctuations $\delta T/T \sim (\gamma-1)\,\delta \rho/\rho \approx (2/3)\,\xi_{t}$ and thermal X-ray emissivity (given $\epsilon_{\nu} \propto n^{2}\,T^{-1/2}$) fluctuations at a given X-ray energy scaling as $\delta \epsilon_{\nu} / \epsilon_{\nu} \sim (5/3)\,\xi_{t}$ (or for broadband-integrated X-rays, $\epsilon \propto n^{2}\,T^{1/2}$, so $\delta \epsilon/\epsilon \sim (7/3)\,\xi_{t}$). Relations like these have been widely used to link SB fluctuations in Perseus to turbulence and to argue for different turbulent heating rates \citep{zhuravleva:2015.turbulence.estimates.in.clusters.from.surface.brightness.fluctuations.perseus,zhuravleva:2018.cluster.turb.props.from.xrays,li:2025.surface.brightness.fluctuations.nearby.clusters.claim.turb.equals.cooling.but.requires.super.extreme.model.pushing.more.like.three.dex.too.low,sanders:2020.bulk.flows.perseus.estimation.density.fluctuations.turbulence,devries:2023.chandra.gas.density.fluctuations.estimation.perseus}, and are plausibly consistent (in observed $\delta \epsilon_{\nu}/\epsilon_{\nu}$ and hence implied $\delta \rho/\rho$), assuming $\xi_{t}$, with velocity fluctuation measurements from Hitomi and XRISM \citep{hitomi:2016.perseus.weak.turb,hlavacek.larrondo:2025.xrism.perseus.preview}. 

Our analytic models simplify by assuming spherical symmetry so we do not model SB fluctuations or turbulence. But as shown in \paperone\ and \papertwo, both simple dimensional arguments and numerical simulations of CR-pressure-dominated CGM/ICM conditions exhibit similar, significant CR density fluctuations \citep{ji:fire.cr.cgm,butsky:2020.cr.fx.thermal.instab.cgm,su:turb.crs.quench,su:2021.agn.jet.params.vs.quenching,ruszkowski.pfrommer:cr.review.broad.cr.physics,weber:2025.cr.thermal.instab.cgm.fx.dept.transport.like.butsky.study}, which would manifest as CR-IC SB fluctuations and behave similarly to hydrodynamic fluctuations. To understand this simply, note that while some of these can be driven by other physics (e.g.\ shock acceleration), the CR diffusion time $t_{\rm diff} \sim \lambda^{2}/\kappa$ is much longer than the turbulent timescale $\lambda/ \delta u_{\rm rms}$ on large scales $\lambda \gtrsim \kappa/\delta u_{\rm rms} \sim 3\,{\rm kpc}\,(\kappa/10^{29}\,{\rm cm^{2}\,s^{-1}})\,(100\,{\rm km\,s^{-1}}/\delta u_{\rm rms})$, so CRs behave quasi-adiabatically in local velocity fluctuations. 

From \S~\ref{sec:equations}, the advective/adiabatic term $p_{\rm cr} \nabla \cdot {\bf u}$ gives rise to $\delta e_{\rm cr}/e_{\rm cr} \sim \gamma_{\rm cr}\,\xi_{t} \sim (4/3) \xi_{t}$, with the CR number density $\delta n_{\rm cr}/n_{\rm cr} \sim \xi_{t}$ and Lorentz factor $\delta \gamma/\gamma \sim (\gamma_{\rm cr}-1)\,\xi_{t} \sim (1/3)\,\xi_{t}$, in terms of the identical definition of $\xi_{t}$. Since the effective temperature of the CR-IC, for photons at the peak of the CMB, scales $\propto \gamma^{2}$, this means the effective $\delta T_{\rm IC}/T_{\rm IC} \sim (2/3)\,\xi_{t}$, while the emissivity ($\epsilon_{\nu} \propto n_{\rm cr}(\gamma[\nu_{\rm obs}])$ in a narrow band or $\epsilon \propto n_{\rm cr} \gamma^{2}$ band-integrated) will scale $\delta \epsilon_{\nu}/\epsilon_{\nu} \sim (4/3)\,\xi_{t}$ (near the spectral peak) or $\delta \epsilon/\epsilon \sim (5/3)\,\xi_{t}$. Thus the scalings with velocity are expected to be basically identical up to a normalization constant that differs by just tens of percent (which is arbitrary anyways in the models owing to its degeneracy with $\eta_{t}$ and assumptions about the isotropy and spectral shape of turbulence). On small scales ($\lesssim 1-10\,$kpc), diffusion will suppress CR density fluctuations (effectively decreasing $t_{c} \sim {\rm MIN}[t_{\rm turb},\,t_{\rm diff}]$), but this is where damping, anisotropy (magnetic), and viscous effects modify the scalings in any case.

Briefly, numerical simulations have shown that CR-dominated CGM/ICM halos tend to show weaker, more sub-sonic (primarily bouyancy-driven) turbulence and weaker non-thermal line broadening, compared to simulations with negligible CR pressure in the halo \citep{butsky:2022.cr.linewidth.effects.cgm}. Therefore, adding CR pressure within the SCC is expected to bring the magnitude of velocity fluctuations and $\xi_{t}$ into better agreement with the recent measurements from microcalorimeters like Hitomi/XRISM of apparently quite weak turbulence in the core.

\begin{figure}
	\centering
	\includegraphics[width=0.99\columnwidth]{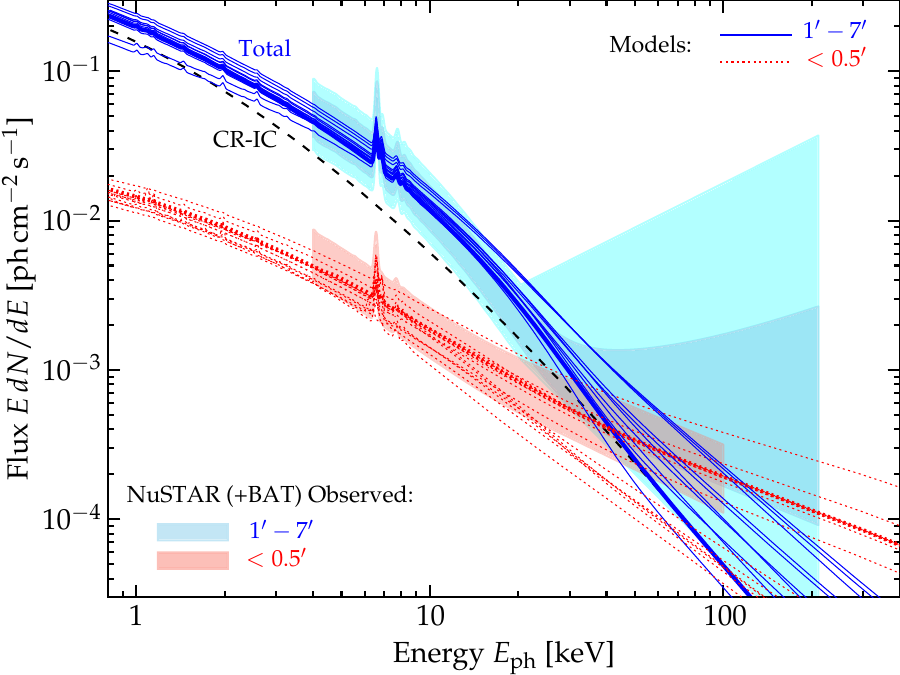}
	\caption{Model hard X-ray spectra (\S~\ref{sec:obs.hard:xr}), compared to NuSTAR+Swift BAT observations of Perseus integrating the central $<0.5^{\prime}$ (including the AGN; \S~\ref{sec:obs.hard:xr}) and true diffuse emission in an annulus from $1^{\prime}-7^{\prime}$. 
	The models are all consistent with the data. All show substantial spectral curvature and would not be fit by the kind of hard power-laws used to search for CR-IC in previous searches ($\Gamma_{X} \sim 1.5-2.5$; \citealt{creech:2024.nustar.perseus.obs}). Most are quite soft at $>20\,$keV, outside $>1\,$kpc. 
	Hitomi data are consistent as well, but significantly less constraining. 	
	\label{fig:hardXR}}
\end{figure} 

\subsection{Hard X-ray and $\gamma$-Ray Observations}
\label{sec:obs.hard}

\subsubsection{Hard X-rays}
\label{sec:obs.hard:xr}

Figs.~\ref{fig:xr.profiles}-\ref{fig:xr.profiles.Z} mostly refer to observations at $<10\,$keV, and Figs.~\ref{fig:xrspec}-\ref{fig:vsthermal} show thermal-like profiles up to $\sim 20\,$keV. Fig.~\ref{fig:hardXR} compares harder X-ray constraints from a few to $\sim 400\,$keV (even higher energies will be considered shortly). Over this range the best constraints come from NuSTAR, in \citet{rani:2018.nustar.perseus.center.obs.agn} and \citet{creech:2024.nustar.perseus.obs}, including additional SWIFT-BAT data at $\sim 100-200\,$keV \citep{oh:2018.swift.bat.hardxr.survey}. We have also compared to the Hitomi results \citep[e.g.][]{hitomi:2018.perseus.core.turbulence.temperatures,hitomi:2018.perseus.temperature.structure.hard.emission}, but these are fully consistent with the NuSTAR data, cover a smaller dynamic range in photon energy and with significantly larger error bars, so it is not helpful to show here. 

Theoretical spectra in this energy range are shown at smaller angular scales in Fig.~\ref{fig:allspec}. However hard X-ray instruments like NuSTAR, Hitomi, and BAT have much more limited angular resolution ($\sim 0.5$, $2$, $0.4$ arcminute HPD for the conditions and energies observed, or $\sim 10-50\,$kpc). As such it has only been recently possible to constrain the diffuse emission in the central regions, measured in two intervals (a central region $<0.5^{\prime}$ projected, or $<11\,$kpc, and a $1^{\prime}-7^{\prime}$ or $22-153\,$kpc annulus projected). For the central annulus we include a standard spectral template for the coronal and/or other ``injection'' X-ray AGN emission from radii much smaller than we model \citep{hopkins:bol.qlf,aird:2015.xr.qso.lf.evol,shen:bolometric.qlf.update}, but its contribution is small at $\lesssim 20\,$keV given the point-like X-ray luminosity of AGN alone at the nucleus of NGC 1275 at $<10\,$keV from Chandra being $\lesssim 10^{43}\,{\rm erg\,s^{-1}}$ (likely potentially contaminated by line-of-sight hot gas emission; \citealt{hitomi:2018.perseus.1275.xray.agn.constraints,reynolds:2021.1275.xr.constraints,imazato:2021.1275.xr.spectrum.variability.constraints,fedorova:2023.perseus.separating.agn.diffuse.xrays}). For the observations we show the $\pm 1\sigma$ range allowed by the best-fit models, and for the larger aperture the systematic range allowed depending on the quoted systematic uncertainty in background subtraction. We also highlight the CR-IC in the fiducial model.

Three things are clear. (1) Overall the predicted spectra agree well in shape and normalization, with order-unity contributions from CR-IC. (2) If anything, many of the predicted models are softer above $\gg 20\,$keV than the median of the observations (though still within their range given uncertainties) -- this means one could allow an even-shallower source or AGN (point-like) emission contamination than that modeled by the NuSTAR or Hitomi teams. And (3) spectral curvature is obvious. 

This is critical, as the NuSTAR and Hitomi studies above claim to put strong upper limits on IC at these radii. But first (1) their IC model was a pure power-law in X-rays, i.e.\ did not allow for any curvature. This is very different from what we actually predict, where even the highest-energy non-thermal component shows significant curvature so would formally be better fit in their models by additional, somewhat-hotter gas (which they do indeed fit). And second (2) the power-law slopes they considered were extremely shallow/hard, with photon indices $\Gamma_{\rm HX}$ even for the $1^{\prime}-7^{\prime}$ region between $\sim 1.5-2$ (motivated by the synchrotron spectral indices of the CRs at much smaller radii, $<10\,$kpc, as we show below, and corresponding to CRs with non-trivially different energies from those important at these X-ray wavelengths) equivalent to an assumed power-law distribution of CRs with $E^{2}\,dN/dE \propto E^{-1}$ to $E^{0}$, while the predicted photon index from CR-IC here at $\gtrsim 30\,$keV is significantly harder/steeper (for the diffuse spectra at $\sim 10-100\,$keV, the effective $\Gamma_{\rm HX}$ [though the spectrum is curved] ranges from $\sim 2.9-3.8$, and in the injection zone ranges $\sim 2-2.7$). If one allows for more realistic CR-IC models with curvature, Fig.~\ref{fig:hardXR} clearly shows the measured NuSTAR spectra are expected.

\begin{figure}
	\centering
	\includegraphics[width=0.99\columnwidth]{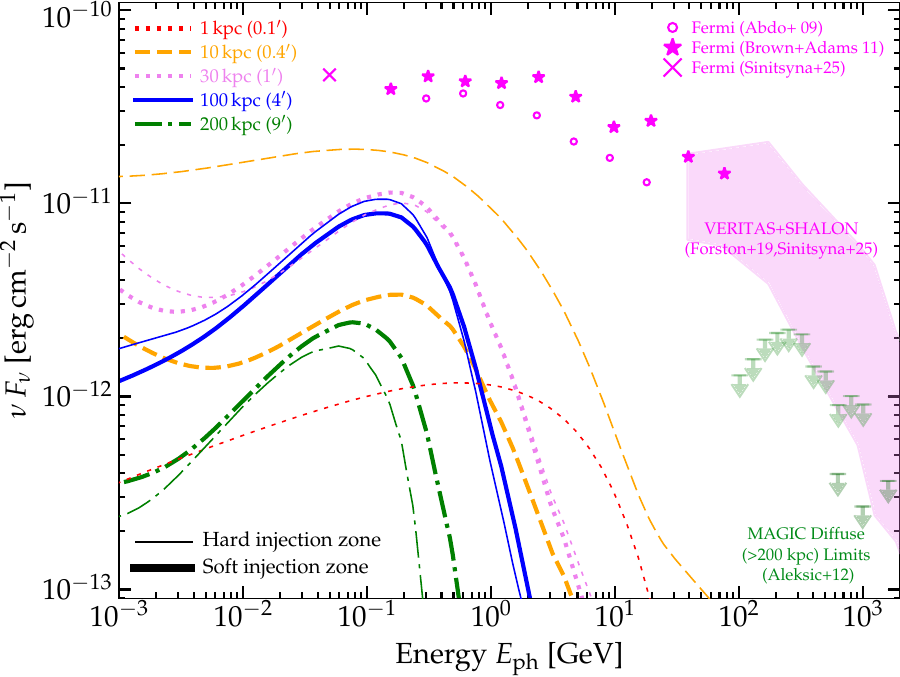}
	\caption{Model diffuse MeV-TeV $\gamma$-ray emission spectra (\S~\ref{sec:obs.hard:gamma}), as in Fig.~\ref{fig:allspec}. 
	We show two extremes of softest-to-hardest injection spectra, and show the flux at different radii (and corresponding size). 
	We compare different Fermi observations, but note the beam is $\sim 1^{\circ}$ ($\sim 1.3\,$Mpc), much larger than the Perseus CC, plus compiled data from MAGIC, SHALON, \&\ VERITAS. For MAGIC we show the upper limits for diffuse extended emission (based on spatial resolution \&\ variability constraints) coming from outside $> 200\,$kpc ($>9^{\prime}$) from NGC 1275. 
	All predicted emission from the CC would appear point-source like to these instruments, and the leptonic $\gamma$-rays contribute at most $\mathcal{O}(10\%)$ to the detected emission in the lowest-energy bins of Fermi ($\sim 100\,$MeV).
	\label{fig:gamma}}
\end{figure}

\begin{figure}
	\centering
	\includegraphics[width=0.99\columnwidth]{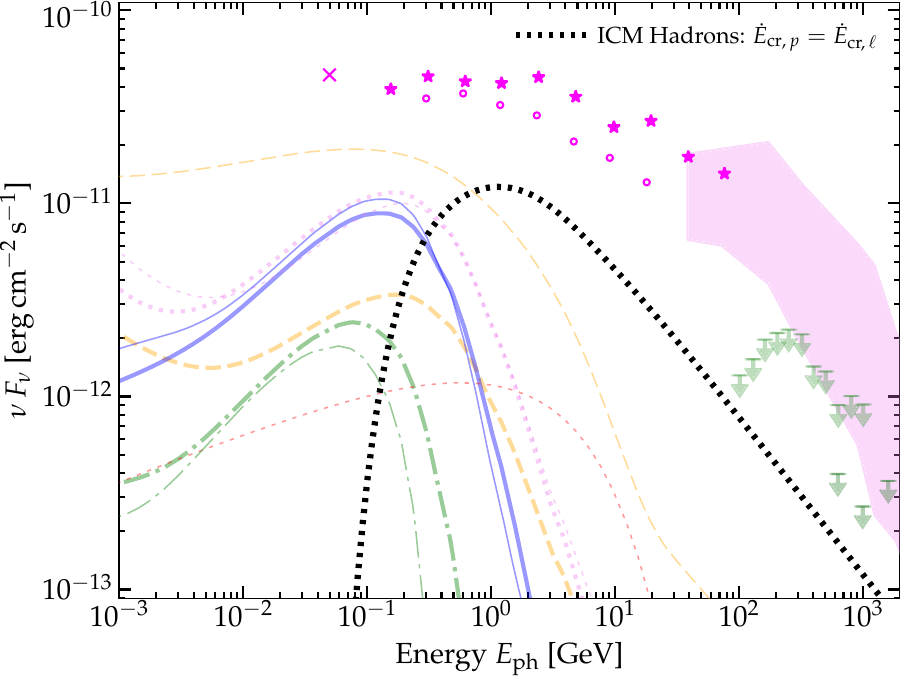}
	\caption{Model $\gamma$-ray emission as Fig.~\ref{fig:gamma}, if we arbitrarily add a population of diffuse/escaping CR hadrons to the model, with effective injection energy (by the time the CRs reach $\sim1\,$kpc) comparable to the leptonic energy ($\dot{E}_{{\rm cr},\,p} = \dot{E}_{{\rm cr},\,\ell}$) and an injection spectrum $\propto dN_{p}/dE_{p} \propto E^{-2.3}$ (modulated by transport so the effective spectrum is LISM-like). We integrate the emission over the entire cluster. The observations allow for up to a couple times the leptonic energy in hadrons in a diffuse component. 	While we are agnostic to the source of the leptons, the NGC 1275 $\gamma$-ray spectrum itself could, in principle, come from a \textit{compact} hadronic population of a couple times the leptonic energy, acting as the lepton source via pionic production.
	\label{fig:gamma.hadrons}}
\end{figure}

\subsubsection{$\gamma$-Rays}
\label{sec:obs.hard:gamma}

Fig.~\ref{fig:gamma} extends the comparison from Fig.~\ref{fig:hardXR} to $\gamma$-rays from MeV to TeV. We compare compilations of data at $\lesssim 100\,$GeV from Fermi (\citealt{abdo:2009.fermi.detection.1275.perseus.strong.gamma.ray.source.1e45.luminosity,brown.adams:2011.fermi.1275.monitoring.variability.spectra}: note the $\sim 40\,$MeV point from \citealt{sinitsyna:2025.ngc.1275.radio.through.gamma.ray.compilation} included here is excluded because of significance and background issues in \citealt{abdo:2009.fermi.detection.1275.perseus.strong.gamma.ray.source.1e45.luminosity,brown.adams:2011.fermi.1275.monitoring.variability.spectra} and other Fermi collaboration papers, so some caution is needed), as well as higher energy $\sim$\,TeV emission detected by SHALON \citep{sinitsyna:2025.ngc.1275.radio.through.gamma.ray.compilation} and VERITAS \citep{fortson:2019.veritas.data.update}, similar to the MAGIC detection of the central regions in e.g.\ \citet{aleksic:2012.magic.1275.central.galaxy.detection}, as well as the upper limit on ``extended'' diffuse TeV emission from outside $\gtrsim 10^{\prime}$ ($\gtrsim 200\,$kpc) from NGC 1275 modeled in \citet{aleksic:2012.magic.gamma.ray.1275.perseus.extended.halo.tev.upper.limits}. 

In the models, we first consider the spectrum produced just by the leptons needed for CR-IC to explain the X-rays. This primarily owes to relativistic bremsstrahlung, which leads to a scaled version of the CR lepton spectrum at a fraction of the CR spectral peak, here $\sim 100\,$MeV, with IC scattering of X-rays providing the additional broader component and ``tails.'' Note that it is important here that we account for multiple-IC scattering as we do, since the X-rays are primarily IC themselves. But for the same reasons as for the initial IC in the X-rays above, none of these form particularly hard spectra. 

Once again we see that outside the injection region, the spectra rapidly converge to very similar predictions. Cases with point-like and/or ISM like CR injection produce steeply cut-off $\gamma$-ray spectra (with photon index $\gtrsim 3.6$ at $>100\,$MeV, but clearly non-power-law spectra). If we explicitly make the injection-zone spectrum harder and extend it to $\gtrsim 10\,$kpc then as expected the $\gamma$-ray spectra harden at the same radii. But even in our most extreme case (the harder injection with explicit $\sim 10\,$kpc injection zone), the predicted $\gamma$-ray spectrum is below that observed in the Perseus center and the MAGIC upper limits on diffuse emission (with a photon index $\sim 2.8$ at $\sim 100$\,MeV-TeV), allowing for point-like emission from the NGC 1275 core or lobes. 

The diffuse leptonic emission associated with CR-IC and the CC primarily emerges at $\sim 1-100\,$MeV (just below the range probed at present), and identifying it observationally requires resolving the $\sim 0.4^{\prime}-4^{\prime}$ region well enough to clearly separate the contribution of the much brighter central source, so would require something like $\sim 10''$ angular resolution ($\sim 100-1000$ times better resolution than Fermi), with correspondingly orders-of-magnitude deeper surface brightness sensitivity.

Next, consider what would occur if there were a comparable hadronic component. Specifically Fig.~\ref{fig:gamma.hadrons} shows the results of assuming a hadronic injection rate equation for the leptonic one, on the same grid (with the same diffusion/streaming speed as a function of rigidity), using the LISM-like injection (scaling the hadronic injection likewise to the LISM proton spectrum observed). We propagate the protons including all the salient loss processes from \citet{hopkins:cr.multibin.mw.comparison}: Coulomb, ionization, and pionic/catastrophic/collisional \citep{1972Phy....60..145G,Mann94}, and calculate their $\gamma$-ray spectrum directly from the pionic rates as in e.g.\ \citet{Mann94,strong:2010.milky.way.sub.calorimetric.by.factor.hundred,dermer:cr.gamma.rays.vs.protons}. The resulting spectrum is still below observed limits. A ratio of $\sim 10-30:1$ of protons-to-leptons in the diffuse ICM, as in the LISM, would be clearly ruled out in the CR-IC scenario, but there could actually be a couple times the CR-IC implied $e_{\rm cr,\,\ell}$ in hadrons, without violating any $\gamma$-ray limits. 

This implies that instead of a leptonic jet, the jets could be ``initially'' hadronic, but on small scales (e.g.\ within the acceleration/injection zone), hadrons are converted to leptons via pionic processes, which requires a column density on \textit{small} scales (injection/acceleration scales) $\sim 10^{22}\,{\rm cm^{-2}}\,(v_{\rm st,\,eff}/100\,{\rm km\,s^{-1}})$ that is very plausible within those regions. As shown above, losses shape the lepton spectrum  (already strongly peaked at $\sim 0.1-1\,$GeV, the cutoff for $e^{\pm}$ produced by $\pi^{0}$ from initial CR $p$) rapidly outside the injection region, so this changes nothing about our scenario (and would feature no diffuse hadronic emission). This is effectively just different injection physics, to which we are agnostic, though interestingly would give a spectrum similar to observed at $\gtrsim 100\,$MeV in the injection region.

The more radical scenario is to argue that the leptons responsible for the minihalo (see \S~\ref{sec:obs.radio:minihalo}) and CR-IC are produced continuously or ``in situ'' throughout the ICM via pionic processes \citep{pfrommer.enslin:2004.hadronic.minihalos.in.cluster.centers}. In these models, if the CR transport is sufficiently slow, then the leptons are effectively calorimetric at each radius (with a local $\dot{e}_{\rm inj} \sim e_{{\rm cr},\,p}/t_{\rm pion} \sim \dot{e}_{\rm loss} \sim e_{{\rm cr},\,\ell}/t_{\rm IC}$, where  $t_{\rm pion} \sim 0.05\,{\rm Gyr}\,({\rm cm^{-3}}/n_{\rm gas})$ and $t_{\rm IC} \sim $\,Gyr at $\sim$\,GeV) , tracing the deposition rate from protons (as those have lower loss rates determined by the gas density, so drift slowly through the gas to larger radii). The integral hadron-to-lepton ratio in Fig.~\ref{fig:gamma.hadrons} clearly constrains this, but does allow for a comparable hadronic content at small radii to the leptons. However, any in-situ hadronic model does face a couple of challenges. (1) Since $\sim 1/2$ of the pionic lepton production rate from protons also goes directly into $\gamma$-rays, this would require the Fermi hadronic emission is extended to $\gtrsim 100\,$kpc ($\gtrsim 5^{\prime}$), not compact. While allowed by the angular resolution of Fermi, this appears to contradict constraints from the observed year-timescale variability. 
(2) A locally-calorimetric model where leptons are sourced by protons (where higher-energy protons have slightly \textit{longer} loss timescales, opposite of leptons) would not obviously predict the very strong observed steepening of the radio spectrum with increasing distance $R$ from NGC 1275 (\S~\ref{sec:obs.radio} below). This appears to be a clear sign of leptons aging as they travel. But it is possible that a model with local hadronic injection plus lepton diffusion/streaming could capture this (though the limit of centrally-concentrated hadronic production just becomes identical to our default models here), or that a strongly-radially-varying magnetic field with the right input spectrum could reproduce the trend. 
Moreover in a local hadronic model the profile of IC and radio would no longer be agnostic to the true/underlying gas density profile, since in such a scenario (unlike jet/core acceleration/injection/production) the ``source region'' is extended to $\sim 200\,$kpc and source production rate at a given radius (from pionic processes) scales $\propto n_{\rm gas}$, and so for a given CR transport model, explaining the X-rays requires the ``correct'' underlying gas density model. And the implied total CR pressure $\propto e_{{\rm cr},\,p} + e_{{\rm cr},\,\ell} \propto e_{{\rm cr},\,\ell}\,(1 + 5\,(0.01\,{\rm cm^{-3}}/n_{\rm gas}))$ would be even larger, though this is not necessarily a serious issue. So it may be possible to construct a viable in-situ hadronic model which gives effectively the same lepton spectra as invoked here to explain the radio and X-ray observations, but we will not consider those models further in this paper.

\begin{figure}
	\centering
	\includegraphics[width=0.98\columnwidth]{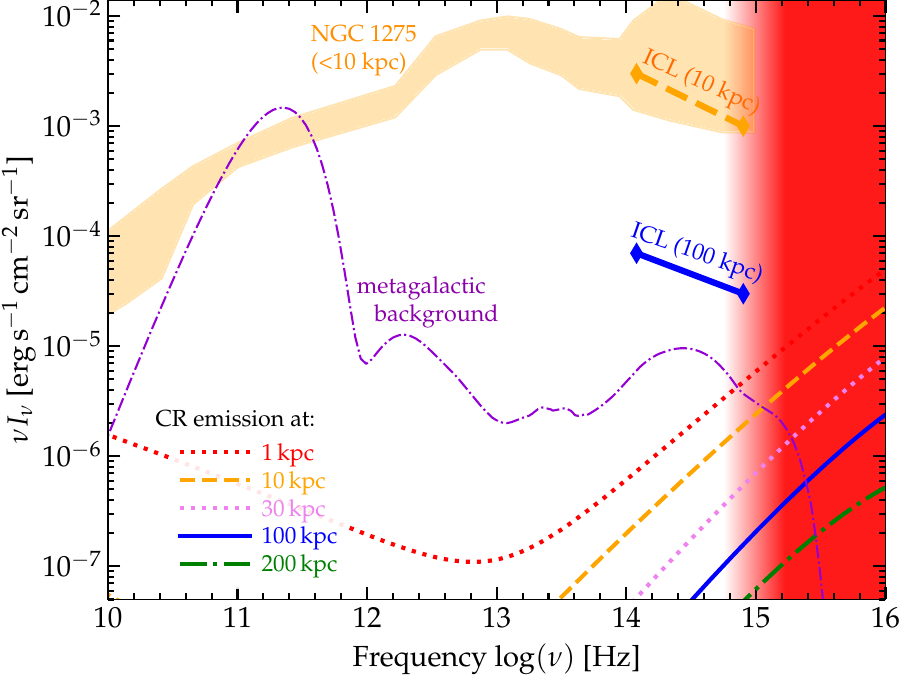}
	\caption{Model (unattenuated) CR emission from high-frequency radio (10\,GHz/3\,cm) through extreme UV ($10^{16}$\,Hz/300\,\AA) , as Fig.~\ref{fig:allspec} (\S~\ref{sec:obs.uvoir}). 
	Red shading shows where the observed $\sim 1.4\times10^{21}\,{\rm cm^{-2}}$ Galactic column towards Perseus \citep{sanders:2007.perseus.profiles.claimed.hardxr.cr.vs.thermal,sanders:2020.bulk.flows.perseus.estimation.density.fluctuations.turbulence} will strongly absorb the spectra. 
	We show the model surface brightness versus frequency at different radii. Note we show just one case here like Fig.~\ref{fig:allspec} but the conclusions are the same in all cases. 
	We compare the metagalactic background surface brightness (\textit{dot-dashed}); the diffuse ICL spectrum where measured at $\sim 10\,$kpc and $\sim 100\,$kpc (\textit{points}); and the spectrum of NGC 1275 integrated within a $\sim 10\,$kpc aperture (\textit{shaded}). 
	In this wavelength range, the direct CR emission is undetectable below cosmic backgrounds and the (usually much brighter) ICL+dust emission+scattering+extended emission from NGC 1275 itself. 
	\label{fig:uvoir}}
\end{figure}

\subsection{Infrared, Optical, and UV Wavelengths}
\label{sec:obs.uvoir}

Briefly, Fig.~\ref{fig:uvoir} compares the intermediate-wavelength EUV-through-FIR/mm/cm emission ($10^{10}-10^{16}\,$Hz, or $3$\,cm to $300$\,\AA) predicted in these models to the backgrounds and other cluster sources described in \S~\ref{sec:variations} as part of our assumed model for calculating IC emission, including: (1) isotropic meta-galactic backgrounds \citep{cooray:2016.extragalactic.background.light.compilations.review,khaire:2019.extragalactic.background.light.spectra}, and (2) other diffuse emission in Perseus, for which we compile gas free-free (using our same gas profiles), starlight from the diffusive ICL (taken from \citealt{kluge:2025.euclid.icl.properties.of.gc.and.icl.in.perseus}), and the observed central NGC 1275 spectrum at $<10\,$kpc \citep[as compiled from the literature in][see references therein]{sinitsyna:2025.ngc.1275.radio.through.gamma.ray.compilation}. 
There is no process whose direct emission from CRs peaks in these wavelengths, and as a result, the predicted direct CR emission in these wavelengths is negligible ($\lesssim 1\%$ of total, and often much smaller) compared to the known diffuse sources (e.g.\ ICL, dust and thermal free-free, jet emission) and isotropic metagalactic backgrounds.

There could be \textit{indirect} effects detectable at these wavelengths. For example, we discuss the indirect effects of comparing SZ (measured at these wavelengths) to X-ray pressures below. In a companion paper we also show that the molecular and atomic line excitation observed (in FIR-through-UV line emission from those species within the central $\sim 30\,$kpc of NGC 1275) is precisely that expected for the CR populations in the CR-IC models here. But given the large uncertainties in modeling these other backgrounds and diffuse emission processes, scattered light from the AGN and central source, and continuum gas emission from the multi-phase medium, it is hard to imagine that a sub-percent fraction of ``diffuse CR-IC emission'' at these wavelengths could be meaningfully isolated and observed.

\subsection{Radio: The Mini-Halo and Giant Halo}
\label{sec:obs.radio}

We now consider frequencies $<10^{10}\,$Hz, i.e.\ the Perseus mini and giant radio halos. This is the synchrotron peak in Fig.~\ref{fig:allspec}, with most of the emission from the extended ACRH at $\sim 1-10\,$MHz. Unfortunately $<10\,$MHz is unobservable at present owing to ionospheric and Galactic absorption, and at $\sim 10-100\,$MHz observations do not exist at present with the combination of low surface-brightness sensitivity, resolution, and dynamic range to map the extended radio, while as we showed above the diffuse/extended emission at $\gtrsim$\,a few GHz is well below diffuse radio backgrounds. Thus the most constraining comparisons with radio data at present are in the range from $\sim 100-1000\,$MHz. At these frequencies, Perseus has radio emission detected at all scales from $\lesssim 1\,$kpc to $\sim$\,Mpc. These are empirically often divided into the radio mini-halo on $\lesssim 100\,$kpc scales (sometimes separating further the ``central'' emission from NGC 1275) and giant halo out to $\sim$\,Mpc -- we will consider both of these in turn, but more broadly we model the entire radial profile range since they should all be connected to the CR lepton population. 

\begin{figure*}
	\centering
	\includegraphics[width=0.48\textwidth]{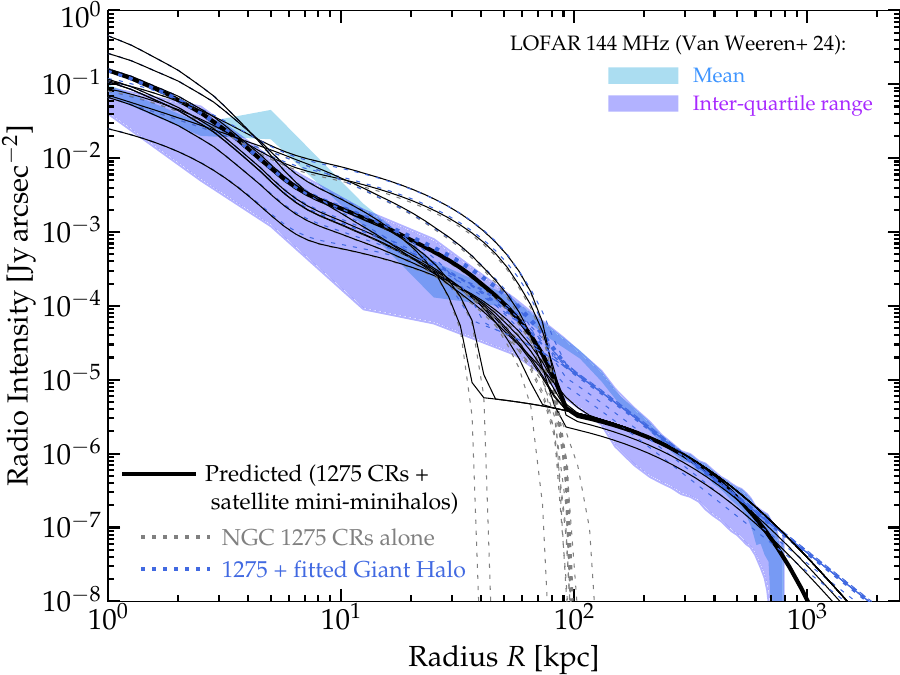}
	\includegraphics[width=0.48\textwidth]{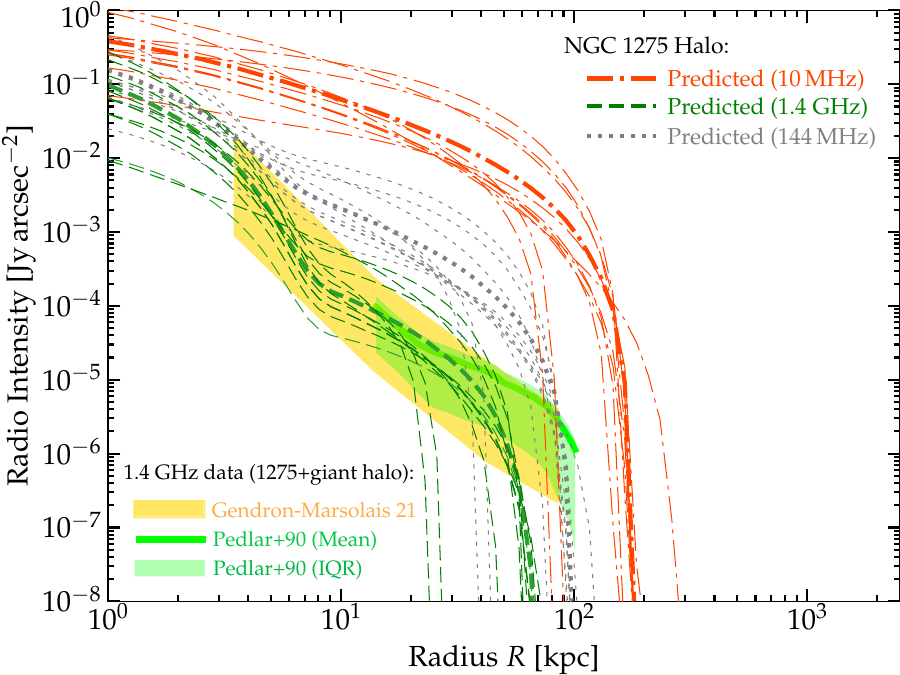}
	\caption{Radio surface brightness profiles (\S~\ref{sec:obs.radio}).
	\textit{Left:} Observations at 144\,MHz from LOFAR \citep{vanweeren:2024.perseus.giant.radio.halo.filled.electrons.just.tiny.fraction.high.energy}, the only to resolve the halo from kpc-Mpc. We plot both the mean $S_{\nu}$ at each projected radius $R$, and the interquartile range of $S_{\nu}$ in pixels at that $R$. We compare the models (\textit{lines}; as Fig.~\ref{fig:xr.profiles}), specifically decomposing the contribution from CRs just from NGC 1275 (traveling outwards), and with a simple model for the giant halo (\S~\ref{sec:obs.radio:gianthalo}) which accounts for injection of CRs by other galaxies in the cluster. 
	\textit{Right:} Same, showing the NGC 1275 CRs at $10\,$MHz and $1.4-1.5$\,GHz. For the latter we show observations (same style) from \citet{pedlar:1990.radio.1.4ghz.radial.profile} and \citet{gendron.marsolais:2021.vla.1.5ghz.perseus.imaging}. 
	The minihalo profile at multiple wavelengths is naturally explained by a population of CRs streaming from NGC 1275, with additional sources or shocks or re-acceleration spread throughout the cluster explaining the low-frequency giant halo (only detected at $144$\,MHz). 
	\label{fig:radio}}
\end{figure*}

\begin{figure}
	\centering
	\includegraphics[width=0.48\textwidth]{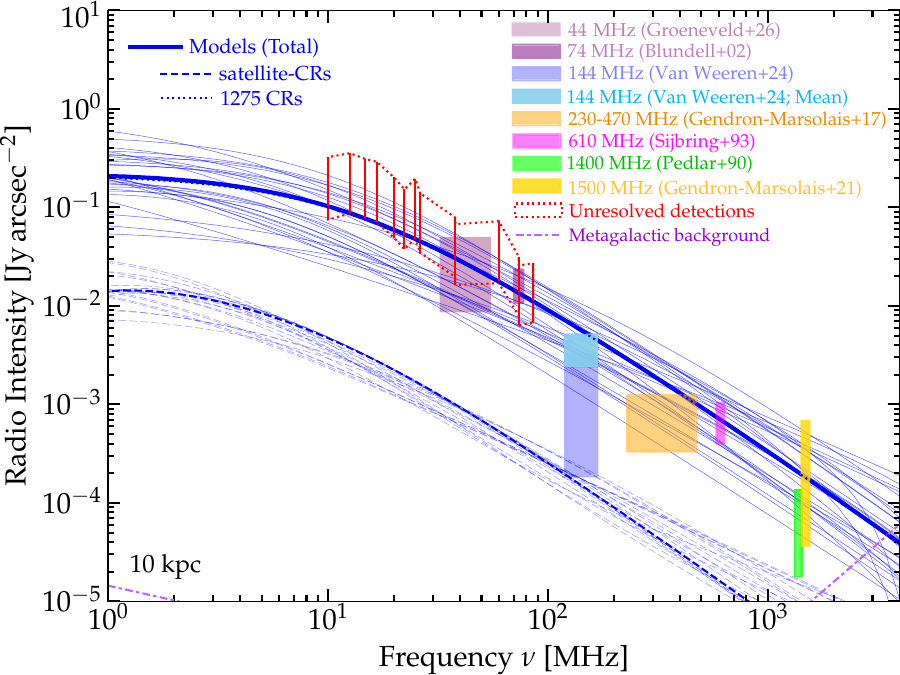}	
	\includegraphics[width=0.48\textwidth]{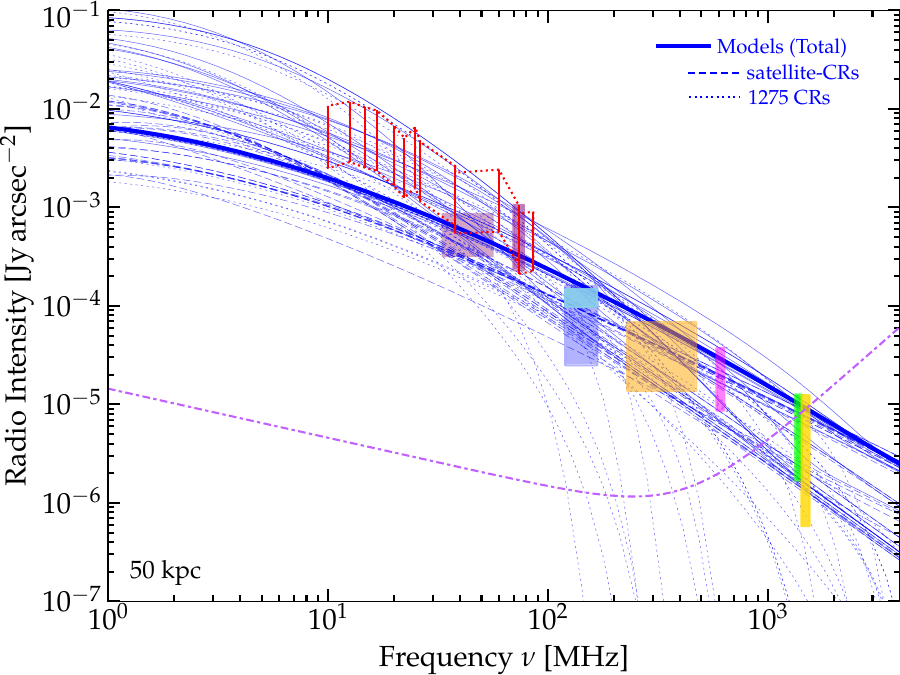}	
	\caption{Radio spectra (\S~\ref{sec:obs.radio}) in the minihalo. We compile spatially-resolved observations (\textit{shaded}, labeled) at radii $10\,$kpc (\textit{top}) and $50\,$kpc (\textit{bottom}), where there is maximal coverage across wavelengths. We also add spatially-unresolved detections at $<76\,$MHz assuming they trace the $74$\,MHz profile (see \S~\ref{sec:obs.radio:minihalo}), and metagalactic backgrounds. For models we show the total and contribution from CRs accelerated in NGC 1275 (streaming out) and satellites (mini-minihalos; \S~\ref{sec:obs.radio:gianthalo}). At $\sim 10\,$kpc, NGC 1275 CRs dominate at all frequencies. By $\sim 50-100\,$kpc, CRs from NGC 1275 dominate at $\lesssim 100\,$MHz but CR-IC losses have suppressed higher-energy CRs, so in most of the model variants plotted the $\gtrsim $\,GHz is dominated by satellite-accelerated CRs. Spectral steepening is clearly evident between $10-50\,$kpc, in both models and observations.
	\label{fig:radio.spectrum}}
	\vspace{-0.2cm}
\end{figure}

\begin{figure}
	\centering
	\includegraphics[width=0.48\textwidth]{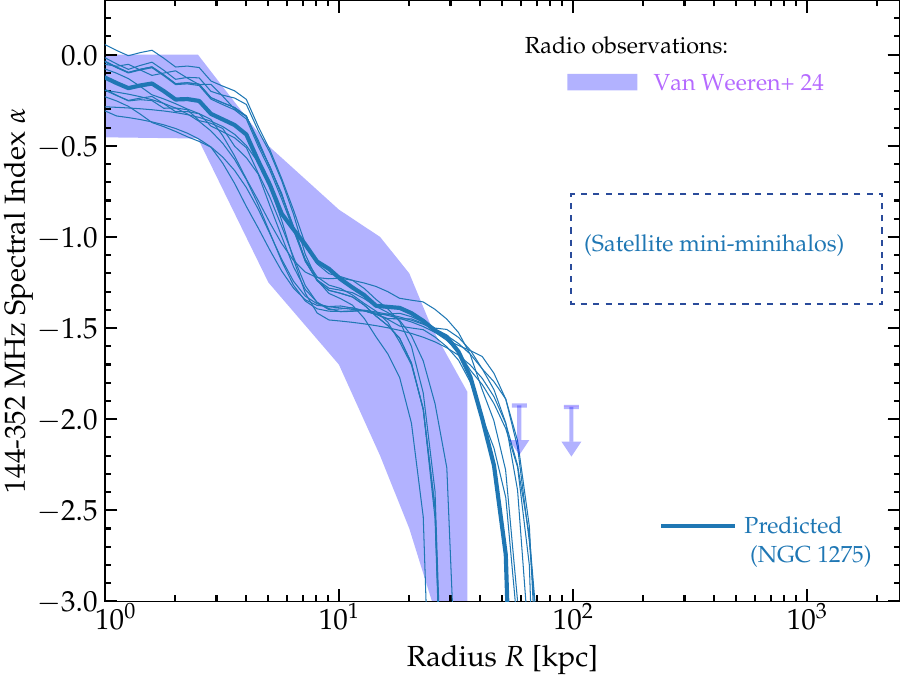}
	\includegraphics[width=0.48\textwidth]{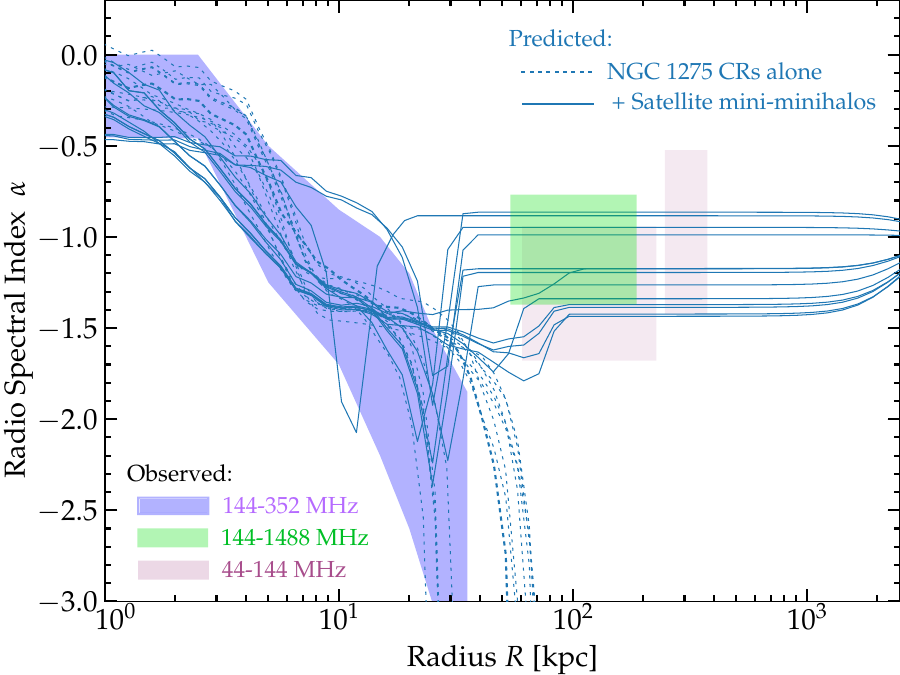}
	\caption{\textit{Top:} Radio spectral index ($S_{\nu} \propto \nu^{\alpha}$) from $144-352$\,MHz. Observed (\textit{shaded}) range shows the interquartile range of indices observed in individual pixels at a given radius $R$ in spatially-resolved 2D images (from the most sensitive LOFAR+VLA data compiled and synthesized to the same $5^{\prime\prime}$ or $1.8$\,kpc beam in \citealt{vanweeren:2024.perseus.giant.radio.halo.filled.electrons.just.tiny.fraction.high.energy}). The lack of data at $\gg 30\,$kpc owes to pixels without a $352$\,MHz detection at this resolution (hence the upper limit). This will differ from the spectral index of the mean spectrum. The steepening/softening with distance from NGC 1275 is evident, and agrees well with that predicted for CRs streaming from the source (though the index within our injection region at $\lesssim 1-10\,$kpc depends on our assumed injection model). At radii $\gtrsim 100\,$kpc we predict these frequencies are dominated by satellite mini-minihalos, and so the index becomes $\sim$\,constant at a shallower $\alpha \sim -1$ as it reflects the mean of mini-minihalos rather than older streaming CRs from NGC 1275.
	\textit{Bottom:} Same, showing the index calculated using the mini-minihalo model from Figs.~\ref{fig:radio}-\ref{fig:radio.spectrum}. We show the $144-1488\,$MHz indices from the same observed maps at the largest radii with area-filling detection at $1488$\,MHz.
	\label{fig:radio.spectral.index}}
\end{figure}

\subsubsection{The Mini-Halo and Its Spectrum}
\label{sec:obs.radio:minihalo}

Fig.~\ref{fig:radio} shows a number of different observations of the radio mini-halo and giant halo. We specifically compile and compare to spatially-resolved radial profiles of the diffuse gas emission from 
$\sim 30-57.5$ ($\sim 44$) MHz LOFAR LBA \citep{groeneveld:2026.investigatingradioemissionperseus}, 
$\sim 74\,$MHz VLA \citep{blundell:2002.perseus.74mhz.vla.obs}, 
$\sim 270-430$\,MHz VLA \citep{gendron.marsolais:2017.perseus.300mhz.vla.obs.and.compilation,gendron.marsolais:2020.perseus.images.300mhz},
$\sim 610$\,MHz WSRT \citep{sijbring:1993.perseus.610mhz.obs}, 
as compiled and cleaned in \citet{gendron.marsolais:2017.perseus.300mhz.vla.obs.and.compilation}, 
plus 
$\sim 144$\,MHz LOFAR HBA \citep{vanweeren:2024.perseus.giant.radio.halo.filled.electrons.just.tiny.fraction.high.energy},
and $1.4-1.5$\,GHz VLA \citep{pedlar:1990.radio.1.4ghz.radial.profile,gendron.marsolais:2021.vla.1.5ghz.perseus.imaging}, as presented in those papers. 
Note that \citet{vanweeren:2024.perseus.giant.radio.halo.filled.electrons.just.tiny.fraction.high.energy} did not fit their ``minihalo'' profile at $\lesssim 30\,$kpc in an attempt to separate this from any ``central'' components associated with 3C 84 (NGC 1275) jets/cavities/lobes -- but here we simply want the total emission profile, since in our models these are the same (continuous) lepton population, so we show the entire measured profile down to $\sim $\,kpc (the highest resolution of the maps therein). 

Fig.~\ref{fig:radio.spectrum} shows the corresponding radio spectrum at a couple different radii, with the meta-galactic backgrounds and models as well. In the spectrum plots, we also include \textit{unresolved} literature data compiled from \citep{roger:1968.22mhz.radio.source.compilation,aslanian:1968.60.mhz.radio.compilation,braude:1969.radio.compilation.decameter,kellermann:1969.3c.catalog.radio.sed.updates,laing:1980.radio.compilation.3c.multifreq,kuehr:1981.3c.update.highfreq.sources.bright,viner:1975.26.mhz.radio.source.compilation,kassim:2007.74.mhz.sources.compilation} at $\le 76\,$MHz. {We stress these do not have radial profiles; instead we convert the total flux detected to a surface brightness here either (1) by assuming the radial profile shape (up to a normalization) is given by the same shape as the mean 144\,MHz data (with most of the integrated emission from $\sim 5-50\,$kpc); or (2) our default approach, simply renormalizing all detections by the same constant at a given $R$ so that the 74\,MHz integrated data agrees with the spatially-resolved 74\,MHz point from \citet{blundell:2002.perseus.74mhz.vla.obs}. These give similar results to within a factor of $\sim 3$, so we add in quadrature a systematic factor $\sim 3$ $\pm1\,\sigma$ uncertainty to the points to represent this (and the spatial scatter in the emission at a given radius as we have done for the resolved profiles, as compared to the much smaller errors in the total flux which are quoted in the literature).} This is not as rigorously constraining, but gives a reasonable consistency check (the spectrum should not deviate too much from these points at lower frequencies) and allows us to extend the data to $\sim 10\,$MHz. 
To show the evolution of the spectrum where the deepest and highest dynamic-range spatial data is available, we also plot in Fig.~\ref{fig:radio.spectral.index} the radial profile (and range at a given radius) of the spectral index from the $\sim 144-430\,$MHz LOFAR+VLA profiles. 

The models here (recalling these are not fits) with emission just from NGC 1275-injected CRs reproduce the radio data from $\sim 10-1500\,$MHz well out to $R \gtrsim 50\,$kpc. At present, the LOFAR measurements at 144 MHz probe the most sensitively (with both the highest spatial resolution and surface brightness sensitivity/dynamic range in radius), and these show a steady decline in intensity/SB as $\propto R^{-(1.5-2)}$ over this radius range. This is steeper than the soft X-ray/CR-IC SB ($\propto R^{-(0.5-1)}$), over a similar range, despite both being proportional to CRs. There are two important effects. (1) The radio \textit{at a given frequency} is proportional to a steep power of $B$ (e.g.\ $\sim B^{4}$), as both the total emitted synchrotron ($\propto B^{2}$) and the characteristic frequency for a given CR energy ($\propto B$) shift,  so even a weak radial dependence of $B(r)$ is non-negligible here. And (2) The two are sensitive to somewhat different CR energies, with $144\,$MHz probing $\sim 4\,{\rm GeV}\,B_{\rm \mu G}^{-1/2}$ CRs (for a diffuse mean-field $B_{\rm \mu G}\,{\rm \mu G}$, discussed below) while a $\sim0.5-1\,$keV soft X-rays IC scattered from the CMB probe $\sim 0.5\,$GeV CRs. So the radio is influenced more strongly by losses and aging of the CRs as they propagate.

These are also directly reflected in the spectral index evolution with radius, shown explicitly in Fig.~\ref{fig:radio.spectral.index} for the best-measured values over the range of minihalo size in \citet{vanweeren:2024.perseus.giant.radio.halo.filled.electrons.just.tiny.fraction.high.energy}, as well as implicitly in our comparison with $144\,$MHz and $1.4$\,GHz data and radio SEDs in Figs.~\ref{fig:radio} \&\ \ref{fig:radio.spectrum}. Like in many other minihalos \citep[see][]{savini:2018.lofar.ultrasteep.radio.emission.surrounding.minihalo.larger.radii.steepening.as.expected.in.coolcore,cuciti:2021.diffuse.cluster.radio.halo.fluxes.brightness.lofreq.gmrt.steeper.slopes.larger,ignesti:2022.lofar.zdrop.cluster.central.ultrasteep.radio.relic.lofar.losing.energy.outside.of.center,edler:2022.abell1033.cluster.case.study.lofar.decaying.crs.super.steep.radio.rejuvenated.at.special.location.older.at.larger.r.as.expected}, there is a clear spectral steepening with distance $r$ from the central source, out to $\sim 50-100\,$kpc.\footnote{This is also consistent with the trend seen in the LOFAR 45-144\,MHz LBA-HBA comparison at $\lesssim 50\,$kpc, with the steepening even more prominent toward the ``ghost cavities'' \citep{groeneveld:2026.investigatingradioemissionperseus}.} This is of course qualitatively expected if CRs are injected by a dominant central source and propagate outwards, as in the models here. However we caution that even in the models here which (for simplicity) assume a constant CR injection rate and propagation speed, there is not a one-to-one correspondence between spectral index and CR age. After all (1) with diffusion+streaming, even with constant source injection rates, there is no single ``age'' observed at a given radius $R$; (2) the CR spectra have significant curvature, and indeed must have curvature if they are affected by losses, so there cannot be a single $\alpha$-age relation at any $r$; and (3) again $B$ depends on $r$, so the portion of the CR spectrum sampled will depend on radius. For example, in the central few kpc, $B$ can be as large as $\sim 10\,{\rm \mu G}$, meaning the 144-352 MHz spectral index probes CRs with energies as low as $\sim 1-2\,$GeV, which near the injection zone can be at or even below the peak in the CR spectrum (Fig.~\ref{fig:crspec}), meaning that the portion of the CR spectrum sampled is very hard, even if the ages are relatively old (e.g.\ this would be true even in the LISM with known CR ages $\sim 10^{7}\,$yr at these energies; see \citealt{bisschoff:2019.lism.cr.spectra} and references therein), and even if the higher-energy CR spectrum at these radii is much softer. As we discuss below (\S~\ref{sec:obs.radio:gianthalo}), the spectrum ``re-hardens'' at larger radii as the contribution of CRs that have not purely streamed from NGC 1275 becomes more important especially at the highest energies (where the loss times are shortest). 

This also means there is some degeneracy between the profile of $B(r)$ and injection spectrum and propagation speeds (loss time versus propagation time), evidenced by the different models here. The radio alone does not disambiguate these, and we show below that all the models considered are within the range allowed by independent constraints on the magnetic field. 
Importantly though, these models \textit{simultaneously} fit the radio mini-halo spectrum, its radio profile, the hard X-rays, $\gamma$-rays, \textit{and} soft X-rays with most of the soft X-ray coming from CR-IC. Thus it appears that the mini-halo in Perseus itself follows naturally from the ACRH hypothesis for the cooling flow. We also do not appear to require any re-acceleration here, and in fact including the maximum possible reacceleration within the context of the models here makes very little difference to any of the predictions shown (\S~\ref{sec:stream.reacc}). Nor is there any continuous hadronic injection or new acceleration from e.g.\ shocks assumed here. Thus, re-acceleration and/or extended (e.g.\ hadronic) injection, while they could occur, are not necessary to explain the minihalo. In \S~\ref{sec:differences}, we discuss why this differs from some claims in the literature, which made very different assumptions about CR transport that do not apply here. 

Note that in Figs.~\ref{fig:radio}-\ref{fig:radio.spectral.index} it is important to consider the range of values in different positions/pixels/beams at a given $R$ in the 2D maps, especially for the spectral index. This is because we are interested in the properties of the diffuse, volume-filling medium. At higher frequencies especially, the total (azimuthally-averaged) emission even by $R \gtrsim 30-40\,$kpc can be strongly dominated by a couple of ``hot spots'' in the map (e.g.\ NGC 1272 and its associated stripped/shocked structures, as shown explicitly at $350$ and $1.5$\,GHz in \citealt{gendron.marsolais:2017.perseus.300mhz.vla.obs.and.compilation,gendron.marsolais:2021.vla.1.5ghz.perseus.imaging}). If the median spectral index $\alpha$ at a given $R$ is very steep (i.e.\ there is very little high-frequency emission), then even a tiny volume which is bright at these wavelengths can strongly bias the spectral index from the mean/integrated/unresolved spectrum (akin to the effects of gas clumping in X-ray emission).

\begin{figure*}
	\centering
	\includegraphics[width=0.98\textwidth]{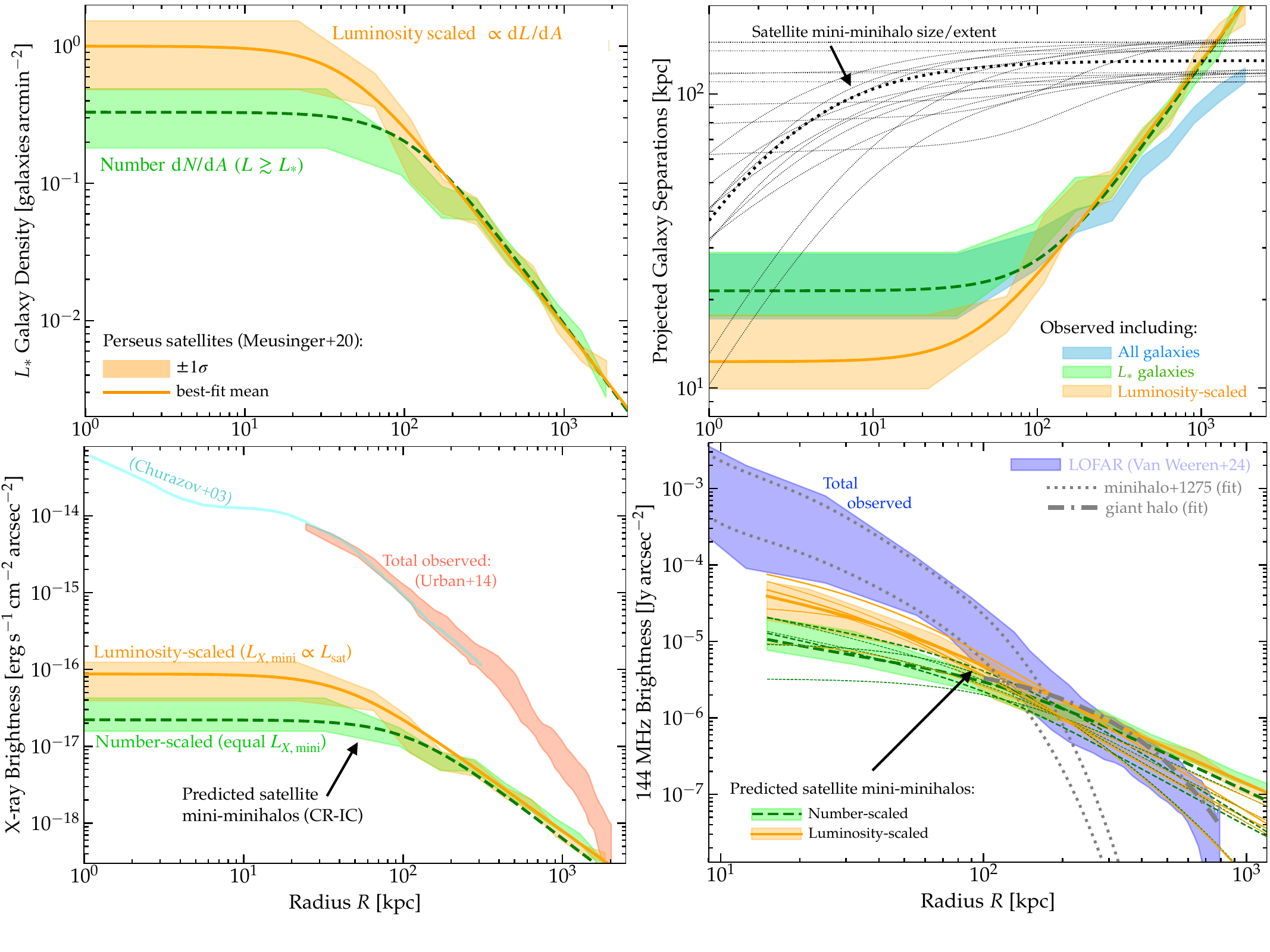}
	\vspace{-0.5cm}
	\caption{Contribution of satellites to the giant radio halo (\S~\ref{sec:obs.radio:gianthalo}). 
	{\em Top left:} Observed Perseus satellite galaxy number density profile and stellar ($i$-band) luminosity profile (scaled to number by median luminosity in the sample), from \citet{meusinger:2020.galaxy.bh.population.in.perseus}.
	{\em Top right:} Same, converted to mean projected separation on the sky between satellites (excluding NGC 1275). The ``all galaxies'' range uses a fainter sample which includes dwarf galaxy companions. We compare the predicted typical size of the ACRH or ``mini-minihalo'' around each satellite, defined as in \paperthree\ (see also see \S~\ref{sec:obs.radio:gianthalo}). Out to $\approx$\,Mpc, this is larger than the inter-galaxy separation, so the mini-minihalos are volume-filling and diffuse.  
	{\em Bottom left:} Predicted collective X-ray CR-IC SB from satellite ACRHs/mini-minihalos, assuming each is a scaled copy of the central ACRH with the total injection rate of satellites given by their total estimated BH accretion rate or mass time-averaged over $\sim$\,Gyr. This is negligible compared to the total X-ray SB observed in Perseus at the same radii, dominated by CR-IC plus thermal diffuse gas emission at large $R$.
	{\em Bottom right:} Predicted collective 144 MHz radio SB from the mini-minihalos, with the same assumptions. These are negligible at $\ll 100\,$kpc where the NGC 1275 mini-halo dominates, but the collective, smooth, volume-filling CR ACRH from satellites produces a profile remarkably similar to the inferred LOFAR giant halo and emission at $\gtrsim 100\,$kpc. Lines show model variants as before. 
	\label{fig:satellites}}
\end{figure*}

\begin{figure}
	\centering
	\includegraphics[width=0.48\textwidth]{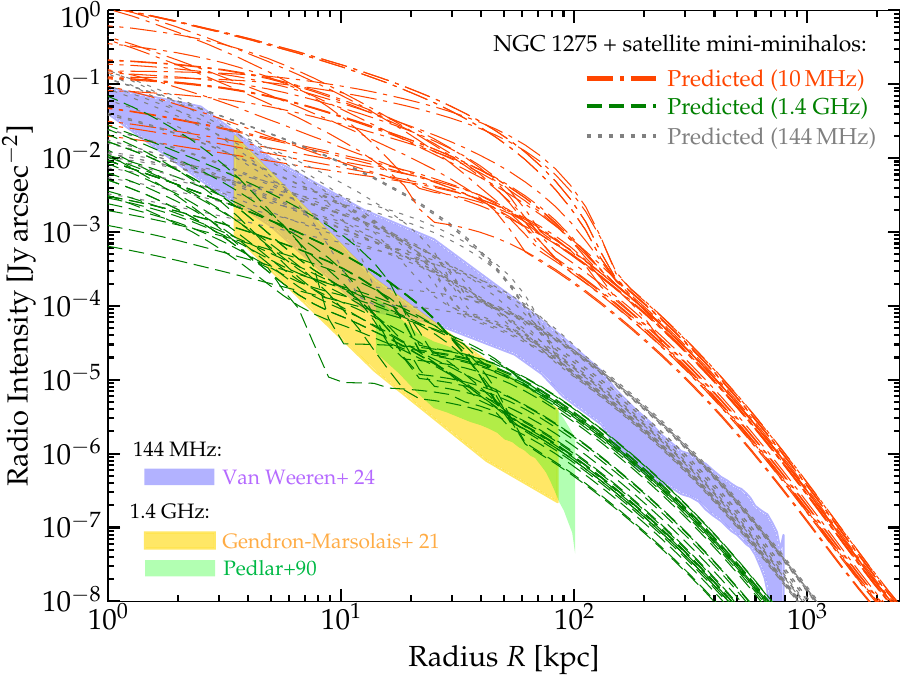}
	\caption{Radio surface brightness profiles (\S~\ref{sec:obs.radio}), as Fig.~\ref{fig:radio.wsats}, but now including both the CRs streaming from NGC 1275 \textit{and} the mini-minihalo contributions from Fig.~\ref{fig:satellites}. 
	The satellite mini-minihalos can naturally explain the radial profile and spectra of the giant halo and its transition from the mini-halo sourced by NGC 1275, though our models in principle allow for other low surface-brightness CR sources at large radii (e.g.\ shocks, reacceleration, etc.) and this has weak effects on the X-rays at those radii (Fig.~\ref{fig:satellites}). 
	\label{fig:radio.wsats}}
\end{figure}

\subsubsection{The Giant Halo: Many Mini-Mini-Halos?}
\label{sec:obs.radio:gianthalo}

Observationally, something does appear to change at $\gg 100\,$kpc. LOFAR measures a giant radio halo at $\sim 144\,$MHz extending to $\sim 1\,$Mpc, but the radial profile changes shape outside of $\gtrsim 100\,$kpc (see Fig.~\ref{fig:radio} and for more detailed analysis, see \citealt{vanweeren:2024.perseus.giant.radio.halo.filled.electrons.just.tiny.fraction.high.energy}). Moreover the spectral slopes appear to change and harden. Some care is needed interpreting this, as there is not much significant detected emission in high-resolution maps at wavelengths other than $144\,$MHz at the largest radii ($\gtrsim$144\,MHz), so the data must be rebinned to a much larger effective beam size at other wavelengths for a significant detection and spectral index measurement at these larger radii, but nonetheless it appears that the 144-1488 MHz spectral index $\alpha^{1488}_{144}$ transitions from steeply-falling with radius (from $\sim 0$ in the central few kpc to $\sim -2$ at $\sim 30\,$kpc) to a much more weakly $r$-dependent $\alpha^{1488}_{144} \sim -1$ at $\gtrsim 100\,$kpc (see \citealt{vanweeren:2024.perseus.giant.radio.halo.filled.electrons.just.tiny.fraction.high.energy} and, for similar conclusions between $\sim 300-1500\,$MHz, \citealt{gendron.marsolais:2021.vla.1.5ghz.perseus.imaging}, or between $\sim 45-144\,$MHz, albeit with larger error bars, see \citealt{groeneveld:2026.investigatingradioemissionperseus}). This transition defines the giant halo detected by LOFAR.

It is extremely challenging to construct an ACRH/CR-IC model which explains the giant halo via \text{only} CRs accelerated in NGC 1275 propagating out. Given the minimum loss time (from just CR-IC losses off the CMB), getting CRs to Mpc would require order-of-magnitude faster bulk CR transport speeds ($\gtrsim 1000\,{\rm km\,s^{-1}}$), which would suppress their energy density at small radii unless we invoked order-of-magnitude larger injection rates (which would be energetically problematic with constraints on NGC 1275), and the very fast transport would then imply much less spectral softening/curvature at small radii (potential contradiction with X-ray and radio spectral shapes). It is instead possible some re-acceleration model or hadronic injection at larger radii could be responsible, but this would have to pick up rather suddenly and strongly outside of $\sim 100\,$kpc, while being weak interior, which seems to contradict what we know about both the strength of turbulence and density profile (which helps determine the hadronic injection rate) with radius. 
It is instead possible that the giant halo simply owes to other physics, e.g.\ shocks induced by cluster mergers accelerating new CRs throughout the cluster, as is widely believed to occur in some other clusters with giant radio halos (\citealt{donnert:2013.modeling.giant.radio.halos,brunetti.jones:2014.cr.diffusion.clusters.minihalo.giant.halo.review.many.solutions.including.faster.transport.and.local.sources,vanweeren:2019.cluster.radio.review}, although see also \citealt{vazza:2015.radio.relics.problems.with.dsa.on.large.scales,whittingham:2024.cluster.Bfield.synch.measurements.biased.to.strongest.B.subregions} for discussion of challenges to this picture). However, it is worth noting that those giant halos tend to be harder/shallower radio sources (most known giant halos coming from $\sim 1.4$\,GHz searches, not $144\,$MHz without a $1.4$\,GHz counterpart at large radii, as in Perseus), and are generally associated with NCCs, not SCCs, and in particular known merging NCCs.\footnote{There are a couple of other Perseus-like exceptions to this rule in SCCs in very massive halos, e.g. \citet{bonafede:2014.giant.radio.halo.radio.quiet.qso.coolingflow,boschin:2018.qso.coolingflow.relaxed.halo.despite.radio.contour.arguments.fully.dynamically.relaxed.radio.cannot.be.from.merger.xray.radio.morph.extent.trace.each.other}.}
Also note that while it has recently been proposed that Perseus had a significant merger at redshift $z \sim 1$ or so, \citealt{hyeonghan:2025.perseus.major.merger.weak.lensing.mass.reconstruction}, the 144 MHz giant halo emission does not show the same morphology or position angle as the putative merger candidate.

But the ACRH models here do make a prediction for a diffuse, extended component. In \S~\ref{sec:obs.radio:minihalo} we focused just on NGC 1275/3C 84 as the CR source, since this is both the BCG and most prodigious radio and $\gamma$-ray and CR source in Perseus, and it is the region on which the global X-ray surface brightness is centered. But these models for ACRHs predict that \textit{every} galaxy should be surrounded by an ACRH from CRs accelerated by both SNe and AGN in the galaxy over the past couple Gyr. Indeed, in \paperone, we specifically argued that the ACRHs around Milky Way and Andomeda-mass galaxies (typical satellite galaxy masses in a massive cluster like Perseus) naturally explained the slowly-falling ($\propto R^{-1}$) soft X-ray emission profiles observed by ROSAT and eROSITA \citep{anderson:2013.rosat.extended.cgm.xray.halos,zhang:2024.hot.cgm.around.lstar.galaxies.xray.surface.brightness.profiles,zhang:2024.erosita.hot.cgm.around.lstar.galaxies.detected.and.scaling.relations} around similar-mass galaxies out to $\sim 100-200\,$kpc around each such galaxy (of order their individual virial radii). 
Moreover, while NGC 1265 is clearly the brightest radio and $\gamma$-ray source, the \textit{total} CR injection from all satellite galaxies must be comparable. For example, many Perseus galaxies are individually bright radio sources themselves: while NGC 1275 (3C 84) is the most luminous, NGC 1264 (3C 83.1B) is within a factor of a couple of its radio luminosity (at both 144 and 1500\,MHz), and IC 310 (also a luminous TeV $\gamma$-ray source,  \citealt{kadler:2012.ic310.luminous.blazar.tev.gamma.ray.source}) within a factor of a few \citep{gendron.marsolais:2021.vla.1.5ghz.perseus.imaging,vanweeren:2024.perseus.giant.radio.halo.filled.electrons.just.tiny.fraction.high.energy}, and both of those are at $\gtrsim 600\,$kpc from the center. Indeed the total radio emission of detected satellites in Perseus at 600\,MHz is similar to the luminosity of NGC 1275, and even excluding the 3 (or 10) brightest detected individual radio satellite galaxies, the next $\sim 20$ brightest radio satellites sum to a luminosity of $\sim 25\%$ ($\sim 10\%$) that of NGC 1275 \citep{gisler.miley:1979.600mhz.perseus.radio.sources}. 

Of course, for these halos we care about $\gtrsim$\,Gyr-averaged, not instantaneous radio CR injection rates, which might be better traced by SMBH mass (or stellar mass, if SNe dominate the CR injection in lower-mass satellites). But, like in most clusters, in Perseus the total SMBH mass and stellar mass in the cluster are dominated by the sum of satellites. NGC 1275 may not even host the largest SMBH in the cluster (\citealt{vandenbosch:2012.ngc1277.blackhole.mass}, but see also \citealt{emsellem:2013.ngc.1277.bh.mass,walsh:2016.overmassive.bh.outlier}), but even if we exclude the 10 largest-mass BHs (plus NGC 1275), the total SMBH mass of Perseus is $\sim 10^{10}\,M_{\odot}$ \citep{meusinger:2020.galaxy.bh.population.in.perseus}, comparable to NGC 1275 and consistent with a Gyr-averaged CR injection rate of $\sim 10^{44}-10^{45}\,{\rm erg\,s^{-1}}$ if we assume average growth timescale $\sim\,t_{\rm Hubble}$ (median Eddington ratio $\sim 0.0007$, similar to the observed values for quenched galaxies with BHs in this mass range, see \citealt{torbaniuk:2024.bhar.sfr.mstar.relations}) and fraction $\sim 10^{-3}$ of accretion energy into CRs (similar to NGC 1275 inference and AGN feedback models from \citealt{su:turb.crs.quench,su:2021.agn.jet.params.vs.quenching,wellons:2022.smbh.growth,byrne:2023.fire.elliptical.galaxies.with.agn.feedback}).
Given the total observed stellar mass and star formation rate of Perseus satellites \citep{cuillandre:2025.euclid.perseus.stellar.mass.functions.bh.mass.functions}, the sum of Ia+core-collapse SNe (assuming standard rates from \citealt{mannucci:2006.snIa.rates,nugrid:yields} and $\sim 10^{50}\,{\rm erg}$ per SNe in CRs with a few to ten percent of that in leptons, as in the LISM; \citealt{higdon:crs.accel.in.superbubbles.in.sne.ejecta.mass.dominated.regions,parizot:2004.superbubble.cr.accel.bulk,becker.tjus:2020.cr.multi.messenger.accel.regions}) likely contributes a significantly smaller CR luminosity ($\lesssim 10^{43}\,{\rm erg\,s^{-1}}$, averaged over $\sim$\,Gyr).

Thus we expect each satellite galaxy in Perseus to contribute its own ACRH or ``mini mini-halo'' (MMH). Individually, the luminosity of one MMH -- set by its time averaged CR injection rate over the last $\sim$\,Gyr -- will be far lower than that of NGC 1275, but their collective luminosity should be a significant fraction of NGC 1275.  However, as shown in \papertwo-\paperthree, and Fig.~\ref{fig:satellites}, the radius and ``effective CR-IC temperature'' (CR spectral shape) and scaled radial profile of the ACRH (MMH, here) emission depends very weakly on (sub)halo mass, magnetic field strength, injection rate, or other properties that could vary galaxy-to-galaxy. The reason is simple: it is set by how far CRs travel in a loss time, itself set by the CMB ($R_{\rm MMH} \sim v_{\rm st,\,eff} \,\Delta t_{\rm loss} \sim 100\,{\rm kpc}\,(v_{\rm st,\,eff}/100\,{\rm km\,s^{-1}})$ or $\Delta t_{\rm loss} \sim $\,Gyr for GeV leptons off the CMB). Because of the slowly-falling profiles (see also Fig.~\ref{fig:xr.profiles}), most of the emission occurs at the largest radii of the MMHs -- they are not strongly peaked around their individual sources. 
In Perseus, the (projected) number density of similar-mass satellite galaxies (${\rm d} N_{\rm sat}/{\rm d} A$) as a function of cluster-centric radius is well-measured, and shown in Fig.~\ref{fig:satellites}. From this one can immediately calculate an effective projected radial distance $\Delta r_{\rm sat}$ (radius of the equivalent circles which would just overlap) between satellites at a given global $R$, as $\Delta r_{\rm sat} \equiv (\pi {\rm d} N_{\rm sat}/{\rm d} A)^{-1/2}$. We see that out to $\sim$\,Mpc, $\Delta r_{\rm sat} \lesssim R_{\rm MMH}$. This means that the MMHs overlap, and the observed emission from them will be  diffuse/volume-filling. Also note that over $\sim$\,Gyr timescales to reach $R_{\rm MMH}$, a more detailed model would account for the actual satellite bulk motion, which can be faster than $v_{\rm st,\,eff}$ and therefore further mix the deposited CRs throughout the halo along the orbit of each satellite, rather than giving a strictly spherical profile around each. Globally, the brightness/luminosity distribution from MMHs should therefore trace the satellite distribution (either weighted by number of massive galaxies, if each contributes comparably, or something like luminosity-weighted if larger/more active galaxies contribute systematically more over Gyr timescales).\footnote{Note that ideas like this for radio halos have certainly been discussed before, going back at least to \citet{jaffe:1977.coma.electrons.radio.emission}, but (1) they considered the giant, high-frequency radio halo in Coma, which is very different from Perseus, and (2) they made very different assumptions about the possible transport physics, magnetic field strength, etc. (reviewed in \S~\ref{sec:differences:reaccel}) which lead to quite different predictions here.}

Fig.~\ref{fig:satellites} therefore considers this minimal model for the giant halo. Specifically we take the CR spectrum of the sum-of-MMHs to be the same as the bolometric emission-weighted sum of the ``primary'' (NGC 1275) minihalo/CR emission (i.e.\ since we are integrating over the entire size of the MMHs and mixing them together, we see some ``averaged'' spectrum), scaled by some per-satellite total CR injection rate relative to NGC 1275 (i.e.\ $L^{\rm bol}_{\rm MMH} \sim \dot{E}_{\rm cr,\,\ell}^{\rm sat}/\dot{E}_{\rm cr,\,\ell}^{1275} \sim (N_{\rm sat}^{-1}\,\sum \dot{E}_{\rm cr,\,\ell}^{\rm sat})/\dot{E}_{\rm cr,\,\ell}^{1275}$). Note our predictions are not particularly sensitive to how we weight this average. We then distribute this emission according to the observed radial distribution of satellites, assuming for specificity that the total CR injection summed from all satellites $\sum \dot{E}_{\rm cr,\,\ell}^{\rm sat}$ is $\sim 10\%$ of the NGC 1275 injection rate. This is a simple model but should give the correct gross scalings without introducing the complication (and extra degress-of-freedom) of considering local variations in $n$, $T$, $B$, etc.\ around every source in Perseus. 

In Fig.~\ref{fig:satellites} we see that this predicts an extended SB profile remarkably similar to the giant halo in Perseus as measured in both \citet{vanweeren:2024.perseus.giant.radio.halo.filled.electrons.just.tiny.fraction.high.energy} and \citet{groeneveld:2026.investigatingradioemissionperseus}. While we have some (factor of a couple in either direction) freedom in the normalization, it is very similar to what we would naively expect from the simple arguments above. In other words, the total giant halo luminosity is roughly what is expected from the sum of MMHs.
More notably the radial profile shape -- the giant halo ``size'' and profile -- is naturally predicted by the satellite distribution,\footnote{Our simple model perhaps cuts off slightly more slowly at $\sim 1\,$Mpc compared to observations, but note in Fig.~\ref{fig:satellites} that this is precisely where the satellite density drops so that the MMHs become non-volume-filling, which we do not account for in the toy model and will move the predictions closer to observations. However this is also where the radio data become more uncertain owing to the very low surface brightness.} without any free parameter. Fig.~\ref{fig:radio.wsats} shows the combination of the mini-halo (CRs from NGC 1275) and these MMH models, at $10$, $144$, and $1400$\,MHz, showing their combination neatly reproduces the observed ``mini+giant'' radio halo profile and spectrum.

And also importantly, this explains the otherwise unusual behavior in the spectral indices versus radius (Fig.~\ref{fig:radio}). At radii $R \ll 100\,$kpc, the radio is dominated by CRs propagating from NGC 1275, both because of its brightness, and because there are no massive satellites interior to this radius, so the CRs are aging (and the synchrotron steepens) as they propagate outwards. But at $R \gg 100\,$kpc, one is seeing the sum of many MMHs around individual galaxies, each averaged over their own propagation distances $<100\,$kpc, so one sees only a globally emission-weighted average spectral index. This should (1) be roughly independent of cluster-centric radius $R$, and (2) be harder/shallower than the outermost radii of any single MMH or the NGC 1275-sourced minihalo (because it integrates over all emission), closer to $\alpha \sim -1$ predicted here. Essentially, the radio giant halo is less steep than the outermost mini-halo because the CRs are younger, on average, since they come from satellites at those cluster-centric radii $R$, not from the BCG CRs having propagated to $R$. 

Note that these satellite MMHs/ACRHs also predict their own CR-IC X-ray halos, as discussed above, so in Fig.~\ref{fig:satellites} we also show the predicted contribution from these to the total X-ray SB profile. Here, unlike the radio, the satellite ACRHs are largely negligible. That is because the true thermal emission from the diffuse hot cluster gas is much larger at these radii than the contributions from small galaxies. As shown in \paperone, X-ray ACRHs around \textit{isolated} (non-cluster) low-mass galaxies are only detectable because their virial temperatures and gas masses are quite low, so they have proportionally much less thermal X-ray emission.

\begin{figure}
	\centering
	\includegraphics[width=0.98\columnwidth]{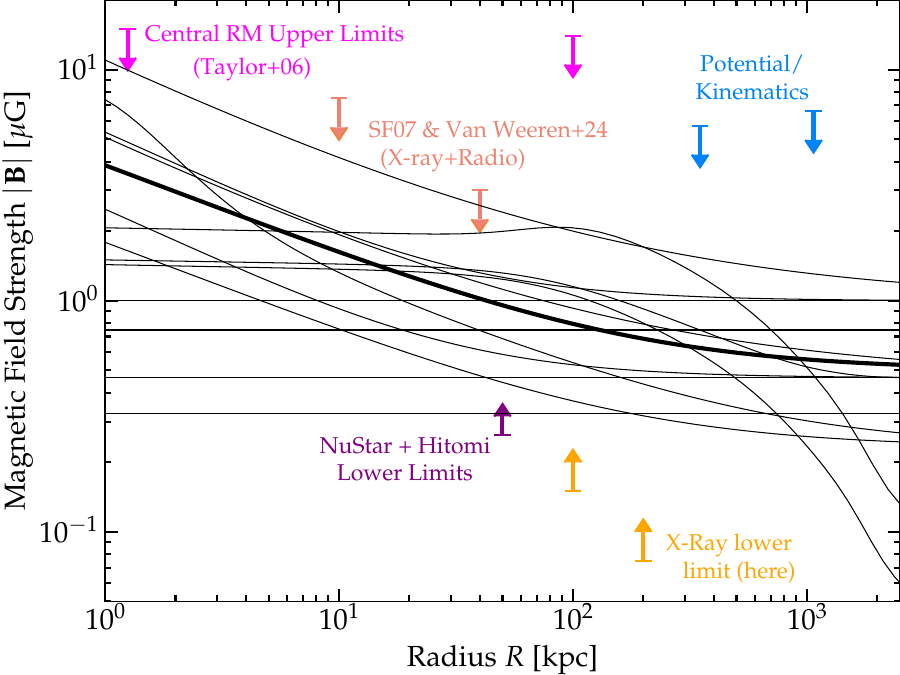}
	\caption{Magnetic fields $B$ (volume-weighted $\langle |{\bf B}|\rangle_{\rm vol}$; \S~\ref{sec:B}) profiles, in the diffuse volume-filling ICM, assumed in the models here (lines, style as in Fig.~\ref{fig:xr.profiles}). We compare upper limits from the RM measurement of the radio core at $\lesssim 0.2\,$pc scales (given the density profile assumed; ``Taylor+06''), from radio-X-ray comparisons placing a lower-limit to IC emission (``SF07 \&\ Van Weeren+24''), and non-thermal pressure effects on kinematics/potential/mass reconstruction (``Potential/Kinematics''), and lower limits from radio+hard X-rays from ``Nustar+Hitomi'' and from the compilation here (``here''). Details of limits in \S~\ref{sec:B}. The models here which reconcile the radio halos and X-ray emission are all consistent with the limits on $B$, with $\sim \mu {\rm G}$ fields (constant or weakly-declining with $R$), and plasma $\beta \sim 100$. 
	\label{fig:B}}
\end{figure} 

\subsubsection{Constraints on the Perseus Magnetic Field}
\label{sec:B}

Given that the radio depends on both the CR leptons \textit{and} magnetic field, it is important to compare the assumed models for the volume-filling mean field $B \equiv \langle |{\bf B}| \rangle_{\rm vol} \sim {{\rm Vol}}^{-1} \int |{\bf B}|\,{\rm d}{\rm Vol}$ in Perseus we assume, to different observational constraints, shown in Fig.~\ref{fig:B}. 

Unfortunately there are no direct constraints on $B$ (e.g.\ from Zeeman-splitting) in Perseus or similar clusters. The closest to a ``clean'' constraint comes from Faraday rotation, specifically detection of a rotation measure 
${\rm RM} \propto  \left( \int_{-\infty}^{+\infty} {\rm d} I_{\rm em} \int_{\ell_{\rm em}}^{\infty} n_{e}({\bf x})\,B_{\|}({\bf x})\,{\rm d}\ell \right) / \int_{-\infty}^{+\infty} {\rm d} I_{\rm em}$ 
along the path of observed photons. Note we include the convolution in ${\rm d}I_{\rm em} ={\rm d}I_{\rm em}({\bf x},\,{\bf B}[{\bf x}])$ (the emissivity at each point) since we cannot assume a point source at infinity. The only observed RMs in Perseus (and in similar SCCs) come from jet-core regions of the nucleus, e.g.\ the $\sim 0.2$\,pc region within the compact radio core (at few-pc separation along the jet from the peak radio intensity) of 3C 84 (the center of NGC 1275's radio AGN emission) described in \citet{taylor:2006.perseus.central.RM.to.core.radio.emission}. 
This is sometimes quoted as a measurement of the cluster $B$, but in addition to depending on the convolutions above, many authors including \citet{taylor:2006.perseus.central.RM.to.core.radio.emission} point out that the strength of the emission, position and correlations with radio emission and spectral indices, strong gradients in RM (RM gradient scale-length $\ll 10\,$pc), and other polarization data (e.g.\ newer VLBI data; \citealt{paraschos:2024.3c84.rm.measurements.vlbi.all.source.local}) all imply this RM comes primarily from the source (i.e.\ the compact radio core itself, gas on $\lesssim 10\,$pc, and certainly $\ll$\,kpc, scales).\footnote{For example, given the measured warm CO and H2 (few thousand K) gas densities in the nuclear disk within $<50\,$pc covering the same position as the observed RM in the NGC 1275 core \citep{scharwachter:2013.nuclear.disk.1275,nagai:2019.1275.co.molecular.nuclear.disk}, and typical magnetic fields in AGN nuclear molecular disks on these scales measured by Zeeman splitting \citep{modjaz:2005.agn.maser.modeling.Bfield.constraints.favor.fluxfrozen.disks,vlemmings:2007.maser.b.limits,mccallum:2007.circinus.maser.B.limits,pesce:2015.agn.maser.upper.limits.B}, it would require a free electron fraction of just $\sim 10^{-8}$ to account for the entire observed RM, and \citet{paraschos:2024.3c84.rm.measurements.vlbi.all.source.local} appear to confirm this with newer EHT data of the 3C 84 core, although debate remains regarding how much of the RM comes from within the jet interior itself \citep{kim:2022.perseus.jet.internal.bfields.within.jet.regions}.}  
Recently, \citet{kam:2026.perseus.faraday.rotation.profiles.all.nuclear.little.large.scale} presented extensive VLBI maps and monitoring data, combined with existing measurements, to show that previous measurements like \citet{taylor:2006.perseus.central.RM.to.core.radio.emission} decrease with radius along the jet extent as ${\rm RM} \sim 1800\,(r/r_{\rm B})^{-2.7}$ (with $r_{\rm B}\sim 11.6\,{\rm pc}$ their estimated Bondi radius). They show that the combination of this trend plus spatial and time variability indicates the RM contribution from large (e.g.\ $\gtrsim$\,kpc) scales must be small, $\lesssim 1000\,{\rm rad\,m^{-2}}$. 
This is already non-trivial to reconcile with a pure-thermal model, because $n_{e}$ in the central regions must be large in those models to reproduce the observed X-rays -- it would require $B \lesssim {\rm \mu G}$ on $1-10\,$kpc scales; and a limit ${\rm RM} \lesssim 100\,{\rm rad\,m^{-2}}$ would definitively rule out those thermal-only models.
But even if we take the most extreme conceivable upper limit, that all of the \citet{taylor:2006.perseus.central.RM.to.core.radio.emission} RM comes from large scales, this sets a firm upper limit on $B$ at different radii from the cluster center. Fig.~\ref{fig:B} shows these, assuming the (more conservative, so higher-upper-limit) $n_{e}$ from the CR-IC model, and random field orientations.

Next, various authors have used the combination of radio and X-rays (synchrotron and limits on IC) to constrain $B$ at larger scales. Of course our models must be consistent with these since they reproduce both observations, but we present more general limits in Fig.~\ref{fig:B}. CR leptons produce synchrotron at frequency $\nu \sim \nu_{c} \sim 10\,{\rm MHz}\,B_{\rm \mu G}\,E_{\rm GeV}^{2}$ and emissivity $\propto E_{\rm GeV}^{2} B^{2} {\rm d} n_{\rm cr}(E_{\rm cr})$ around that $\nu_{c}$, and IC scatters CMB photons to X-ray energies $\sim 3\,{\rm keV}\,E_{\rm GeV}^{2}$ with emissivity  $\propto E_{\rm GeV}^{2} {\rm d} n_{\rm cr}(E_{\rm cr})$. So for a given $B$, there is a correspondence between the IC and synchrotron intensity at the appropriately-matched $\nu$ and $E_{\rm cr}$ if both are observed at the same position.\footnote{The same CRs producing peak CR-IC emission at some X-ray energy $E_{\rm keV}\,{\rm keV}$ will be producing synchrotron at $\sim 3.8\,{\rm MHz}\,B_{\rm \mu G}\,E_{\rm keV}$, so $144$\,MHz radio is associated with $\sim 40\,B_{\rm \mu G}^{-1}\,{\rm keV}$ hard X-rays.} \citet{sanders:2007.perseus.profiles.claimed.hardxr.cr.vs.thermal} claim to detect a lower limit to the CR energy density (pressure) in Perseus at $\sim 10-50\,$kpc, by identifying harder emission at $\sim 3-20\,$keV with CRs (we stress this, if correct, would be strictly a lower limit, because they neglect the possibility of curvature in the CR spectrum and therefore assume no CR contribution to the soft X-rays). If correct, the combination of this with the LOFAR radio spectrum (noting this corresponds to $\sim 10-80\,B_{\rm \mu G}$\,MHz) at the same radii \citep{vanweeren:2024.perseus.giant.radio.halo.filled.electrons.just.tiny.fraction.high.energy} sets an \textit{upper} limit of $B_{\rm \mu G} \sim (7,\,3)$ at $\sim (10,\,50)$\,kpc. We stress that this claimed detection has been controversial, for the reasons reviewed above (\S~\ref{sec:obs.hard} and \citealt{creech:2024.nustar.perseus.obs}); if it is not significant, the upper limit moves \textit{up}, so is less constraining. A more robust \textit{lower} limit, however, can be established by noting that for a given synchrotron intensity, one cannot over-predict the \textit{total} soft+hard X-ray luminosity. Analysis of the NuSTAR \citep{creech:2024.nustar.perseus.obs} and Hitomi \citep{hitomi:2018.perseus.temperature.structure.hard.emission} spectra in the Perseus core give the lower limits shown; we extend this by performing the same exercise given the \citet{vanweeren:2024.perseus.giant.radio.halo.filled.electrons.just.tiny.fraction.high.energy} synchrotron spectra at larger radii, which give typical lower limits of $\mathcal{O}(0.1\,{\rm \mu G})$, typical in SCCs with this method \citep[see also][]{cova:2019.cluster.ic.upper.limits.B.lower.limits,mirakhor:2025.nustar.obs.cluster.lower.limit.bfield.nonthermal.limits.depend.on.assuming.strict.powerlaw.for.ic}. 

Most indirectly, where hydrostatic X-ray mass/potential reconstruction and external constraints from weak lensing and/or satellite kinematics agree well, one obtains an upper limit to the nonthermal pressure, which in turn gives an upper limit to $P_{B} \equiv B^{2}/8\pi$. Using the different mass modeling methods below (\S~\ref{sec:mass:unconstrained}), we estimate this where the two agree best, at $\sim 200-1000\,$kpc (at smaller radii, there are sufficient systematic differences between the X-ray and other mass constraints that the upper limits to $P_{B}$ are well above the plot and completely uninteresting). At larger $R \sim R_{500}$, this implies a non-thermal pressure which must be comparable to or smaller than the thermal pressure (i.e.\ $\lesssim 50\%$ of the total pressure), giving an upper limit $B_{\rm \mu G} \lesssim 7$, while at the best-constrained point (in both X-ray and lensing/kinematics), $\sim 300-400\,$kpc, one recovers something quite close to the commonly-quoted limit that $P_{B}$ must be $\lesssim 10\%$ of the total pressure there, giving $B_{\rm \mu G} \lesssim 6$. 


All our variant models are consistent with all of these upper and lower limits, and indeed fill the range in-between. Thus there is no tension between the models here and any magnetic field constraints. Clearly, the observations imply $\mathcal{O}({\rm \mu G})$ fields, though both models with this $B\sim$\,constant and $B$ decreasing from $\sim 10\,{\rm \mu G}$ at $\lesssim 1\,$kpc to $\sim 0.1\,{\rm \mu G}$ at $\sim R_{\rm 500}$ (e.g.\ $B_{\rm \mu G} \sim (r / 30 {\rm kpc})^{-2/3}$) are allowed. These are all consistent with something like the canonical $\beta \equiv P_{\rm thermal} / P_{\rm B} \sim 100$ often invoked \citep[e.g.][]{walker:2017.perseus.beta.100} in the cluster core.

\subsection{Mass Models of Perseus \&\ CR or ``Non-Thermal'' Pressure}
\label{sec:mass}

The models here imply a non-negligible CR pressure, with $P_{\rm cr} \gtrsim P_{\rm thermal}$ interior to $r \lesssim 50-80\,$kpc, given the extreme SCC luminosity of Perseus (Fig.~\ref{fig:xr.profiles.deriv}; the range of radii depends on the model and ``true'' underlying thermal profile). It is important to consider how this compares to constraints on ``non-thermal pressure'' in Perseus.

\subsubsection{Constraints Which Do Not Apply to CRs Here}
\label{sec:mass:unconstrained}

The majority of ``non-thermal pressure'' constraints in clusters in the literature are really constraints on the \textit{turbulence} in clusters -- e.g.\ constraints on non-thermal line-broadening from microcalorimeters like Hitomi or XRISM \citep{hitomi:2016.perseus.weak.turb,hitomi:2018.perseus.core.turbulence.temperatures,hlavacek.larrondo:2025.xrism.perseus.preview,zhang:2026.xrism.perseus.turb.zmetallicity.profile} or constraints on surface brightness fluctuations and therefore (to some extent) gas density fluctuations \citep{zhuravleva:2015.turbulence.estimates.in.clusters.from.surface.brightness.fluctuations.perseus,sanders:2020.bulk.flows.perseus.estimation.density.fluctuations.turbulence,devries:2023.chandra.gas.density.fluctuations.estimation.perseus,li:2025.surface.brightness.fluctuations.nearby.clusters.claim.turb.equals.cooling.but.requires.super.extreme.model.pushing.more.like.three.dex.too.low}. We stress that these do not constrain the CR pressure or pressure ratio $P_{\rm cr}/P_{\rm thermal}$ -- intrinsically CRs (like magnetic fields) do not contribute to line broadening, and the fluctuations can be larger or smaller in CR pressure dominated halos and have similar power spectra and characteristic scales \citep[see][]{ji:fire.cr.cgm,ji:20.virial.shocks.suppressed.cr.dominated.halos,butsky:2020.cr.fx.thermal.instab.cgm,kempski:2020.cr.soundwave.instabilities.highbeta.plasmas.resemble.perseus.density.fluctuation.power.spectra}. In a highly non-linear sense (which requires ``live'' CR-MHD dynamical simulations to properly model), CR-pressure-dominated halos tend to produce sub-sonic, buoyancy-driven turbulence, independent of the CR-to-gas pressure ratio when $P_{\rm cr} \gg P_{\rm gas}$ (references above and \citealt{2009ApJ...699..348S,Rusz17,Mao18:cr.winds.uplift.from.pressure,Buts18,gurvich:2020.fire.vertical.support.balance,chan:2021.cosmic.ray.vertical.balance,weber:2025.cr.thermal.instab.cgm.fx.dept.transport.like.butsky.study}), and CR streaming actually drives sub-sonic buoyancy instabilities which resemble observed structure in clusters \citep{kempski:2023.cr.bouyancy.instability.streaming.driving.motions.in.clusters}, very similar to the ``standard models'' invoked for cluster turbulence and constraints implied by the observations above. 
Indeed, numerical simulations have shown that for a given potential, CR-dominated halos tend to show \textit{smaller}, more sub-sonic gas motions and narrower kinematic line-broadening, compared to thermal-pressure dominated halos \citep{butsky:2022.cr.linewidth.effects.cgm}, which would alleviate the challenge that most simulations face in explaining the relatively weak line broadening observed in the Perseus core by Hitomi/XRISM. 
In future work, we will study this in more detail with fully-dynamical simulations of these buoyancy effects, but for now simply emphasize they are completely consistent with the turbulence constraints measured. 

It is also sometimes claimed that $\gamma$-rays provide an upper limit of $P_{\rm cr}/P_{\rm thermal} \sim 1\%$ in clusters in general \citep[e.g.][]{huber:2013.stacked.fermi.gamma.ray.clusters.weak.upper.limits.for.hadronic.production,ackermann:2014.cosmic.ray.fermi.gamma.ray.upper.limits.galaxy.clusters.data.not.as.model.dependent}. However we have already shown the models here are well below $\gamma$-ray luminosities in Perseus, so obviously such limits cannot apply here. As discussed below (\S~\ref{sec:differences}), the key issue is that those claimed $\gamma$-ray constraints are highly model-dependent and make a number of assumptions not valid here (like that the CRs are purely hadronic).

\begin{figure}
	\centering
	\includegraphics[width=0.98\columnwidth]{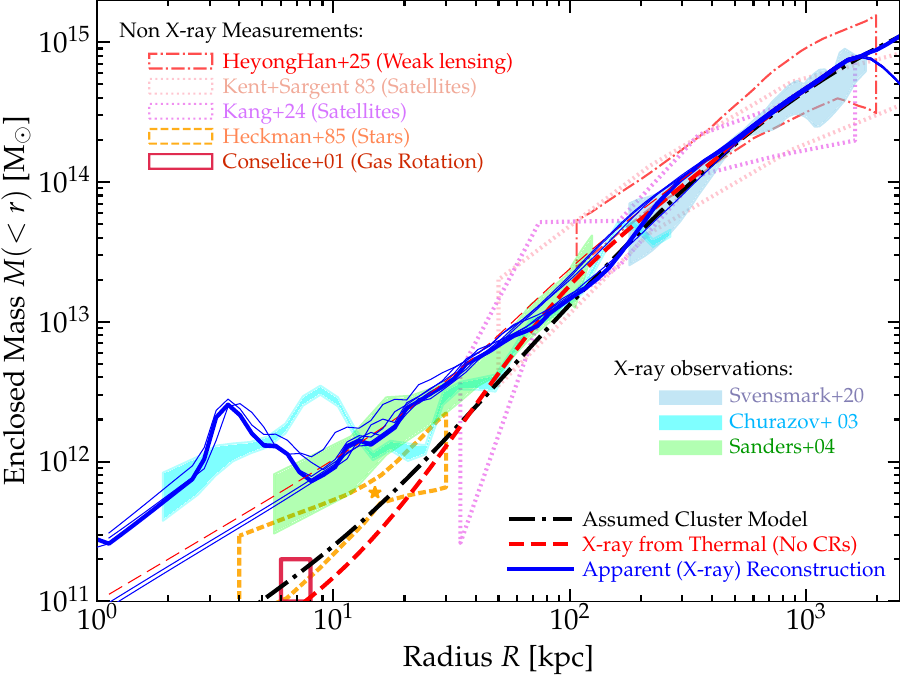}
	\caption{Cluster dynamics/mass model/potential constraints (\S~\ref{sec:mass}). 
	Non X-ray constraints are compiled from weak lensing, satellite kinematics, stellar kinematics of NGC 1275, and gas rotation in NGC 1275 (details in \S~\ref{sec:mass:mass}). 
	We show our assumed cluster true mass model (\textit{black line}), motivated by these. 
	We compare X-ray mass/potential reconstruction (assuming spherical symmetry and hydrostatic equilibrium; range of allowed values \textit{shaded} from \citealt{sanders:2004.perseus.profiles,svensmark:2021.perseus.cluster.mass.reconstruction.anisotropy.marginalization.spatial.2d.info}, and mean with uncertainties from \citealt{churazov:2003.perseus.profiles}), and the model predictions using the ``apparent'' predicted X-ray pressure profiles/temperature profiles in Fig.~\ref{fig:xr.profiles}. 
	At $\gtrsim 50-80\,$kpc agreement is very good, and the cancelling effects of CR-IC influencing CR pressure slopes and $T_{X}$ actually make the ``apparent'' profile closer to reality, but well interior to the CC the enclosed mass is biased high compared to observations, just as observed.
	\label{fig:massmodel}}
\end{figure}

\subsubsection{Constraints Which Do Apply to CRs: Mass Models}
\label{sec:mass:mass}

The constraint on ``non-thermal pressure'' which does potentially apply here is given by the combination of constraints on mass/potential models, comparing X-ray-inferred profiles using the assumption of hydrostatic equilibrium and pure thermal emission to independent constraints from dynamics and/or lensing (all the models here assume spherical symmetry). Fig.~\ref{fig:massmodel} shows such a comparison, including our assumed underlying mass model for our simple analytic cluster, and compiles many different constraints in Perseus. First we collect different constraints independent of the X-rays. Perseus is very well-studied and there are, at present, weak lensing mass maps \citep{hyeonghan:2025.perseus.major.merger.weak.lensing.mass.reconstruction}, satellite kinematics-based mass reconstructions (integrating the spherical Jeans equation with the data from \citealt{kent.sargent:1983.perseus.satellite.kinematics.and.counts} or \citealt{kang:2024.perseus.redshift.survey.satellite.kinematics.and.spatial.distribution}, using \citealt{svensmark:2021.perseus.cluster.mass.reconstruction.anisotropy.marginalization.spatial.2d.info}, or alternatively with similar results \citet{shi:2024.cluster.mass.modeling.satellite.kinematics.jam} to model the distribution of anisotropy parameter vs $R$), stellar dynamics of NGC 1275 \citep{heckman:1985.stellar.kinematics.radio.galaxies.perseus} (this depends on the run of the anisotropy parameter with radius, which we marginalize over following \citealt{wolf:2010.disperson.gal.masses} or \citealt{agnello:2014.jeans.modeling.ellipticals.methods.review}, giving the best-constrained value with just $<20\%$ error bars as expected from \citealt{wolf:2010.disperson.gal.masses}, shown as the smaller point), and gas rotation in atomic+molecular gas disks in NGC 1275 \citep{conselice:2001.ngc.1275.perseus.halpha.emission.broad.properties}. Our assumed mass model is consistent with these constraints by construction, but it is worth noting that all of the data can be plausibly reproduced by a very simple model on these scales of an isothermal sphere central (NGC 1275) potential ($\sigma \sim 250\,{\rm km\,s^{-1}}$) plus NFW halo on the mean concentration-mass relation \citep{comerford:obs.concentrations}.

Now compare this to the inference from X-rays, specifically assuming hydrostatic equilibrium. Then the enclosed mass is given by $M_{\rm enc}^{X}(<r) \approx -(k_{B} T_{X} r / G \mu m_{p})\,\partial \ln P_{X} / \partial \ln r$. We show this for the different observations in Fig.~\ref{fig:massmodel}, as well as our predicted ``apparent'' X-ray profiles from the same \citep{churazov:2003.perseus.profiles,sanders:2004.perseus.profiles,svensmark:2021.perseus.cluster.mass.reconstruction.anisotropy.marginalization.spatial.2d.info}, using the spherically-averaged, de-projected profiles from each. At radii $\gtrsim 50-80\,$kpc, and certainly $\gg 100\,$kpc, there is good agreement to within the model uncertainties (often factor $\sim 2$, but within $\sim 10\%$ at the best-constrained radii at a few hundred kpc). Interestingly, at these radii, even within the SCC but $\gtrsim 50\,$kpc, the ``apparent'' X-ray profile actually agrees better with the true kinematic mass model than the profile we would obtain using the ``true'' thermal quantities, and the bias is very small. The reason for this is shown in \paperthree: the enclosed mass depends primarily on $T_{X}$, which is biased by CR-IC but only by a small amount (in log space) to smaller values, but this is offset by $\partial \ln n / \partial \ln r$ which drives most of $\partial \ln P/\partial \ln r$, towards apparently larger absolute values (steeper apparent profiles). 

At smaller radii well interior to the CF ($\lesssim 50-80\,$kpc), it is obvious that the X-ray reconstruction of Perseus is strongly biased compared to non X-ray models, whether from gas rotation or stellar kinematics or semi-empirical models (e.g.\ simply taking the observed stellar mass of NGC 1275 and any reasonable typical dark matter-to-stellar mass ratios inside $\lesssim 20\,$kpc; see e.g.\ \citealt{borriello03,hopkins:cusps.fp,bolton:fp,koopmans:lenses.isothermality.and.anisotropy,guimaraes:isothermal.mass.profiles.lenses,newman:2013.cluster.mass.profiles.multi.method}). The X-rays imply $\gtrsim 10^{12}\,M_{\odot}$ interior to the central $\sim 10\,$kpc, even though the gas, BH, and stellar masses interior to $<10\,$kpc are $\sim 10^{9}\,M_{\odot}$, $\sim 10^{10}\,M_{\odot}$, $\sim 10^{11}\,M_{\odot}$ -- i.e.\ we would require an order-of-magnitude more dark matter than baryonic matter inside the effective (half-light) radius of NGC 1275 to explain the X-rays, which would not only be inconsistent with the stellar and gas kinematics but also order-of-magnitude higher than essentially any known massive elliptical at this size scale (this would make NGC 1275 a $\sim 10-20\sigma$ outlier in the fundamental plane, for example; \citealt{cappellari:fp,zhu:2023.manga.detailed.fundamental.plane.ell.dynamics.dark.matter.fracs.from.scaling.laws.and.fast.vs.slow.rotation,daddona:2025.stellar.mass.fundamental.plane.hyperplane.update}). This offset is quite similar to other observed SCCs \citep{newman:2013.cluster.mass.profiles.multi.method,simet:2017.weak.lensing.xray.cosmology.masses.agreement.but.large.radii,allingham:2023.clusters.kinematic.lensing.vs.xray.mass.profiles.large.disagreement.qualitative.similar.nfw.profiles}\footnote{Indeed, to our knowledge, the only arguments for $\sim 10\%$-level agreement in mass profiles at small radii $\ll 50\,$kpc (as compared to larger radii which are better studied, see \citealt{sayers:2021.cluster.core.constraints.nonthermal.motions}) apply to some of the weakest known CCs (e.g.\ Fornax and Virgo) in e.g.\ \citet{churazov:2008.virgo.mass.profile.kinematics.vs.xrays.as.measurement.of.nonthermal.pressure}, and even there different models and X-ray measurements gave up to $\sim 60\%$ deviations, while more recent improved measurements and modeling of the same clusters has demonstrated that the newer X-ray and kinematic data and models at the same radii now deviate by more like factors of $\sim 2-3$ \citep{gebhardt:2009.bh.mass.revision.new.models.m87,murphy:2011.new.m87.models,liepold:2023.m87.dynamical.masses}.}. 

We emphasize that the X-rays \textit{alone} clearly indicate that they are unreliable for hydrostatic mass estimation at these radii ($\ll 50-80\,$kpc): non-parametric deprojection (as compared to the by-construction smooth profiles fit with \citet{sanders:2004.perseus.profiles}) gives non-monotonic hydrostatic enclosed masses $M_{\rm enc}(<R)$ out to $\sim 30-50\,$kpc (i.e.\ these features are correlated with the real X-ray $T_{X}$, $n_{X}$ profiles, hence appearing in our models, not simply a deprojection artifact), and Perseus contains obvious X-ray structure at these radii, so radial profiles constructed from the $T_{X}(R,\,\phi)$ \&\ $n_{X}(R,\,\phi)$ measured along different wedges or azimuthal angles $\phi$ disagree systematically at the factor $\gtrsim 2$ level \citep{sanders:2004.perseus.profiles}.\footnote{We should note that, to our knowledge nobody in the field is actually using X-rays for precision mass/potential modeling for large-scale structure at these radii in Perseus, in part because of this and similar features in other SCCs \citep{zhang:2008.cluster.profiles.lensing.xray.good.down.to.0pt2.r500.inner.makes.scatter.much.larger.biases.cosmological.measurements,sanders:2020.bulk.flows.perseus.estimation.density.fluctuations.turbulence,pratt:2022.cluster.density.profiles.also.need.to.excise.cores.to.get.clean.lx.t.mass.relations}, though the caveats in \S~\ref{sec:mass:mass} do imply some cautions to attempts like those in \citet{mantz:2016.cluster.cosmology.testing.nfw.profiles,eckert:sidm.profiles.xcop.clusters} to use hydrostatic masses to constrain dark matter mass profiles and therefore some dark matter models at small radii in clusters.} This point has been better made elsewhere -- our point here is simply to show that the models here are completely consistent with all of the observed kinematic constraints on Perseus and its gas dynamics versus X-ray-implied mass/potential profiles.

We emphasize this because it is sometimes stated that these constraints imply $\lesssim 10\%$ non-thermal pressure. However, as we have shown: (1) at large radii (where such constraints are typically applied), this is easily satisfied by the models here; (2) at intermediate radii, the deviation of the ``apparent'' X-ray profiles from kinematic can actually be \textit{smaller} with CR-IC and large $P_{\rm cr}$ than without it, because of how it jointly influences $T_{X}$ and $n_{X}$; (3) at small radii in SCCs like Perseus, this predicts the same deviations in mass/potential models between X-ray and independent kinematic constraints as observed. Moreover it is worth noting that in this particular case, the mass reconstruction is actually biased to higher masses than kinematics imply, which is opposite the behavior often assumed when modeling sources of non-thermal pressure empirically. But as shown in \paperthree, the sign of the CR-IC effect can go either way, depending on the details of CR profile, underlying true profile, and ratio $P_{\rm cr}/P_{\rm true}$.

\begin{figure}
	\centering
	\includegraphics[width=0.98\columnwidth]{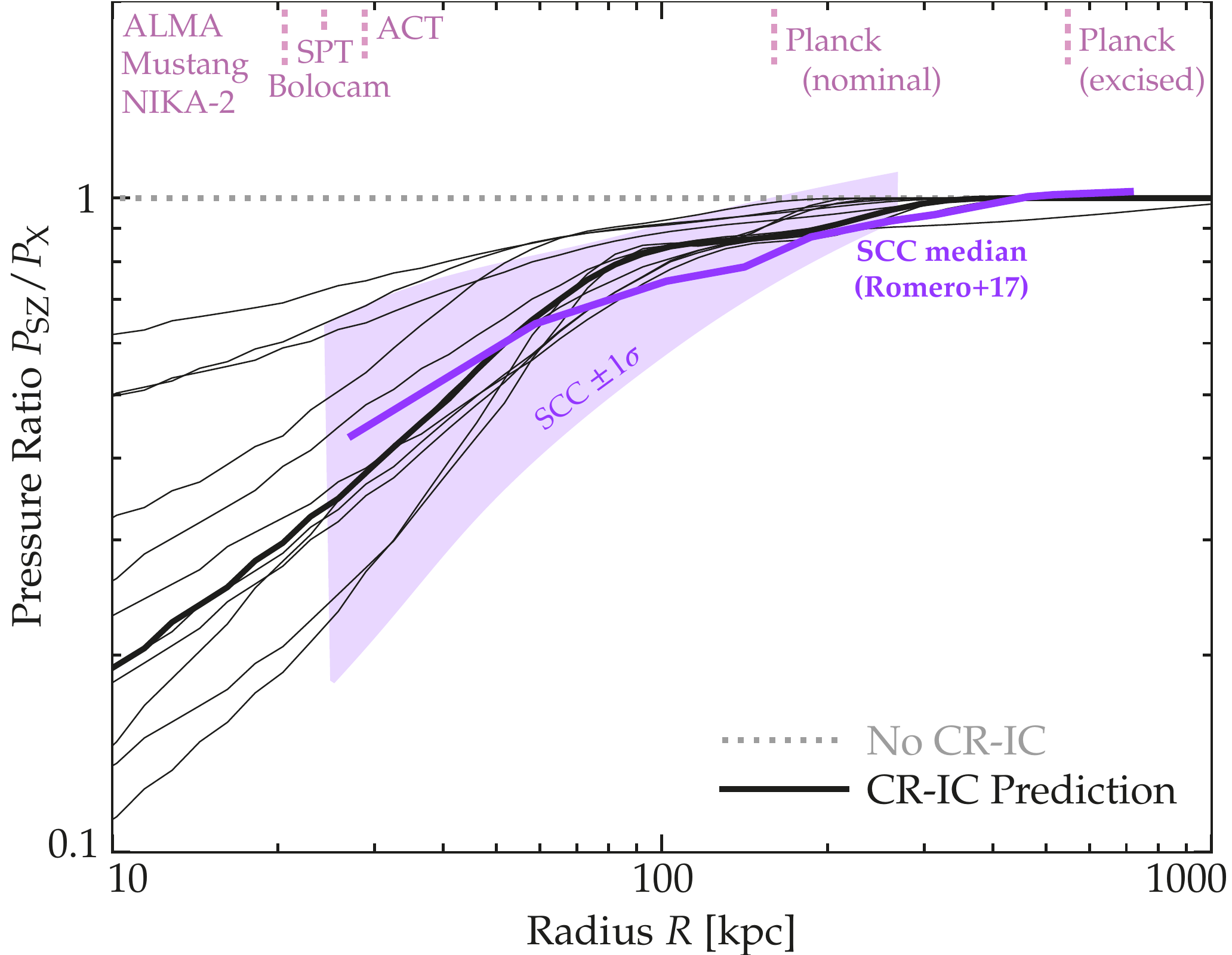}
	\caption{Ratio of true thermal pressure or pressure inferred from the Sunyaev-Zeldovich effect with uncontaminated measurements, $P_{\rm SZ} \approx P_{\rm therm,\,true}$, to X-ray inferred pressure $P_{X}$ (Fig.~\ref{fig:xr.profiles}), as a function of deprojected radius $R$, in the models here (style as Fig.~\ref{fig:B}). 
	Because of CR-IC, $P_{\rm SZ} < P_{\rm X}$ in the CC center ($\lesssim 60-80\,$kpc), though how strongly depends on the true thermal properties we assume. 	
	We compare resolution (HPBW) of various instruments (ALMA, Mustang(2) and NIKA-2 at $<10\,$kpc). 
	Owing to strong contamination from NGC 1275 (luminous at $\sim 100\,$GHz) and the large angular size of Perseus, only Planck data relevant here exists, at radii well outside the radii of predicted effects (and it was still excluded from the main Planck analysis). We therefore compare instead to the median and $\pm 1\sigma$ range measured in \citet{romero:2017.cluster.pressure.profiles.highres.sz.xray.cool.cores.show.central.pressure.deficit} and \citet{silich:2025.cr.ic.tests.in.zw.3146.cluster} from a sample of 8 similarly X-ray luminous (but much less $100\,$GHz-luminous) SCCs for which uncontaminated high-resolution data exists at $\ll 100\,$kpc.
	\label{fig:SZ}}
\end{figure}

\subsubsection{Non-Thermal Pressure in SZ}
\label{sec:sz}

In previous papers we predicted the SZ effect would show a clean signature of CR-IC, as the pressure inferred from SZ would be sensitive to the true thermal pressure $P_{\rm SZ} \approx P_{\rm therm}^{\rm true}$, while that inferred from X-rays traces the ``apparent'' $P_{X} = n_{X} k_{B} T_{X}$ biased by CR-IC (generally to larger values in the CC center), so the ratio $P_{\rm SZ}/P_{X}$ would decrease in cluster centers. Fig.~\ref{fig:SZ} shows this prediction for the models here. Indeed there is a robust signature of CR-IC at $\ll 70\,$kpc or so. The magnitude of the suppression varies widely as it depends on the details of the underlying ``true'' thermal profile we assume (as well as the CR model). 

We show the resolution (HPBW or FWHM) of various instruments at the distance of Perseus in Fig.~\ref{fig:SZ} for comparison. Bolocam/ACT/SPT resolution, let alone Planck, are probably insufficient to measure this robustly, especially if one adopts a more conservative approach (it is important to resolve a given annulus with a few beam-widths separation from the NGC 1275 core). On the other hand, instruments like ALMA, Mustang, and NIKA-2 have $\lesssim 10-20^{\prime\prime}$ resolution so could easily resolve the salient scales in principle, but there are two serious problems limiting them in practice. First, Perseus is far too large on the sky, so the central signal would be extremely difficult to measure without filtering out the modes of interest. Second, the NGC 1275 AGN nucleus is famously bright ($\sim 10$\,Jy) at typical SZ frequencies ($\sim 100$\,GHz), so sidelobe contamination could badly bias any SZ measurement. For e.g.\ Bolocam, this would translate to $\sim 300\,{\rm mK}$ or a Compton $y\sim 0.1$, $\sim 1000$ times brighter than the predicted SZ signal -- i.e.\ sidelobe control at better than $-30\,{\rm dB}$ would be required. For MUSTANG2 this would require control at $-45\,{\rm dB}$. Even for Planck (where this problem motivates the larger Planck excision radius at $>500$\,kpc), this requires control to $-15\,{\rm dB}$ at $\sim R_{500}$. As a result, it is unlikely that this measurement can be made in the near future for Perseus specifically, but we compare to the range observed in other, uncontaminated (and smaller on the sky) SCCs from \citet{romero:2017.cluster.pressure.profiles.highres.sz.xray.cool.cores.show.central.pressure.deficit} and \citet{silich:2025.cr.ic.tests.in.zw.3146.cluster}, which are suggestively similar to the predicted model range.

An alternative to SZ which is not contaminated by NGC 1275 would be something like dispersion measures (DM) of distant background sources like fast radio bursts (FRBs), which directly probe the true line-of-sight column density at a given impact parameter $R$ as ${\rm DM} \propto \int n_{e}(R,\,z)\,dz$. Given the difference in the X-ray inferred density $n_{X}$, assuming pure thermal emission, in the core, compared to the (much lower) densities allowed in the CR-IC picture (Fig.~\ref{fig:xr.profiles}), this predicts the same qualitative effect as in SZ, $N_{\rm H,\,true}/N_{\rm H,\,X} < 1$ within the SCC. Statistically much larger samples of FRBs are needed to obtain multiple sources with impact parameters within $R < 80\,$kpc of the Perseus core, but at least a couple FRBs have been detected in clusters \citep{connor:2023.frb.foreground.cluster.outskirts}.

\section{Discussion: What Is Different From Previous Models?}
\label{sec:differences}

The models here provide a natural and simple explanation for the multi-wavelength properties of Perseus from MHz radio, through soft and hard X-rays up to TeV $\gamma$-rays, with a simple model of CR lepton streaming. As noted throughout the text where relevant, some aspects of this appear, at first, to be at odds with older conclusions. Here we briefly summarize what is different about the models/data/assumptions here, compared to those works. 

\subsection{New Data}
\label{sec:differences:data}

One important point is that there are multiple new and improved datasets which have only recently been published. In particular LOFAR data is tremendously helpful and informative: it extends the radio to $\lesssim 100\,$MHz and much lower surface brightness, showing that (1) the minihalo is indeed ultra-steep, likely with most of the radio emission at $\lesssim 10\,$MHz (below even the LOFAR bands, based on the spectral indices measured), and (2) it steepens rapidly with increasing $R$ (steepening had been seen previously at $600-1400\,$MHz but is now much more clear with the lower frequencies dominating) in the regime dominated by NGC 1275, before (3) re-flattening at large radii where satellites contribute, while also (4) correlating \textit{positively} (outside the very central ``active jet'' cavity) with the X-ray emission \citep{vanweeren:2024.perseus.giant.radio.halo.filled.electrons.just.tiny.fraction.high.energy}. Another important new dataset is the NuSTAR data \citep{creech:2024.nustar.perseus.obs}, which allows for the first unambiguous measurement of the diffuse/extended hard X-ray emission spectrum. On top of these, new measurements of both the metallicity of NGC 1275 in gas and stars, and the kinematics of the cluster, provide new complementary tests of the models discussed above. These make it clear that there is indeed a significant metallicity deficit, i.e.\ the soft X-ray spectral fitting methods appear to infer a metallicity well below what is given by optical/UV estimators. 

\subsection{Why Is Reacceleration (or Hadronic Injection) Not Needed in Radio?}
\label{sec:differences:reaccel}

It is common to see claims that the minihalo in Perseus requires some continuous reacceleration or injection (e.g. continuous hadronic production of leptons) to explain its extent, but this is clearly not the case in the models here. As reviewed elegantly in \citet{brunetti.jones:2014.cr.diffusion.clusters.minihalo.giant.halo.review.many.solutions.including.faster.transport.and.local.sources}, the argument for reacceleration (or some extended hadronic injection) depends on a number of strict assumptions which are not true in the models we consider here, and removing even one of those strict conditions makes it fairly easy to understand how the minihalo could be produced without reacceleration. From most-to-least important, these prior assumptions are: 

(1) Models often assumed CR transport is purely diffusive, so the transport time to distance $r$ scales as $\Delta t \sim r^{2}/\kappa \sim 30\,{\rm Gyr}\,(R/100\,{\rm kpc})^{2}\,(10^{29}\,{\rm cm^{2}\,s^{-1}}/\kappa)$, as compared to including any streaming component giving $\Delta t \sim r / v_{\rm st} \sim 1\,{\rm Gyr}\,(R/100\,{\rm kpc})\,(100\,{\rm km\,s^{-1}})$ \citep{pfrommer.enslin:2004.hadronic.minihalos.in.cluster.centers,vazza:2024.cr.electrons.seeding.mini.review}. That quadratic-versus-linear difference means that even a small streaming speed will almost always dominate bulk transport at large radii $\gtrsim 100\,$kpc, and order-of-magnitude reduces the transport time of CRs (shorter $\Delta t$ at large $R$ by a factor $\kappa/v_{\rm st}\,R \sim 0.003\,(\kappa/10^{29}\,{\rm cm^{2}\,s^{-1}})\,(100\,{\rm km\,s^{-1}}/v_{\rm st})\,({\rm Mpc}/R)$). Note that the ``streaming'' term here can arise from advection/convection, \Alf{ic} or super-\Alf{ic} streaming as in self-confinement models, super-diffusion \citep{liang:2025.leaky.boxes.levy.flights.modeling.crs.transport}, or simply a gradual radial decrease of the CR scattering rate ($\nu_{\rm cr} \propto 1/r$, so $\kappa \propto r$) -- these all have the same effect, for our purposes. 

(2) In addition to neglecting streaming, reacceleration models often assume extremely low diffusion coefficients, as low as $\kappa \sim 10^{27}\,{\rm cm^{2}\,s^{-1}}$ (or even the fully-trapped limit $\kappa+v_{\rm st}\,R \rightarrow 0$, see \citealt{sarazin:1999.cr.electron.emission.cluster.centers.xrays,bruggen:2002.crs.advective.only.in.clusters,ensslin:2007.cosmic.ray.advection.injection.model.in.structure.formation.sim,zuhone:2013.turbulent.reaccel.secondary.electrons.as.potential.minihalo.explanation.radio.emission}), which are based on decades old ``leaky box'' type models for LISM CRs that have been widely shown to be biased and order-of-magnitude incorrect, as compared to modern fits with codes like GALPROP, DRAGON, PICARD, USINE, and others that robustly give $\kappa_{\rm eff}$ (accounting for the weighted-energy-average and for streaming/advective/convective speeds adopted as well) about two orders of magnitude larger \citep[see e.g.][]{delaTorre:2021.dragon2.methods.new.model.comparison,zhao:2021.spatially.dependent.cr.propagation.disk.halo.models,korsmeier:2022.cr.fitting.update.ams02,hopkins:cr.multibin.mw.comparison,hopkins:cr.multibin.mw.comparison,dimauro:2023.cr.diff.constraints.updated.galprop.very.similar.our.models.but.lots.of.interp.re.selfconfinement.that.doesnt.mathematically.work,jacobs:2023.isotope.ratios.large.halo.size.required.constrain.volume.of.inhomogenous.slower.diffusion.zones.in.model,tovar:2024.inhomogeneous.diffusion.cr.spectra,silver:2024.cr.propagation.low.energies.new.data,recchia:2024.cr.spectral.modeling.features.bumps.wiggles,delatorre.luque:2024.gas.models.of.galaxy.key.for.scale.height.but.need.halo.for.crs,ramirez:2024.3d.struct.galaxy.gas.influences.cr.fitting.params.vs.2d}. Assuming a very low $\kappa$ has a double effect: it increases the diffusion time (assuming diffusion only) as $\sim R^{2}/\kappa$, but also increases the reacceleration rate as $\sim (v_{A}-\bar{v}_{A})^{2}/27\,\kappa$, so the ratio of diffusion-to-reacceleration time scales $\propto (R\,v_{A}/\kappa)^{2}$. 

(3) Reacceleration models generally neglect multiple sources, despite the plethora of known radio sources in Perseus and other SCCs. This is important especially for the giant halo and any structure at $\gtrsim 100\,$kpc. Per \S~\ref{sec:obs.radio:gianthalo}, beyond $R\gtrsim 100\,$kpc, if CRs travel more than $\sim 100\,$kpc from their individual galaxy sources, they will ``overlap'' each other's mini-mini-halos. This means that the effective ``travel distance'' required for any CR is really never more then $R^{\rm local}_{\rm mini} \sim 100\,$kpc, even if the galaxies being considered are globally at $R \sim 1\,$Mpc from the cluster center. This reduces the maximum travel time from $\sim  3000\,{\rm Gyr}\,(R/{\rm Mpc})^{2}\,(10^{29}\,{\rm cm^{2}\,s^{-1}}/\kappa)$ for a single source to $\sim 1\,{\rm Gyr}\,(R^{\rm local}_{\rm mini}/100\,{\rm kpc})\,(100\,{\rm km\,s^{-1}} / v_{\rm st,\,eff})$. 

(4) Some of the historical models have assumed unphysically high loss rates for CRs. In particular some (see discussion in \citealt{jaffe:1977.coma.electrons.radio.emission,gitti:2002.perseus.minihalo.xray.inverse.compton,brunetti.jones:2014.cr.diffusion.clusters.minihalo.giant.halo.review.many.solutions.including.faster.transport.and.local.sources}) invoked fields as strong as $B \gtrsim 50\,{\rm \mu G}$ at $R \gtrsim 100\,$kpc, equivalent to $\beta \ll 1$ (or simply assumed $\beta=1$, i.e.\ equipartition fields; \citealt{brunetti:1997.ic.clusters.assuming.super.strong.b.fields.equipartition,abe:2024.future.gamma.ray.perseus.obs.modeling}). Today this is clearly ruled out by multiple observational constraints (see \S~\ref{sec:B}), and is far larger than the predictions of any cosmological models \citep{dolag:2000.cluster.magnetic.fields.early.cosmo.sim.predictions,dolag:2002.cluster.cosmological.sims.magnetic.fields,borgani:2011.cosmo.sims.galaxy.clusters.review,hopkins:mhd.gizmo,hopkins:cg.mhd.gizmo,donnert:2018.cluster.magnetic.fields.cosmo.sims.review,barnes:2018.sphmhd.structure.formation.simulations.review,ponnada:2024.fire.fir.radio.from.crs.constraints.on.outliers.and.transport,nelson:2024.tng.cluster.sims.profiles.basic.properties,whittingham:2024.cluster.Bfield.synch.measurements.biased.to.strongest.B.subregions}. But it would have again the dual effect of making CR loss rates much faster (loss time $\propto B^{-2}$ in this limit) while making reacceleration rates larger ($\propto v_{A}^{2} \propto B^{2}$). 

(5) In the reacceleration term in the CR equations-of-motion (in historical reacceleration models), CR scattering is almost always implicitly assumed to be perfectly isotropic, ``grey'' (pitch angle-independent), and gyro-resonant, with zero streaming motion (zero anisotropy in the distribution function of CRs, effectively $\nabla P_{\rm cr} = \bf{0}$). This gives the traditional expressions for reacceleration \citep{ptuskin:1988.cr.reacceleration.nonresonant.turbulent}.\footnote{Specifically, Eq.~\ref{eqn:onemoment.full} with $v_{\rm st}=0$, $\nabla P_{\rm cr} \rightarrow 0$, $\kappa = \kappa_{\rm gyro,\,\|} = v_{\rm cr}^{2} / (3\,\nu_{\rm cr}^{\rm gyro})$, and $|\bar{v}_{A}| \equiv v_{A} |\nu_{+}-\nu_{-}|/|\nu_{+}+\nu_{-}| \rightarrow 0$, i.e.\ $\nu_{+}=\nu_{-}$ exactly, with $\nu_{\pm}$ the scattering from forward/backward propagating modes along the field line.}. 
But this is actually the maximum-possible reacceleration term. As discussed in  \citet{hopkins:cr.multibin.mw.comparison,hopkins:2021.sc.et.models.incompatible.obs,hopkins:cr.spectra.accurate.integration} at the energies of interest ($\sim$\,GeV), if CR self-confinement plays any significant role in exciting  scattering, or even if it does not (in pure extrinsic turbulence models for scattering) if the CRs are (on average) propagating down gradients in density and $|{\bf B}|$ (which they are, by definition, here, to reach larger $R$) then these should not be symmetric, which means $|\bar{v}_{A}| \sim v_{A}$ and the reacceleration rate is reduced by a significant factor. Moreover as shown in both analytic derivation and direct numerical simulation in e.g.\ \citet{mckenzie.webb:1984.acoustic.cr.plasma.instabilities.kinetic.deriv,begelman.zweibel:1994.cr.acoustic.instability.proto.staircase.model,bustard:2022.reacceleration.sims.and.calcs.strongly.suppressed.with.any.streaming.or.high.beta}, almost any anisotropy/streaming/$\nabla P_{\rm cr}$, or finite-wavespeed effects if $\beta > 1$, will suppress any non-resonant reacceleration by orders of magnitude (so the reacceleration rate scales $\propto 1/\kappa_{\rm gyro} \propto \nu_{\rm cr}^{\rm gyro}$, not $1/\kappa_{\rm tot} \propto \nu_{\rm cr}^{\rm tot}$; see also \citealt{tsung:2021.cr.outflows.staircase,sampson:2025.cr.plasma.coupling.supersonic.turbulence.using.jiang.oh.will.miss.some.effects.of.b.gradients}). But in extrinsic turbulence models (the only models where reacceleration occurs) in clusters, the resonant scattering (proportional to $D_{pp}$) is suppressed by many orders of magnitude owing to a combination of anisotropy and damping 
\citep{chandran00,yan.lazarian.02,yan.lazarian.04:cr.scattering.fast.modes,lazarian:2016.cr.wave.damping,kempski:2021.reconciling.sc.et.models.obs,hopkins:2021.sc.et.models.incompatible.obs}. Consistent with these arguments, LISM CR observations coupled to Galactic CR transport models have found growing evidence against significant reacceleration \citep[e.g.][and references therein]{orlando:2015.multi.wavelength.CR.lepton.constraints.gamma.rays.mostly,2016ApJ...824...16J,gabici:2019.cr.paradigm.challenges.mostly.well.explained.in.galaxy.sims,hopkins:cr.multibin.mw.comparison,silver:2024.cr.propagation.low.energies.new.data,dimauro:2023.cr.diff.constraints.updated.galprop.very.similar.our.models.but.lots.of.interp.re.selfconfinement.that.doesnt.mathematically.work}. 

Taken together, this explains not only why the models here do not ``need'' reacceleration (or extended hadronic injection) to explain the mini and giant radio halos, but in fact we find reacceleration (even in the limits above where it would be strongest) is at most a weak correction to our predictions. We do stress the reacceleration we refer to here is specifically ``diffusive'' or ``turbulent'' reacceleration (the term often parameterized as ``$D_{pp}$'' in CR models): reacceleration by large-scale, powerful convective compression (large compressive $\nabla \cdot {\bf u}$) or first-order Fermi (re)acceleration in shocks (DSA) from e.g.\ cluster mergers, are still plausible sources of large-scale CR acceleration in extreme merging clusters (but obey qualitatively different scalings).

\subsection{Why Isn't Inverse Compton Already Ruled Out in X-Rays?}
\label{sec:differences:cric}

As discussed in \S~\ref{sec:obs.xr}-\ref{sec:obs.hard}, at $\sim 0.1-100\,$keV in X-rays, observations appear to be perfectly consistent with much of the apparent SCC cooling luminosity at $\lesssim 100\,$kpc coming from CR-IC rather than thermal emission. This is despite various prior claims that only a small fraction of the X-rays could come from CR-IC. This conclusion that CR-IC was small, in those analyses, generally depended on a few very specific assumptions which we showed are not true in the models here. 
(1) They assumed power-law CR-IC spectra, i.e.\ no CR spectral curvature. But this is never valid near the energies of interest, even for spectra extremely close to injection, and it only requires $\lesssim $\,Myr ($<1\,$kpc travel distances) for curvature to develop in arbitrarily hard pure-power-law injection spectra. 
(2) They assumed very hard CR spectra (often motivated by much older observations of the radio halos, or their spectral indices at either different radii or at energies which do not correspond to the same CR energies emitting CR-IC), but we find the CR spectra soften rapidly because of losses, as also required by the actual observed radio spectra at the same radii, producing much softer (and more curved) spectra than the softest assumed CR-IC models in previous fits. 
(3) Many past fits actually \textit{assumed} (implicitly) that most of the X-rays were thermal, and implicitly did not allow large CR-IC, because they did things like fit out a sum of thermal components (and sometimes AGN components) and looked for residuals to this to identify as CR-IC. But this will by definition ``fit out'' (and classify as thermal) any CR-IC component which resembles the predicted spectra here. 
(4) These fits often assumed CRs were spatially smooth and uniform, so (by assumption) assumed they could not produce brightness fluctuations or multi-component spectra. But given any time-variability in the source (as observed!), or finite streaming speeds (which must be the case at the radii of interest), the natural expectation is multi-component spectra, with large-scale variability in surface brightness features very similar to those observed.

We reiterate the point made in \paperthree, that strictly speaking there is no X-ray measurement alone (of spectral shape or line ratios or excitation or broadening) that can rule out a significant CR-IC component of the continuum, if one allows for the full physical degrees of freedom fitting the problem (multi-phase gas with variable abundances and absorption and an arbitrary CR spectral shape, at each position). One can constrain the energetics and properties needed in the CRs in order for CR-IC to be important (as we do), but ultimately determining whether CR-IC is important or negligible requires multi-wavelength constraints of the sort we present here.

\subsection{Why Don't the $\gamma$-Rays Rule Out Strong CR Pressure?}
\label{sec:differences:gamma}

As noted above, there are a number of claims in the literature that $\gamma$-rays rule out $\gtrsim 1\%$ of the pressure in clusters coming from CRs. But these make several very strong and model-dependent assumptions which are not true in any of the models considered here. Specifically, they assume 
(1) that the CRs are entirely hadronic (most of these studies did note that leptonic CRs would be essentially invisible to the $\gamma$-rays; e.g.\ \citealt{funk:2013.fermi.cta.sensitivities.mev.gev.even.cta.not.close.to.needed.for.leptonic.cluster.mw.halo.gamma.rays.detection,ackermann:2014.cosmic.ray.fermi.gamma.ray.upper.limits.galaxy.clusters.data.not.as.model.dependent,keshet:2025.stacked.cluster.gamma-ray.detection.claim.large.r.flat.spectrum.possible.shock.acceleration.cr.spectrum.hadronic.signature}); 
(2) that CR pressure $P_{\rm cr}$ varies in a specific power-law fashion with gas density according to strictly ``adiabatic CR transport'' (no streaming/diffusion) models \citep{ensslin:2007.cosmic.ray.advection.injection.model.in.structure.formation.sim,pinzke.pfrommer:2010.cluster.gamma.ray.emission.simple.scalings.for.specific.advection.acceleration.models}, completely unlike the streaming+diffusion models here; 
(3) that the gas densities are much higher (using the ``apparent'' values $n_{X}$ derived from X-rays assuming no CR-IC) than they must be with significant CR-IC, which boosts the $\gamma$-ray emission for a given hadronic energy density (potentially by more than an order of magnitude; see Fig.~\ref{fig:xr.profiles} and \paperthree); 
(4) that one drops or neglects clusters with known central $\gamma$-ray sources (precisely those we expect to be brightest in CR-IC), including Perseus specifically (excluded from the constraints in those studies explicitly; see \citealt{abdo:2009.fermi.detection.1275.perseus.strong.gamma.ray.source.1e45.luminosity,huber:2013.stacked.fermi.gamma.ray.clusters.weak.upper.limits.for.hadronic.production,ackermann:2014.cosmic.ray.fermi.gamma.ray.upper.limits.galaxy.clusters.data.not.as.model.dependent,manna:2024.stacking.clusters.sz.detected.gives.clear.gamma.ray.signal.but.could.be.dominated.by.strong.radio.sources.not.diffuse}), because the bright central source makes it difficult or impossible to detect diffuse extended $\gamma$-rays at an interesting level; and 
(5) that the $\gamma$-ray emission comes from much larger radii $\sim R_{500}$ to $\sim R_{\rm vir}$ (see references above and \citealt{colafrancesco:2010.fermi.1275.perseus.gamma.rays.centrally.dominated.large.diffuse.flux.allowed}), not the central $\sim 100\,$kpc as predicted here, because the Fermi angular resolution ($\sim$\,degree) corresponds to $\sim$\,Mpc scales in Perseus (and larger in more distant clusters). 
Dropping even one of these assumptions removes any significant tension between $\gamma$-ray observations and large CR pressure in SCC centers, and clearly the models here (where \textit{none} of these assumptions apply) produce strong CR pressure in the SCC center while being well below the observed $\gamma$-ray limits and luminosities.

\section{Conclusions}
\label{sec:conclusions}

We show that a simple model of CRs injected by the central radio galaxy (NGC 1275/3C 84) in Perseus can explain the observed multi-wavelength diffuse emission spectra from $10^{6}-10^{26}\,$MHz and radii $\sim$\,kpc to Mpc, specifically including the long-and-short wavelength radio, optical/UV/IR, soft and hard X-rays, and $\gamma$-rays, while simultaneously reproducing independent constraints on magnetic field strengths, kinematics/dynamics/mass/potential reconstruction, ``non-thermal pressure,'' metallicity, and the SZ effect. As discussed for SCC clusters in general in \papertwo-\paperthree, as CRs propagate to $\gtrsim$\,kpc scales, losses introduce significant curvature into the CR spectrum, which means CMB photons IC scattered by the CRs appear very similar to $\sim$\,keV thermal emission in soft X-rays, and low-frequency radio sources (dropping out of $\gamma$-rays). We apply these models for the first time to an individual cluster, focusing on the best-studied and brightest X-ray SCC, Perseus. The basic scalings expected for CRs streaming from the center, plus IC emission, naturally predict the X-ray inferred or ``apparent'' density/temperature/entropy/pressure/cooling time/metallicity/mass-deposition rate within the SCC ($\lesssim 100\,$kpc), with extended hard X-ray and $\gamma$-ray emission in excellent agreement with observations. 

These also immediately predict the properties of the observed low-frequency radio mini-halo, including its radial surface brightness profile and spectrum from $\sim 20-10^{4}$\,MHz (and spectral evolution with radius). Notably, no re-acceleration or extended hadronic injection is necessary: the minihalo properties are predicted by the same population of streaming CRs. Future $\sim 10\,$MHz observations could test predictions here. But if we (more speculatively) extrapolate these models to the satellite population with simple scalings, we also predict a giant radio halo out to $\sim$\,Mpc scales with the observed radial profile and spectrum, coming from the contribution of many individually lower-luminosity ``mini-mini-halos'' around each satellite galaxy in the cluster. The CR halos predicted around NGC 1275 should exist from all galaxies in the cluster, just with much lower individual luminosities/brightness, but with similar sizes $\sim 100\,$kpc (set by IC loss physics), which means their diffuse emission would overlap in a volume-filling-diffuse component which dominates over the surviving CRs from NGC 1275 at radii $\gtrsim 100\,$kpc, while being smoothed in time over $\gtrsim$\,Gyr (the CR lifetime) so they do not trace current AGN but rather some much smoother time-and-population-and-space-average. If the satellites are the ultimate source of the giant halo CRs, we note that the projected surface brightness would trace the satellite distribution (as observed), the spectral index would be weakly-dependent on radius (as observed) and reflect the emission-weighted average of the younger CRs from each mini-mini-halo (also as observed), and the total giant halo luminosity (relative to NGC 1275+mini-halo) would be similar to the total satellite source injection/radio luminosity relative to NGC 1275 ($\sim 10\%$, also as observed). This is distinct from some other giant radio halos which are much brighter at higher CR energies ($\gtrsim$\,GHz), as in e.g.\ Coma, and appear to be associated with phenomena like mergers.

The CR-IC scenario, as discussed in \papertwo-\paperthree, provides an immediate explanation for a number of outstanding puzzles in Perseus and similar SCCs. Not only does it explain the minihalo and (potentially) giant radio halo (without any fine-tuned models for reacceleration or hadronic injection), and the quasi-universal X-ray density/entropy/pressure profiles, and the otherwise anomalous (high-temperature) harder X-ray emission, but it also naturally explains why the classical cooling flow mass deposition rate ($\dot{M}_{\rm cool} \sim 500-1000\,{\rm M_{\odot}\,yr^{-1}}$) is so much larger (by a factor $\sim 100$ in Perseus) than any directly observed gas cooling or lower-temperature gas reservoir or star formation would allow, and why the apparent cooling luminosity $\sim 10^{45}\,{\rm erg\,s^{-1}}$ (from $\sim 100\,$kpc) is so similar to the apparent jet/cavity power and $\gamma$-ray luminosity and scales so tightly with the compact radio luminosity (from $\ll$\,kpc; see \paperthree\ for details). These are all just different manifestations of the same leptonic source, rather than some fine-tuned balance between thermal cooling and AGN heating. It also explains the observed deviations between mass/potential models and metallicities estimated via X-rays versus other methods at radii well inside the SCC ($\ll 100\,$kpc).

\subsection{Model Assumptions \&\ Differences from Previous Models}
\label{sec:conclusions:assumptions}

The key assumptions in our models are: 
\begin{itemize}

\item The total effective leptonic injection rate $\dot{E}_{\rm cr,\,\ell}$ from the NGC 1275 AGN+jet+cavity system is something like $\sim 10^{45}\,{\rm erg\,s^{-1}}$, as indicated by standard models of the compact radio core, jets, AGN and $\gamma$-rays \citep{bottcher:2013.blazar.modeling.almost.all.blazars.better.fit.by.leptonic.cr.models.not.hadronic,tavecchio:2014.jet.leptonic.luminosity.ngc.1275.2e45,keenan:2021.jet.leptonic.power.1e41to1e45.easily.produced.from.modest.agn.bursts.or.steady.jets,hodgson:2021.perseus.jet.minimum.kinetic.luminosity.gamma.rays,foschini:2024.blazar.agn.jet.power.favor.leptonic.large.power.energy.much.more.than.kinetic.lobe.cavity.power}. 

\item The magnetic field in the volume-filling, X-ray emitting gas at $\sim 100\,$kpc is something like $0.5-5\,{\rm \mu G}$, as indicated by present upper and lower limits (from RMs, kinematics, and prior radio+X-ray constraints; \S~\ref{sec:B} \&\ Fig.~\ref{fig:B}; \citealt{taylor:2006.perseus.central.RM.to.core.radio.emission,sanders:2007.perseus.profiles.claimed.hardxr.cr.vs.thermal,hitomi:2018.perseus.temperature.structure.hard.emission,svensmark:2021.perseus.cluster.mass.reconstruction.anisotropy.marginalization.spatial.2d.info,vanweeren:2024.perseus.giant.radio.halo.filled.electrons.just.tiny.fraction.high.energy,creech:2024.nustar.perseus.obs,kam:2026.perseus.faraday.rotation.profiles.all.nuclear.little.large.scale}).

\item The ``effective'' transport speeds ($v_{\rm st} + v_{\rm adv} + \kappa/R + ...$) for CRs are something like $\sim 40-200\,{\rm km\,s^{-1}}$ at $\sim$\,GV in the diffuse gas at $\sim 100\,$kpc, order-of-magnitude similar to observed values in the ISM and CGM \citep{karwin:2019.fermi.m31.outer.halo.detection,recchia:2021.gamma.ray.fermi.halos.around.m31.modeling,butsky:2022.cr.kappa.lower.limits.cgm,hopkins:2025.crs.inverse.compton.cgm.explain.erosita.soft.xray.halos}, as well as observed buoyant/bulk transport speeds within the SCC \citep{hitomi:2018.perseus.core.turbulence.temperatures,kempski:2023.cr.bouyancy.instability.streaming.driving.motions.in.clusters,hlavacek.larrondo:2025.xrism.perseus.preview}, and within an factor of a few of their \Alf\ speeds.

\end{itemize}

With these conditions met, we have shown that most of our qualitative predictions are not extremely sensitive to the details of the underlying ``true'' $B$, $n$, $T$, or $Z$, the source/injection spectrum of injection/acceleration volume, how exactly the transport speeds scale with radius or plasma properties, or the presence of turbulent reacceleration or streaming losses. Where they are more sensitive, these parameters often enter in a degenerate fashion, so with modest exploration of parameter space many distinct but qualitatively similar solutions are plausible.

As discussed in detail in \S~\ref{sec:differences}, these assumptions explain why (1) turbulent re-acceleration is neither necessary nor significant in the models here, to explain the radio mini-halo (and even giant halo); (2) previous claims of strong upper limits IC in X-rays (soft and hard), which assumed qualitatively different CR spectra than the models above predict, could not distinguish the extremely thermal-like CR-IC spectra predicted; and (3) prior claims of limits to non-thermal pressure (from $\gamma$-rays or turbulent or kinematic constraints) do not apply to the models here (and those measurements are well-reproduced).

We stress that we considered intentionally extremely-simple models. We assumed spherical symmetry, time-steady-state with a constant $\dot{E}_{\rm cr,\,\ell}$, constant $\kappa$ and $v_{\rm st}$ in space and time, and simple empirical fits for the gas properties. If we relaxed these assumptions, it gives many new degrees of freedom. For example, given a more realistic time-and-space variable CR injection rate and CR transport physics, one expects a multi-age distribution of CRs to reside at a given radius (just like one would expect more than a single gas temperature in a multi-phase gas system). We could then fit the observed spectra to effective ``multi-temperature CR-IC'' plus multi-temperature gas models (effectively, fitting a model for the best-fit CR spectrum plus gas phases as a function of position to the X-ray and radio multi-wavelength maps simultaneously). Given the similarity of the simple models here to the observed spectra already, and the already-known degeneracies fitting multi-wavelength and/or multi-component spectra to any of these datasets \citep[see][]{matsushita:2002.m87.virgo.cluster.obs.metallicity.drop.temperature.fitting.challenges,mazzotta:2004.xray.temperature.measurement.modeling.and.caveats,avestruz:2014.cluster.mocks.from.sims.sensitivity.temperature.measurements,vijayan:2022.cluster.cgm.multitemperature.fit.challenges,zhuhone:2023.cluster.temperature.fitting.sensitivities.simulations}, it is almost trivially guaranteed that one could find formally ``good fits'' throughout Perseus with such an approach. The problem is that fitting the true degrees-of-freedom we expect (a multi-component CR spectrum with arbitrary spectral curvature and sum-of-components plus a multi-phase gas distribution with arbitrary temperature and density distributions) is highly degenerate, so the best fit solutions would almost certainly be non-unique. 

\subsection{Alternative Models and Future Work}
\label{sec:conclusions:future}

We emphasize that nothing we have argued strictly rules out the ``traditional'' (no CR-IC) interpretation of the Perseus SCC, but it is challenging to reconcile the full ensemble of observations with that scenario for reasons we review here. The cleanest way to recover that limit is to make the following assumptions. 
(1) Assume order-of-magnitude faster CR transport speeds $v_{\rm st} \gg 1000\,{\rm km\,s^{-1}}$ outside the AGN (the CR source luminosity of NGC 1275 must be broadly similar to explain the AGN and $\gamma$-ray properties), so that CRs escape rapidly and maintain a much lower $e_{\rm cr}$ and therefore much weaker CR-IC (Fig.~\ref{fig:xrspec}-\ref{fig:hardXR}).
(2) To explain the radio with much lower $e_{\rm cr}$, assume factor of $\gtrsim 5-10$ stronger magnetic fields $B$ everywhere.
This will create tension with observed Faraday rotation measurements (one must assume that, despite being measured in the sub-pc radio core, the RMs measured actually come from gas at $\sim 100\,$kpc, with plasma $\beta \sim 1$ around the mini-halo radius $\sim 100\,$kpc; see Fig.~\ref{fig:B}), and with the non-thermal pressure upper limits from potential/hydrostatic/kinematic constraints, but these tensions are not impossible to reconcile. 
(3) Re-introduce strong re-acceleration to explain the giant radio halo (Fig.~\ref{fig:radio}). This would require something like $\kappa$ being very low, despite streaming speeds $v_{\rm st}$ being very large, so the CR transport would have to be some unusual form of sub-diffusion or advection. 

On top of this, without CR-IC, there is of course the usual cooling flow problem in the X-rays with $99\%$ of the apparent cooling luminosity at $\sim 100-200\,$kpc being offset by an AGN at $\ll 1\,$kpc without producing much visible cooling; some other physics must explain how the CF sets up the quasi-universal apparent soft-X-ray density $n_{X}$ and temperature $T_{X}$ profiles; additional new physics like dust depletion (i.e.\ large dust populations in $\sim 10^{8}\,{\rm K}$ gas) must be invoked to explain why the X-ray-inferred metallicities $Z_{X}$ are suppressed relative to UV/optical metallicities at $\ll 100\,$kpc (Fig.~\ref{fig:xr.profiles.Z}; \citealt{panagoulia:2015.cluster.metal.profiles.with.drops,liu:2019.cluster.z.drop.compilation.model.discussion,ng:2024.cluster.detailed.chandra.profiles.coolcore.zdrops.sizes.profiles.nonuniversal.norm}); the ``intermediate'' hard X-ray components require a small, undetected-at-other wavelengths component of highly super-virial $\sim 20-60\,$keV gas (Fig.~\ref{fig:hardXR}); reacceleration models must be fine-tuned (fitting numbers like $\kappa(R)$) to give the correct halo extent while also explaining why the radio spectrum softens from $1-100\,$kpc instead of hardening (the expectation for a re-acceleration-dominated spectrum; \S~\ref{sec:differences:reaccel}); and different physics such as shocks must be invoked to explain the higher-frequency component and/or giant halo, whose similarity to the galaxy profile observed is coincidental in this case (Fig.~\ref{fig:satellites}); and the observed strong deviations of the central kinematics and X-ray mass/potential reconstruction must still imply the gas is highly out-of-hydrostatic equilibrium, but in this case without any other source of pressure (Fig.~\ref{fig:massmodel}). All of this is, in principle, still plausible. But an appealing aspect of the CR-IC models here is that all of this phenomenology instead has a single, simple physical explanation.

In a companion paper, we will investigate another prediction of the models here: ionization and excitation of neutral (atomic/molecular) gas within the apparent SCC. Owing to the large implied CR ``bath'' in the models here, strong ionization and excitation is predicted, and indeed this is observed \citep{johnstone:2007.warm.h2.emission.in.clusters.ccs,ferland:2008.molecular.emission.in.cooling.flow.filaments.shielding.excitation.models,ferland:2009.particle.ionization.needed.for.molecular.line.emission.in.cc.perseus,salome:2011.extended.molecular.gas.around.perseus.cluster.core,vantyghem:2017.13co.detection.molecular.gas.clusters.direct.column.and.mass}, consistent with the other observational phenomena discussed here, and challenging to understand in a model without significant CR-IC.

Perseus is typical of many SCC clusters, but there are regimes where we might expect qualitatively different behaviors. First is at high redshifts. Because the CMB energy density scales $\propto (1+z)^{4}$, if CR streaming speeds are universal/constant and injection is leptonic (streaming and diffusing from a central source), then CR-IC halos would become systematically brighter in the emission frame (similar brightness \textit{observed}), but also more compact, at higher-redshift, with the effects becoming potentially noticeable as early as $z \gtrsim 0.5-1$ or so. This could be very important for the extreme SCCs known at intermediate redshifts, like H1821+643 
and the Phoenix cluster \citep{walker:2014.case.study.cluster.with.extremely.luminous.2e47.qso.super.low.central.entropy.looks.like.very.strong.coolingflow.central.100kpc.but.modeling.quasar.as.causing.cooling.with.thermal.leads.to.factor.30.discrepancy,mcdonald:2019.most.relaxed.spt.clusters.xray.profiles.properties}, 
$z\gtrsim 1$ (e.g.\ 3C 186, SPT J2215+3537, and others; \citealt{siemiginowska:2010.3c186.radio.loud.z1.cluster.xrays.profiles,santos:2012.z1.cluster.xray.coolcore.profiles,girardini:2021.evolution.cluster.profiles.highz.with.uncertainties,calzadilla:2023.highz.relaxed.scc.cluster.starburst.radio.galaxy.xray.profiles}), and even $z\gtrsim 2$ (Spiderweb, IDCS J1426.5+3508, MQN01, XLSSC 122; \citealt{brodwin:2016.massive.highz.cluster.profile.xray,mantz:2018.xxl.clusters.highz.xray.profiles.z2.cluster,lepore:2024.spiderweb.cluster.properties.highz.emission.from.jets.cr.ic,travascio:2025.hyperluminous.z3.qso.cr.ic.evidence,duffy:2022.xxl.cluster.xray.profiles.highz})
which seem to show very high-surface-brightness SCCs (comparable brightness observed independent of redshift, as predicted for CR-IC, see \paperthree)  with sometimes sharp edges and apparent gradients in X-ray properties (like $Z_{X}$) and more compact central peaks at the highest redshifts (see \paperthree).

At the opposite extreme are WCCs, including systems like Virgo and Centaurus (as well as smaller groups like HGC 62). These have orders-of-magnitude lower X-ray inferred cooling and mass deposition rates than Perseus, and show essentially no evidence for any neutral or multi-phase gas within the SCC at any radii outside the BCG nucleus ($\gtrsim 1\,$kpc; \citealt{salome:2016.centaurus.molecular.atomic.gas.low.sfe,boselli:2019.virgo.ionized.gas.halpha.filament.measurements.low.mass.sfr,boizelle:2025.m87.co.survey.very.strong.upper.limits.almost.no.molecular.gas}). But they still appear to show some of the signs of a CR-IC ``boosted'' CC: they have strong central radio sources (albeit orders-of-magnitude less luminous than NGC 1275); their central X-ray-inferred metallicities are  much smaller than those measured in UV/optical at the same radii \citep{russell:2018.m87.inner.kpc.multi.temp.and.strong.agn.maps.with.zdrop.many.signatures.of.cr.ic,xrism:2025.m87.xray.temperature.metallicity.velocity.central.regions.projected.estimate}; they show an apparent rising X-ray surface brightness \citep{churazov:2008.virgo.mass.profile.kinematics.vs.xrays.as.measurement.of.nonthermal.pressure,mccall:2024.erosita.virgo.profiles.xr.sb.along.octants}; and they have ultra-steep radio ``mini-mini-halos'' around the BCG \citep{mtshweni:2022.meerkat.ghz.images.spectral.indices.m87,wu:2025.m87.radio.multifrequency.maps.and.spectral.hardening.strongly.outside.core,degasperin:2025.virgo.50.to.1500.mhz.radio.halo.images}. This suggests similar CR processes may be occurring there too, but at an overall reduced level.

\subsection{Future Observational Tests}
\label{sec:conclusions:obs}

We have attempted to compile extensive multi-wavelength Perseus diffuse-gas measurements, to test the models here. Of course, future measurements could further constrain and inform the models here, and potentially definitively rule in or out proposals here like the role of CR-IC in X-ray emission, aging CRs in the minihalo, or satellite mini-mini halos contributing to the giant halo. Here we summarize the most straightforward tests proposed in the text.

\begin{enumerate}

\item Better \textit{spatial} and \textit{spectral}-resolved X-ray measurements that can obtain detailed de-projected multi-phase (multi-$Z_{X}$, $T_{X}$) measurements well within the SCC, of the sort possible with future instruments like ATHENA, could be very constraining \textit{if} coupled to new independent measurements of gas-phase $Z \approx Z_{\rm true}$ from non X-ray (radio/IR/optical/UV) measurements. Gas clearly exists out to $\gtrsim 30\,$kpc in phases (ionized $\sim 10^{4-5}$\,K, neutral, molecular) where such measurements are conceivable. The variant models here differ widely in their predicted behaviors for $Z_{X}$ given some true $Z_{\rm true}$, so this combination is particularly powerful for breaking those degeneracies, with $Z_{X}(R) \lesssim Z_{\rm true}(R)$ strongly favoring significant non-thermal emission.

\item Spatially-resolved \textit{low-frequency} radio at $\sim 1-20\,$MHz would enable one to directly probe the same electrons responsible for X-ray CR-IC. The X-rays are sensitive to the spectrum of $\sim 0.3-2\,$GeV leptons, while current radio with required spatial resolution and sensitivity (to map diffuse emission within the core, away from the jets/lobes) corresponds to $\sim 4-30$\,GeV leptons (while reaching $\sim 1-20\,$MHz would probe $\sim1$\,GeV leptons). Then one could directly ask whether the spectral shapes are potentially consistent with CR-IC contributing significantly to the X-ray continuum, and how the local variations in emissivity do or do not correlate within the diffuse CC gas. To avoid significant CR-IC emission, while reproducing existing radio, there would need to be a sharp cutoff in the radio spectrum between those frequencies and the frequencies already observed (e.g.\ just below $\sim 70\,$MHz). 

\item As noted above, a pure-thermal X-ray explanation would require larger gas densities and order-of-magnitude larger $B \gtrsim 10\,{\rm \mu G}$ within the outer CC ($\sim 30-100\,$kpc), in order to still explain the radio (for any reasonable CR spectrum). As such, \textit{independent} Faraday rotation measurements from \textit{background} sources, like fast radio bursts (FRBs), with impact parameters within the inner CC $\ll 100\,$kpc of Perseus, would be especially valuable. Given the gas density profiles implied by the standard pure-thermal fits, plus the minimum $B$ required to reproduce the radio mini-halo spectrum, pure-thermal models would imply $|{\rm RM}(R \sim 30-80\,{\rm kpc})| \gtrsim 30,000\,{\rm rad\,m^{-2}}$. Upper limits to RMs below this would strongly favor ACRH models like those here, and detections or upper limits below $\lesssim 1000\,{\rm rad\,m^{-2}}$, like the $\lesssim 110\,{\rm rad\,m^{-2}}$ found in the one measured FRB RM in a cluster (in Abell 2311; \citealt{connor:2023.frb.foreground.cluster.outskirts}), would definitively rule out ``thermal only'' models for the soft X-rays (given the observed radio and hard X-rays). Lacking background sources, improved measurements of the spatial and temporal variability of RMs from the cluster radio core and jets can improve the RM upper limits from the diffuse gas (with the upper limit from diffuse gas at $\gg $\,kpc corresponding to the minimum RMs measured in space-and-time across the jets being monitored). In Virgo, for example, this has shown that the RM from the diffuse CC gas must be $\lesssim 130\,{\rm rad\,m^{-2}}$ \citep{park:2019.m87.rms.detailed.modeling.huge.spatial.temporal.datasets.strong.upper.limits.outside.bondi.radius}, an order of magnitude smaller than the contribution from the (rapidly space-and-time varying) jet core+sheath. 

\item Future SZ measurements resolving the CC would provide a very clear and robust distinguishing method between models with significant CR-IC (more low-energy leptons) and those without, and place a robust prior on the true thermal pressure $P_{\rm true} \approx P_{\rm SZ}$. As discussed above however, this is challenging, as it requires a combination of high resolution and dramatic improvements in sidelobe control to avoid contamination by the core of NGC 1275. Future samples of fast radio burst (FRB) or dispersion measure (DM) measurements probing the central CC could accomplish the same (measuring the true column $N_{\rm H,\,true}$) without contamination concerns.

\item Future high-energy constraints from $\gamma$-rays and very hard X-rays can also be a powerful constraint. However especially for $\gamma$-rays, which could most strongly constrain the hadron-to-lepton ratios, or leptonic spectrum near $\sim$\,GeV, directly testing the predictions here would require orders-of-magnitude improvements in angular resolution (to few-arcsecond) and background control/extended low-surface brightness sensitivity (ideally a factor $\sim 200-1000$ improvement over GeV Fermi backgrounds), with peak sensitivity between $\sim 10-1000\,$MeV. 

\item Measurements of CR ionization of dense atomic/molecular gas within the cooling filamentary structure in the Perseus CC can provide a powerful and unique bolometer of the pressure and number density of precisely the low-energy CRs that matter for CR-IC and for the ACRH and aging diffuse synchrotron minihalo. As noted above, making quantitative predictions for these measurements will be the subject of future work (in preparation).

\end{enumerate}

All of these represent important future topics for study.

\begin{acknowledgements}
We thank Eliot Quataert, Peng Oh, Daisuke Nagai, and Irina Zhuravleva for helpful discussions. Support for PFH was provided by a Simons Investigator Grant.
\end{acknowledgements}

\bibliographystyle{mn2e}
\bibliography{ms_extracted}

\clearpage

\end{document}